\documentclass{PoS}

\newcommand{\z}{&&\hspace*{-1cm}}

\newcommand{\bea}{\begin{eqnarray}}
\newcommand{\eea}{\end{eqnarray}}
\newcommand{\be}{\begin{equation}}
\newcommand{\ee}{\end{equation}}

\newcommand{\ar}{\overline a_s}

\title{
Parton 
distributions 
at low $x$ and gluon- and quark average multiplicities
}

\ShortTitle{
Parton 
distributions
}

\author{\speaker{Anatoly Kotikov}
         \thanks{The work was supported in part by RFBR grant 
No. 13-02-01005-a.
}\\
        BLThPh, Joint Institute for Nuclear Research, Dubna\\
        E-mail: \email{kotikov@theor.jinr.ru}}

\abstract{We
shown the general approach for $Q^2$ evolution of parton 
densities and fragmentation functions at low $x$ based on the diagonalization.
The diagonalization leads to the two components in the $Q^2$ evolution,
each of which contains a nonperturbative parameter. 
The values of the parameters can be found by fits of the experimental data
for the deep-inelastic scattering structure function $F_2$ and for average jet 
multiplicities.
One of the components
contains the all large logarithms $\ln(1/x)$ and produce 
the basic contribution at small $x$ region. The second one is regular 
at low $x$ but its contribution is very important to have a good agreement 
with experimental data.\\

}

\FullConference{XXII International Baldin Seminar on High Energy Physics Problems,\\
		15-20 September 2014\\
		JINR, Dubna, Russia}

\begin{document}

\section{Introduction}

The evaluation of the cross-sections for hadron-hadron iteractions needs the 
sufficiently precise knowledge
%ruther accurate knowledgement 
of parton distribution functions (PDFs) and
parton fragmentation functions (FFs), which are the important part of any
%in their 
cross-section. The properties of PDFs and FFs 
%theirselves 
can be taken from processes of the deep-inelatic scattering (DIS) 
and $e^+e^-$-collisions,
respectively. In this report we will concentrate only for the high-energy 
limits  of PDFs and FFs, which are needed
%important 
for modern experiments studied on LHC collider.

\subsection{PDFs}
The experimental data from
HERA on the DIS
%deep-inelastic scattering (DIS) 
structure function
(SF) $F_2$ \cite{H197}-\cite{Aaron:2009aa},
%\cite{H197,ZEUS01,Aaron:2009aa},
%and 
its derivative
%s $\partial F_2/\partial \ln(Q^2)$ \cite{H197,Surrow} and 
$\partial \ln F_2/\partial \ln(1/x)$ \cite{Surrow}
%-\cite{DIS02} 
%\cite{H1slo,Surrow,DIS02} 
and the heavy quark parts $F_2^{cc}$ and $F_2^{bb}$ 
\cite{Collaboration:2009jy}
%-\cite{Lipka:2009zza}
enable us to enter into
a very interesting kinematical range for
testing the theoretical ideas on the behavior of quarks and gluons carrying
a very low fraction of momentum of the proton, the so-called small-$x$
region. In this limit one expects that 
the conventional treatment based on the
Dokshitzer--Gribov--Lipatov--Altarelli--Parisi (DGLAP) equations \cite{DGLAP}
does not account for contributions to the cross section which are
leading in $\alpha_s \ln(1/x)$ and, moreover, the parton 
%distribution unction (PDFs),
densities
% (PD), in particular the gluon ones, 
are becoming large and need to develop a 
high density formulation of QCD.
 However, the
reasonable agreement between HERA data and the next-to-leading-order (NLO) and
next-to-next-to-leading-order (NNLO)
approximations of
perturbative
QCD has been observed for $Q^2 \geq 2$ GeV$^2$ (see reviews in \cite{CoDeRo}
and references therein) and, thus,
perturbative QCD could describe the
evolution of $F_2$ and its derivatives
%structure functions
up to very low $Q^2$ values,
traditionally explained by soft processes.
%It is of fundamental importance to find out the kinematical region where
%the well-established perturbative QCD formalism
%can be safely applied at small $x$.

The standard program to study the $x$ behavior of
quarks and gluons
is carried out by comparison of data
with the numerical solution of the DGLAP
%Dokshitzer-Gribov-Lipatov-Altarelli-Parisi (DGLAP)
equation
\cite{DGLAP}\footnote{ At small $x$ there is another approach
based on the Balitsky--Fadin--Kuraev--Lipatov (BFKL) equation 
\cite{BFKL}, whose application 
%will be dicussed below in Appendix A.
is out of the scope of this work. 
} by
fitting the parameters of the PDF
$x$-profile 
%of partons 
at some initial $Q_0^2$ and
the QCD energy scale $\Lambda$ \cite{fits}-\cite{Kotikov:2010bm}.
%\cite{fits,GRV,Ourfits} [KKS].
However, for analyzing exclusively the
%small
low-$x$ region, there is the alternative of doing a simpler analysis
by using some of the existing analytical solutions of DGLAP evolution
in the 
%small
low-$x$
limit \cite{BF1}--\cite{HT}.
This was done so in \cite{BF1}
where it was pointed out that the HERA small-$x$ data can be
interpreted in 
terms of the so-called doubled asymptotic scaling (DAS) phenomenon
related to the asymptotic 
behavior of the DGLAP evolution 
discovered many years ago \cite{Rujula}.

The study of \cite{BF1} was extended in \cite{Munich,Q2evo,HT}
to include the finite parts of anomalous dimensions
of Wilson operators
% and Wilson coefficients
\footnote{ 
In the standard DAS approximation \cite{Rujula} only the singular
parts of the anomalous dimensions were used.}.
This has led to predictions \cite{Q2evo,HT} of the small-$x$ asymptotic PDF
form 
%of parton distributions (PD)
in the framework of the DGLAP dynamics
%equation 
starting at some $Q^2_0$ with
the flat function
 \begin{eqnarray}
f_a (Q^2_0) ~=~
A_a ~~~~(\mbox{hereafter } a=q,g), \label{1}
 \end{eqnarray}
where $f_a$ are the parton distributions multiplied by $x$
and $A_a$ are unknown parameters to be determined from the data.\\

\subsection{FFs and average jet multiplicities}
Collisions of particles and nuclei at high energies usually
produce many hadrons
and their production 
%of hadrons 
is a typical process where  nonperturbative phenomena are involved.
However, for particular observables, this problem can be avoided.
In particular, the {\it counting} of hadrons in a jet that is initiated at a
certain scale $Q$ belongs to this class of observables.
%In this case, one can adopt with quite high accuracy the hypothesis of Local
%Parton-Hadron Duality (LPHD), which simply states that parton distributions are
%renormalized in the hadronization process without changing their shapes
%\cite{Azimov:1984np}.
Hence, if the scale $Q$ is large enough, this would in principle allow
perturbative QCD to be predictive without the need to consider phenomenological
models of hadronization.
Nevertheless, such processes are dominated by soft-gluon emissions, and it is a
well-known fact that, in such kinematic regions of phase space, fixed-order
perturbation theory fails, rendering the usage of resummation techniques
indispensable.
As we shall see, the computation of average jet multiplicities (AJMs) indeed requires
small-$x$ resummation, as was already realized a long time ago
\cite{Mueller:1981ex}.  
In Ref.~\cite{Mueller:1981ex}, it was shown that the singularities for
$x\sim 0$, which are encoded in large logarithms of the kind $1/x\ln^k(1/x)$, 
spoil perturbation theory, and also render integral observables in $x$
ill-defined, disappear after resummation.
Usually, resummation includes the singularities from all orders according to a
certain logarithmic accuracy, for which it {\it restores} perturbation theory.

Small-$x$ resummation has recently been carried out for timelike splitting
fuctions in the $\overline{\mathrm{MS}}$ factorization scheme, which is
generally preferable to other schemes, yielding fully analytic expressions.
In a first step, the next-to-leading-logarithmic (NLL) level of accuracy has
been reached \cite{Vogt:2011jv,Albino:2011cm}.
In a second step, this has been pushed to the
next-to-next-to-leading-logarithmic (NNLL), and partially even to the
next-to-next-to-next-to-leading-logarithmic (N$^3$LL), level \cite{Kom:2012hd}.
Thanks to these results, we were able in   \cite{Bolzoni:2012ii,Bolzoni:2013rsa}
to analytically compute the NNLL
contributions to the evolutions of the 
%average 
gluon and quark AJMs
%jet multiplicities 
with normalization factors evaluated to NLO
and approximately to next-to-next-to-next-to-order (N$^3$LO) in the
$\sqrt{\alpha_s}$ expansion.
The previous literature contains a NLL result on the small-$x$ resummation of
timelike splitting fuctions obtained in a massive-gluon scheme.
Unfortunately, this is unsuitable for the combination with available
fixed-order corrections, which are routinely evaluated in the
$\overline{\mathrm{MS}}$ scheme.
A general discussion of the scheme choice and dependence in this context may
be found in Refs.~\cite{Albino:2011bf}.

The 
%average 
gluon and quark AJMs,
%jet multiplicities, 
which we denote as  
$\langle n_h(Q^2)\rangle_{g}$ and $\langle n_h(Q^2)\rangle_{q}$, respectively,
represent the average numbers of hadrons in a jet initiated by a gluon or a
quark at scale $Q$.
In the past, analytic predictions were obtained by solving the equations for
the generating functionals in the modified leading-logarithmic approximation
(MLLA) in Ref.~\cite{Capella:1999ms} through N$^3$LO in the expansion
parameter $\sqrt{\alpha_s}$, i.e.\ through $\mathcal{O}(\alpha_s^{3/2})$.
However, the theoretical prediction for the ratio
$r(Q^2)=\langle n_h(Q^2)\rangle_g/\langle n_h(Q^2)\rangle_q$ given in
Ref.~\cite{Capella:1999ms} is about 10\% higher than the experimental data at
the scale of the $Z^0$ boson.
%, and the difference with the data becomes even
%larger at lower scales, although the perturbative series seems to converge
%very well.
An alternative approach was proposed in Ref.~\cite{Eden:1998ig}, where a
differential equation for the 
%average 
gluon-to-quark 
%jet multiplicity 
AJM ratio was
obtained in the MLLA within the framework of the colour-dipole model, and the
constant of integration, which is supposed to encode nonperturbative
contributions, was fitted to experimental data.
A constant offset to the 
%average 
gluon and quark AJMs
jet multiplicities 
was also introduced in Ref.~\cite{Abreu:1999rs}. 

Recently, we proposed a new formalism \cite{Bolzoni:2012ed,Bolzoni:2012ii,Bolzoni:2013rsa} that
solves the problem of the apparent good convergence of the perturbative series
and does not require any ad-hoc offset, once the effects due to the
mixing between quarks and gluons are fully included. 
Our result is a generalization of the result obtained in
Ref.~\cite{Capella:1999ms}.
In our new approach, the nonperturbative informations to the gluon-to-quark AJM
%jet multiplicity 
ratio are encoded in the initial conditions of the evolution
equations. \\
%Motivated by the excellent agreement of our results with the experimental data
%found in Ref.~\cite{Bolzoni:2012ii}, we proposed in  \cite{Bolzoni:2013rsa}
%%here 
%to also use our approach
%to extract the strong-coupling constant $\alpha_s(Q_0^2)$ at some reference
%scale $Q_0$ and thus extend our analysis by adding an apropriate fit parameter.

%\subsection{Structure of the review}
This contribution
%The paper 
is organized as follows.  Section 2 contains 
general formulae for the $Q^2$-evolution of PDFs and FFs. The 
generalized DAS approach is presented in Section 3. Sections 4 and 5 contain
basic formulae of $Q^2$-dependence of FFs at low $x$ and the AJMs,
%average jet multiplicities, 
respectively.
In Section 6 we compare our formulae with the experimental data for the DIS 
SF $F_2$ and the AJMs
%average jet multiplicities
%experimental data 
and present
%discuss 
the obtained results.
Some discussions can be found in the conclusions.
The procedure of the diagonalization and its results for PDF and SF Mellin moments
%Some preliminary results accounting for BFKL corrections in our analysis 
can be found  in AppendixA.

\section{ Approach }

Here we breifly touch on some points concerning theoretical part of our analysis.
%For a bit detailed account see~\cite{KK2001}.

%%%%%%%%%%%%%%%%%%%%%%%
\subsection{Strong coupling constant}

The strong coupling constant is determined from the renormalization group equation. 
Moreover,
%Note here that 
the perturbative coupling constant $a_s(Q^2)$ is different at
the leading-order (LO), NLO and NNLO approximations. 
Indeed, from the renormalization group equation
we can obtain the following equations for the coupling constant
%\begin{subequations}
%\label{as:LO&NLO}
\begin{eqnarray}
 \frac{1}{a_s^{\rm LO}(Q^2)} -\frac{1}{a_s^{\rm LO}(M_Z^2)} \, = \, \beta_0 
% \ln{\left(\frac{Q^2}{\Lambda^2_{\rm LO}}\right)}
\ln{\left(\frac{Q^2}{M_Z^2}\right)}
\label{as:LO} 
\end{eqnarray}
at the LO approximationm and
\begin{eqnarray}
 \frac{1}{a_s^{\rm NLO}(Q^2)} - \frac{1}{a_s^{\rm NLO}(M_Z^2)} \, + \,
% \frac{\beta_1}{\beta_0} 
b_1 \ln{\left[
 %\frac{\beta_0^2 a_s(Q^2)}{\beta_0+ \beta_1 a_s(Q^2)}\right]} \, = \, 
\frac{a_s^{\rm NLO}(Q^2)(1+ b_1 a_s^{\rm NLO}(M_Z^2))}{
a_s^{\rm NLO}(M_Z^2)(1+ b_1 a_s^{\rm NLO}(Q^2))}\right]} \, = \, 
\beta_0 \ln{\left(\frac{Q^2}{M_Z^2}\right)}
% \beta_0 \ln{\left(\frac{Q^2}{\Lambda^2}\right)}
\label{as:NLO}
\end{eqnarray}
at the NLO approximation.
%%\end{subequations}
%Usually at the NLO level ${\rm \overline{MS}}$-scheme is used, so we apply
%$\Lambda = \Lambda_{\rm \overline{MS}}$ below.
%%in the Eqs.~(\ref{an:NLO}) and (\ref{as:NLO}).

At NNLO level $a_s^{\rm NNLO}(Q^2) \equiv a_s(Q^2)$
%the latter 
is more complicated and it is given by
%\bea
\be
\frac{1}{a_s(Q^2)} - \frac{1}{a_s(M_Z^2)} +
%&+& 
b_1 \ln{\left[\frac{a_s(Q^2)}{a_s(M_Z^2)}
\sqrt{\frac{1 + b_1a_s(M_Z^2) + b_2 a_s^2(M_Z^2)}
{1 + b_1a_s(Q^2) + b_2 a_s^2(Q^2)}}
\right]}
% \nonumber \\ \nonumber &+& 
+ \left(b_2-\frac{b_1^2}{2}\right)\cdot I = \beta_0 \ln{\left(\frac{Q^2}{M_Z^2}\right)}\,.
\label{1.co}
\ee
%\eea
The expression for $I$ looks:
$$
I=\cases{
\displaystyle{\frac{2}{\sqrt{\Delta}}} \left(\arctan{\displaystyle{\frac{b_1+2b_2a_s(Q^2)}{\sqrt{\Delta}}}} 
-\arctan{\displaystyle{\frac{b_1+2b_2a_s(M_Z^2)}{\sqrt{\Delta}}}}\right)&for $f=3,4,5; \Delta>0$,\cr
\displaystyle{\frac{1}{\sqrt{-\Delta}}}\ln{\left[
\frac{b_1+2b_2a_s(Q^2)-\sqrt{-\Delta}}{b_1+2b_2 a_s(Q^2)+\sqrt{-\Delta}}
\cdot
\frac{b_1+2b_2 a_s(M_Z^2)+\sqrt{-\Delta}}{b_1+2b_2 a_s(M_Z^2)-\sqrt{-\Delta}} \right]}&for $f=6;\quad\Delta<0$, \cr
}
$$
where $\Delta=4b_2 - b_1^2$ and $b_i=\beta_i/\beta_0$ are read off from the QCD $\beta$-function:
\be
\beta(a_s) ~=~ -\beta_0 a_s^2 - \beta_1 a_s^3 - \beta_2 a_s^4 +\ldots =  -\beta_0 a_s^2 \Biggl(1+
b_1 a_s + b_2 a_s^2 +\ldots \Biggr) \, ,
\label{alpha}
\ee
where
%\begin{eqnarray}
\begin{equation}
\beta_0 = 11 - \frac{2}{3} f, ~~ \beta_1 = 102 - \frac{38}{3} f, ~~
\beta_2 = \frac{2857}{2} - \frac{5033}{18} f + \frac{325}{54} f^2,
\label{alpha1}
\end{equation}
%\end{eqnarray}
with 
%$C_A=3$, $C_F=4/3$, and $T_F=n_f/2$ being colour factors and $n_f$ 
$f$ being the
number of active quark flavours.

%A normalization point is taken at the gauge boson peak $M_Z^2$. There is a number of reasons behind that choice, amongs them:
%fewer heavy quark thresholds have to be crossed over to reach a normalization point,
%and a perturbative approach must be applicable at the value of $Q^2_0$. Besides, 
%impact of higher order corrections derived from PDF normalization conditions is the more negligible the higher normalization point is.

\subsection{PDFs and DIS SF $F_2$}

The DIS SF can be represented as a sum of two terms:
\be
F_2(x,Q^2)= F_2^{NS}(x,Q^2) + F_2^{S}(x,Q^2)\,,
\label{3.aa}
\ee
the nonsinglet (NS) and singlet (S) parts. At this point let's introduce PDFs, the gluon distribution function $f_g(x,Q^2)$
and the singlet and nonsinglet quark distribution functions $f_S(x,Q^2)$ and $f_{NS}(x,Q^2)$~\footnote{Unlike the standard case, here PDFs are multiplied by $x$.}:
\bea
{\bf f}_{S}(x,Q^2) &\equiv& \sum_{i}^{f} {\bf f}_{i}(x,Q^2)= V(x,Q^2) + S(x,Q^2)\,, \nonumber \\ 
{\bf f}_{NS}(x,Q^2) &\equiv& 
%{\bf f}_{NS}^{em}(x,Q^2) =~ 
\sum_{i}^{f} e^2_i \, {\bf f}^{NS}_{i}(x,Q^2),~~
% \nonumber \\
{\bf f}^{NS}_{i}(x,Q^2) = {\bf f}_{i}(x,Q^2)-\frac{1}{f} \,
{\bf f}_{S}(x,Q^2) \, ,
%{\bf u}_v(x,Q^2) - {\bf d}_v(x,Q^2)\,,
\label{3.aa1}
\eea
where 
%$f$ is the number of quark flavors (${\bf u}$p, ${\bf d}$own, ${\bf s}$trange,$\ldots$), 
$V(x,Q^2)={\bf u}_v(x,Q^2)+{\bf d}_v(x,Q^2)$ is the distribution of valence quarks and $S(x,Q^2)$ is
a sum of sea parton distributions set equal to each other.

There is a direct relation between SF moments~ $M_n(Q^2)$ and those of PDFs
\be
M^{j}(n,Q^2) ~=~\int_0^1 dx x^{n-2} F^{j}_2(x,Q^2),~~
{\bf f}_{j}(n,Q^2) ~=~\int_0^1 dx x^{n-2} {\bf f}_{j}(x,Q^2),
~~~(j=NS,S,G),
\label{3.aa2}
\ee
which has
%For example, in the nonsinglet case it looks~\cite{Buras}:
%The moments of the DIS SF $F_2$ have 
the following form
\bea
M_n^{NS}(Q^2) &=& 
%R_{NS}(f)\, 
C_{NS}(n,\ar(Q^2))\cdot {\bf f}_{NS}(n,Q^2)\,,
\label{3.a} \\
M_n^{S}(Q^2) &=& e \,
%R_{S}(f)\, 
\biggl(
C_{S}(n,\ar(Q^2))\cdot {\bf f}_{S}(n,Q^2)
+ C_{g}(n,\ar(Q^2))\cdot {\bf f}_{G}(n,Q^2) \biggr) \,,~~~~~~
\label{3.ab}
\eea
with
%\be
%\ar(Q^2)=\frac{\alpha_s(Q^2)}{4\pi} \label{as}
%\ee
%and 
$C_{j}(n,\ar(Q^2))$ $(j=NS,S,G)$ are the Wilson coefficient 
functions.
The constant $e$
%$R_{NS}(f)$ and $R_{S}(f)$  
depends on weak and 
electromagnetic charges and is fixed for electromagnetic charges
to 
\be
e
%R_{S}(f) 
~=~ \frac{1}{f} \, \sum_{q}^{f} e^2_q \, .
%\frac{1}{6},~~ R_{S}(f) ~=~ \frac{5}{18}
\label{RNS}
\ee
%be one sixth 
%for $f=4$~\cite{Buras}.

Note that the NS and valence quark parts are negledgible at low $x$ and, thus,
%, $F_2^{NS}(x,Q^2) \sim o$ when $x \to 0$, and it
%will be absend below in our analysis, t.e. below 
$F_2^S(x,Q^2) = F_2(x,Q^2)$ and $S(x,Q^2)={\bf f}_{S}(n,Q^2) \equiv {\bf f}_{q}(n,Q^2)$.

%%%%%%%%%%%%%%%%%%%%%%%
\subsection{$Q^2$-dependence of SF moments}

The coefficient functions 
%$C_{NS}^{twist2}(n,\ar(Q^2))$, 
$C_{q}(n,a_s(Q^2)) \equiv C_{S}(n,a_s(Q^2))$ and 
$C_{g}(n,a_s(Q^2))$ are further expressed through
the functions 
%$B^j_{NS}(n)$, 
$B^{(i)}_{q}(n)$ and $B^{(i)}_{g}(n)$, respectively, 
which are known 
exactly~\cite{MVV2005,Buras:1979yt}~\footnote{For the integral and even complex $n$ values, the 
coefficients $B^{(i)}_{a}(n)$ and $Z^{(i)}_{a,b}(n)$ $(a,b=q,g)$ 
can be obtained using the analytic continuation~\cite{Kazakov:1987jk}.}
\be
C_{a}(n, a_s(Q^2)) = 1 - \delta^g_a 
+ a_s \cdot B^{(1)}_{a}(n) 
+ a_s^2 \cdot B_{a}^{(2)}(n) + {\cal O}(a_s^3)\, , ~~~ (a=(q,g)) \,
\label{1.cf} 
\ee
where $\delta^g_a $ is the Kroneker symbol.

The $Q^2$-evolution of the PDF moments can be calculated within a framework 
of perturbative QCD (see e.g.~\cite{Buras:1979yt,Yndu}).
After diagonalization (see Appendix A),
we see that the quark and gluon densities contain the so called $"+"$- and 
$"-"$-components
\be
{\bf f}_{a}(n,Q^2) = {\bf f}_{a}^+(n,Q^2) +  {\bf f}_{a}^-(n,Q^2) ~~~(a=q,g),
\label{3S.0}
\ee
which in-turn evaluated already independently:
\be
{\bf f}_{a}^{\pm}(n,Q^2) = {\bf \tilde{f}}_{a}^{\pm}(n,Q_0^2) 
\cdot \left[\frac{a_s(Q^2)}
{a_s(Q^2_0)}\right]^{\frac{\gamma_{\pm}^{(0)}(n)}{2\beta_0}}
\cdot H_{a}^{\pm}(n,Q^2)\,, % \nonumber \\
\label{3S.1}
\ee
where
\be
\gamma_{\pm}^{(0)}(n) ~=~ \frac{1}{2} \biggl(
\gamma_{qq}^{(0)}(n)+\gamma_{gg}^{(0)}(n) \pm
\sqrt{\Bigl(\gamma_{qq}^{(0)}(n)-\gamma_{gg}^{(0)}(n)\Bigr) +
4 \gamma_{qg}^{(0)}(n)\gamma_{gq}^{(0)}(n) } \biggr)
\label{3S.1}
\ee
is the anomalous dimensions of the $"+"$- and 
$"-"$-components, which are obtained from the elements of the martix of the
LO anomalous  quark and gluon anomalous dimensions.

At LO, the normalization coefficients ${\bf \tilde{f}}_{a}^{\pm}(n,Q_0^2)$
have the form
\be
{\bf \tilde{f}}_{a}^{\pm}(n,Q_0^2) ~=~ {\bf f}_{a}^{\pm}(n,Q_0^2) , 
\label{3S.2}
\ee
where \footnote{To conrary \cite{Buras:1979yt} we replace $\tilde{\alpha}_n$
by $\beta_n$. Another expressions for the projectors $\alpha_n$, 
$\beta_n$ and $\epsilon_n$ can be found in \cite{Kotikov:1992qx}.}
\bea
{\bf f}_{q}^{-}(n,Q_0^2) &=&
 {\bf f}_{S}(n,Q_0^2) \cdot \alpha_n +
{\bf f}_{g}(n,Q_0^2) \cdot \beta_n,~~
%{\bf \tilde{f}}_{S}^{+}(n,Q_0^2) ~=~ {\bf \tilde{f}}_{S}(n,Q_0^2)
%- {\bf \tilde{f}}_{S}^{-}(n,Q_0^2), 
%\nonumber \\
{\bf f}_{g}^{-}(n,Q_0^2) ~=~
%&=&
 {\bf f}_{g}(n,Q_0^2) \cdot (1-\alpha_n) +
{\bf f}_{q}(n,Q_0^2) \cdot \epsilon_n,~~
\nonumber \\
{\bf f}_{a}^{+}(n,Q_0^2) &=& {\bf f}_{a}(n,Q_0^2)
- {\bf f}_{a}^{-}(n,Q_0^2)
\label{3S.3}
\eea
and
%where
\be
\alpha_n \, = \, \frac{\gamma_{qq}^{(0)}(n)-
\gamma_{+}^{(0)}(n)}{\gamma_{-}^{(0)}(n)-\gamma_{+}^{(0)}(n)},~~
\beta_n \, = \, \frac{\gamma_{qg}^{(0)}(n)}{\gamma_{-}^{(0)}(n)
-\gamma_{+}^{(0)}(n)},~~
\epsilon_n \, = \, \frac{\gamma_{gq}^{(0)}(n)}{\gamma_{-}^{(0)}(n)
-\gamma_{+}^{(0)}(n)} .
\label{3S.4}
\ee

Above LO, the normalization factors ${\bf \tilde{f}}_{a}^{\pm}(n,Q_0^2)$
become to be
\bea
{\bf \tilde{f}}_{a}^{\pm}(n,Q_0^2) &=& 
{\bf f}_{a}^{\pm}(n,Q_0^2) \cdot \biggl(1 
- a_s(Q^2_0) Z_{\pm\pm}^{(1)}(n) -a_s^2(Q^2_0) Z_{\pm\pm}^{(2)}(n) \biggr)
\nonumber \\
&&+ {\bf f}_{a}^{\mp}(n,Q_0^2) \cdot a_s(Q^2_0)  \, \biggl( 
Z_{\mp\pm,a}^{(1)}(n) +a_s(Q^2_0) Z_{\mp\pm,a}^{(2)}(n) \biggr), 
\label{3S.5}
\eea
where (see Appendix A, where $Z^{(i)}_{\pm\pm} = - V^{(i)}_{\pm\pm}$, 
$Z_{\mp\pm,a}^{(i)} = - V_{\mp\pm,a}^{(i)}$, $(i=1,2)$)
\begin{eqnarray}
Z_{\pm\pm}^{(1)}(n) &=& \frac{1}{2\beta_0} \, 
\biggl[ \gamma_{\pm\pm}^{(1)}(n) -
\gamma_{\pm}^{(0)}(n)\, b_1\biggr]\,, ~~
%\nonumber \\
Z_{\pm\mp,q}^{(1)}(n) ~=~
%&=& 
\frac{1}{2\beta_0+\gamma_{\pm}^{(0)}(n)-
\gamma_{\mp}^{(0)}(n)}  \, \gamma_{\pm\mp}^{(1)}(n) \,, ~~ \label{3S.6} \\
%\nonumber \\
Z_{\pm\pm}^{(2)}(n)&=& \frac{1}{4\beta_0} \, \biggl[
%2\beta_0 Z^2_{NS}(n) + 
\gamma^{(2)}_{\pm\pm}(n)
-\Bigl(\gamma^{(1)}_{\pm\pm}(n) -
\gamma^{(0)}_{\pm}(n)Z_{\pm\pm}^{(1)}(n)\Bigr)b_1 + 
\gamma^{(0)}_{\pm}(n)(b^2_1-b_2) 
%\nonumber \\&&
-\sum_{i=\pm} 
\gamma^{(1)}_{\pm i}(n)Z_{i\pm,q}^{(1)}
\biggr]\,, \nonumber \\
Z_{\pm\mp,q}^{(2)}(n)&=& \frac{1}{4\beta_0+\gamma_{\pm}^{(0)}(n)-
\gamma_{\mp}^{(0)}(n)} \, \biggl[
%2\beta_0 Z^2_{NS}(n) + 
\gamma^{(2)}_{\pm\mp}(n)
-\Bigl(\gamma^{(1)}_{\pm\mp}(n) - \gamma^{(0)}_{\pm}(n) Z_{\pm\mp}^{(1)}(n)
\Bigr)b_1 
%\nonumber \\
%%+ \gamma^{(0)}_{\pm}(n)(b^2_1-b_2) 
%&&
-\sum_{i=\pm} 
\gamma^{(1)}_{\pm i}(n) Z_{i\mp,q }^{(1)} \biggr]
\nonumber
%\label{3S.6}
\end{eqnarray}
and
\be
Z_{\pm\pm,g}^{(i)}(n) ~=~ Z_{\pm\pm,q}^{(i)}(n) ~=~ Z_{\pm\pm}^{(i)}(n),~~
Z_{\pm\mp,g}^{(i)}(n) ~=~ Z_{\pm\mp,q}^{(i)}(n) \cdot 
\frac{\gamma_{qq}^{(0)}(n)-
\gamma_{\mp}^{(0)}(n)}{\gamma_{qq}^{(0)}(n)-\gamma_{\pm}^{(0)}(n)},~~(i=1,2)
%, \nonumber \\
%Z_{NNLO,G}^{\pm\mp}(n) &=& Z_{NNLO,S}^{\pm\mp}(n) \cdot 
%\frac{\gamma_{SS}^{(0)}(n)-
%\gamma_{\mp}^{(0)}(n)}{\gamma_{SS}^{(0)}(n)-\gamma_{\pm}^{(0)}(n)}.
\label{3S.7}
\ee
Here $\gamma_{\pm\pm}^{(k)}(n)$ and $\gamma_{\pm\mp}^{(k)}(n)$
are the elemens of matrices of anomalous dimensions, which have been obtained
after diagonalization
from $\gamma_{ab}^{(k)}(n)$ $(a,b=q,g)$
%factors before $\ar$ in the expansion 
%with respect to the latter of the anomalous dimensions $\gamma_{NS}(n,\ar)$ 
(latter taken in the exact form from ~\cite{MVV2004}):
\bea
\gamma_{--}^{(k)}(n) &=&  \gamma_{qq}^{(k)}(n) \cdot \alpha_n +
 \gamma_{qg}^{(k)}(n) \cdot \epsilon_n + \gamma_{gq}^{(k)}(n) \cdot \beta_n +
\gamma_{gg}^{(k)}(n) \cdot (1-\alpha_n), \nonumber \\
\gamma_{-+}^{(k)}(n) &=&  \gamma_{--}^{(k)}(n) - \left(  \gamma_{qq}^{(k)}(n)
+ \frac{1-\alpha_n}{\beta_n} \, \gamma_{qg}^{(k)}(n) \right), ~~
%\nonumber \\
\gamma_{++}^{(k)}(n) =
%&=&  
 \gamma_{qq}^{(k)}(n) +  \gamma_{gg}^{(k)}(n) -
\gamma_{--}^{(k)}(n) , \nonumber \\
\gamma_{+-}^{(k)}(n) &=&  \gamma_{++}^{(k)}(n) - \left(  \gamma_{qq}^{(k)}(n)
- \frac{\alpha_n}{\beta_n} \, \gamma_{qg}^{(k)}(n) \right).
\label{3S.7a}
\eea

The function $H_a^{\pm}(n, Q^2)$ 
up to NNLO may be represented as
%~\footnote{There is an option to proceed
%with calculating by using the exact exponential expressions for both
%$H^{NS}(n,Q^2,Q^2_0)$ and coefficient functions. We've checked that the difference in results originating
%from utilizing this alternative is found to be ${\cal O}(10^{-5})$ and
%can safely be neglected to the accuracy we work with, thus justifying the actual use 
%of the expansion forms, given in the text, in the evaluations throughout.}
\be
H_a^{\pm}(n, Q^2) ~=~
%&=& 
1 + a_s(Q^2) \biggl(Z_{\pm\pm}^{(1)}(n)- 
Z_{\pm\mp,a}^{(1)}(n) \biggr) 
%\nonumber \\
%%\nonumber \\
%%&-& \left[\ar(Q^2)-\ar(Q^2_0)\right]\ar(Q^2_0) [Z^{NLO}_{NS}(n)]^2 
%%&\nonumber \\ \nonumber&+& \left[
%&&
+ a_s^2(Q^2)  \biggl(\tilde{Z}_{\pm\pm}^{(2)}(n)- 
\tilde{Z}_{\pm\mp,a}^{(2)}(n) \biggr)
%-\ar^2(Q^2_0)\right] Z^{NNLO}_{NS}(n)
+ {\cal O}\left(a_s^3(Q^2)\right)\,,
\ee
where~(see \cite{Buras:1979yt} and Appendix A)
\begin{eqnarray}
\tilde{Z}_{\pm\pm}^{(2)}(n) &=& Z_{\pm\pm}^{(2)}(n) + 
\sum_{i=\pm} Z_{\pm i,q}^{(1)} Z_{i\pm,q }^{(1)}, ~~
%\nonumber \\
\tilde{Z}_{\pm\mp,q}^{(2)}(n) ~=~
%&=& 
Z_{\pm\mp,q}^{(2)}(n) + 
\sum_{i=\pm} Z_{\pm i,q}^{(1)} Z_{i\mp,q }^{(1)}, \nonumber \\
\tilde{Z}_{\pm\mp.g}^{(2)}(n) &=& \tilde{Z}_{\pm\mp,q}^{(2)}(n) \cdot 
\frac{\gamma_{qq}^{(0)}(n)-
\gamma_{\mp}^{(0)}(n)}{\gamma_{qq}^{(0)}(n)-\gamma_{\pm}^{(0)}(n)}.
\label{3S.7b}
\eea

\subsection{
%PDFs and FFs
Fragmentation functions 
and their evolution}
\label{ffs}

When one considers AJM
%average multiplicity 
observables, the basic equation is the
one governing the evolution of FFs 
%of the fragmentation functions 
$D_a(x,\mu^2)$ for
the gluon--quark-singlet system $a=g,q$.
In Mellin space, it reads:  
\begin{equation}
\mu^2\frac{\partial}{\partial \mu^2} \left(\begin{array}{l} D_q(\omega,\mu^2) \\ D_g(\omega,\mu^2)
\end{array}\right)
=\left(\begin{array}{ll} P_{qq}(\omega,a_s) & P_{gq}(\omega,a_s) \\
P_{qg}(\omega,a_s) & P_{gg}(\omega,a_s)\end{array}\right)
\left(\begin{array}{l} D_q(\omega,\mu^2) \\ D_g(\omega,\mu^2)
\end{array}\right),
\label{ap}
\end{equation}
where $P_{ab}(\omega,a_s)$, with $a,b=g,q$, are the timelike splitting
functions,
\footnote{$P_{a,b} = -\gamma^{\tau}_{a,b}/2$, where $\gamma^{\tau}_{a,b}$ are the timelike anomalous 
dimensions.}  $\omega=n-1$, with $n$ being the standard Mellin moments with
respect to $x$.
%, and $a_s(\mu^2)=\alpha_s(\mu)/(4\pi)$ is the coupling constant.
%couplant.
The standard definition of the hadron AJMs
%multiplicities in terms of the FFs
%fragmentation functions 
%is given by their integral over $x$, which clearly 
corresponds to the first Mellin moment, with $\omega=0$
(see, e.g., Ref.~\cite{Ellis:1991qj}):
\begin{equation}
\langle n_h(Q^2)\rangle_{a}\equiv \left[\int_0^1 dx \,x^\omega
D_a(x,Q^2)\right]_{\omega=0}=D_a(\omega=0,Q^2),~~~ (a=g,q) \, .
\label{multdef2}
\end{equation}
%where $a=g,q$ for a gluon and quark jet, respectively.

The timelike splitting functions $P_{ab}(\omega,a_s)$ in Eq.~(\ref{ap}) may be
computed perturbatively in $a_s$,
\begin{equation}
 P_{a,b}(\omega,a_s)=
\sum_{k=0}^\infty a_s^{k+1} P_{ab}^{(k)}(\omega).
\end{equation}
The functions $P_{ab}^{(k)}(\omega)$ for $k=0,1,2$ in the
$\overline{\mathrm{MS}}$ scheme may be found in
Refs.~\cite{Gluck:1992zx,Moch:2007tx,Almasy:2011eq} through NNLO and in 
Refs.~\cite{Vogt:2011jv,Albino:2011cm,Kom:2012hd} with small-$x$ resummation
through NNLL accuracy.

\subsection{Diagonalization of FFs}

As it was in the spacelike case (see subsection 2.3 and Appendix A), it
%It 
is not in general possible to diagonalize Eq.~(\ref{ap}) because the
contributions to the timelike-splitting-function matrix do not commute at
different orders.
The usual approach is then to write a series expansion about the LO
%leading-order(LO) 
solution, which can in turn be diagonalized.
One thus starts by choosing a basis in which the timelike-splitting-function
matrix is diagonal at LO (see, e.g., Ref.~\cite{Buras:1979yt} and Appendix A),
\begin{equation}
P(\omega,a_s)=
\left(\begin{array}{ll} P_{++}(\omega,a_s) & P_{-+}(\omega,a_s) 
\\ P_{+-}(\omega,a_s) & P_{--}(\omega,a_s)\end{array}\right)
=a_s\left(\begin{array}{ll} P^{(0)}_{++}(\omega) & 0 
\\ 0 & P^{(0)}_{--}(\omega)\end{array}\right)+a_s^2 P^{(1)}(\omega)
+\mathcal{O}(a_s^3),
\label{pmbasis}
\end{equation} 
with eigenvalues $P_{\pm \pm}^{(0)}(\omega)$.
In one important simplification of QCD, namely ${\mathcal N}=4$ super
Yang-Mills theory, this basis is actually more natural than the $(g,q)$ basis
because the diagonal splitting functions $P^{(k){\mathcal N}=4}_{\pm\pm}(\omega)$ may there be
expressed in all orders of perturbation theory as one universal function 
$P_{\rm uni}^{(k)}(\omega)$
with shifted arguments \cite{Kotikov:2002ab}.
%, i.e. $P^{(k){\mathcal N}=4}_{\pm\pm}(\omega) = 
%P_{\rm uni}^{(k)}(\omega \mp 1)$).
%\footnote{Really it has a place in spin-dependent case.  
%The situation in the spin-averaged case slightly more complicated, because in this case, the 
%equation (\ref{ap}) must be added to the contribution of scalars.}

It is convenient to represent the change of FF basis 
%for the fragmentation functions 
order by order for $k\geq 0$ as \cite{Buras:1979yt}
\footnote{The difference in the diagonalization to compare with the spacelike case considered above is following:
$\gamma^{(i)}_{qg} \leftrightarrow -2 P^{(i)}_{gq}$ and, thus, $\beta_\omega \leftrightarrow
\epsilon_\omega $.}
:
%\begin{eqnarray}
\begin{equation}
D^+(\omega,\mu_0^2) = (1-\alpha_{\omega})D_s(\omega,\mu_0^2) 
- \epsilon_\omega D_g(\omega,\mu_0^2),~~
%\nonumber\\
D^-(\omega,\mu_0^2) = \alpha_{\omega}D_s(\omega,\mu_0^2) + 
\epsilon_\omega D_g(\omega,\mu_0^2). 
\label{changebasisin}
\end{equation}
%\end{eqnarray}

This implies for the components of the timelike-splitting-function matrix that
\begin{eqnarray}
P^{(k)}_{--}(\omega) &=& \alpha_\omega  P^{(k)}_{qq}(\omega) 
+ \epsilon_\omega P^{(k)}_{qg}(\omega) + 
\beta_\omega  P^{(k)}_{gq}(\omega)
+ (1-\alpha_\omega)  P^{(k)}_{gg}(\omega), \nonumber \\
 P^{(k)}_{-+}(\omega) &=& P^{(k)}_{--}(\omega) - 
\left(P^{(k)}_{qq}(\omega) +
\frac{1-\alpha_\omega}{\epsilon_\omega}  P^{(k)}_{gq}(\omega) \right), ~~
%\nonumber \\
P^{(k)}_{++}(\omega) 
%&=&  
P^{(k)}_{qq}(\omega) +  P^{(k)}_{gg}(\omega) 
- P^{(k)}_{--}(\omega),
\nonumber \\
 P^{(k)}_{+-}(\omega) &=& P^{(k)}_{++}(\omega) - \left(P^{(k)}_{qq}(\omega) -
\frac{\alpha_\omega}{\epsilon_\omega}  P^{(k)}_{gq}(\omega) \right)
= P^{(k)}_{gg}(\omega) - 
\left(P^{(k)}_{--}(\omega) - \frac{\alpha_\omega}{\epsilon_\omega}  
P^{(k)}_{gq}(\omega) \right),\quad
\label{changebasis}
\end{eqnarray}
where $\alpha_\omega $, $\beta_\omega $ and $\epsilon_\omega $ are given in Eq. (\ref{3S.4}).
%\begin{equation}
%\alpha_\omega=\frac{P_{qq}^{(0)}(\omega)-P_{++}^{(0)}(\omega)}
%{P_{--}^{(0)}(\omega)-P_{++}^{(0)}(\omega)},\qquad
%\epsilon_\omega=\frac{P_{gq}^{(0)}(\omega)}
%{P_{--}^{(0)}(\omega)-P_{++}^{(0)}(\omega)},\qquad
%\beta_\omega=\frac{P_{qg}^{(0)}(\omega)}{P_{--}^{(0)}(\omega)-P_{++}^{(0)}(\omega)}.
%\label{elements}
%\end{equation}

Note, howerver, that the approach (\ref{pmbasis}) is not so conveninet in FF case, 
because we would like to keep the diagonal part of $P(\omega,a_s)$ matrix without an expansion
%with respect to 
on $a_s$. So. below our
%Our 
approach to solve Eq.~(\ref{ap}) differs from the usual one (see  \cite{Buras:1979yt}) 
%in that we
We write the solution expanding about the diagonal part of the all-order
timelike-splitting-function matrix in the plus-minus basis, instead of its LO
contribution. 
For this purpose, we rewrite Eq.~(\ref{pmbasis}) in the following way:
\begin{equation}
P(\omega,a_s)=
\left(\begin{array}{ll} P_{++}(\omega,a_s) & 0 
\\ 0 & P_{--}(\omega,a_s)\end{array}\right)
+a_s^2 \left(\begin{array}{ll} 0 & P^{(1)}_{-+}(\omega) 
\\ P^{(1)}_{+-}(\omega) & 0\end{array}\right)\
+\left(\begin{array}{ll} 0 & \mathcal{O}(a_s^3) 
\\ \mathcal{O}(a_s^3) & 0\end{array}\right).
\label{sfdec}
\end{equation}

In general, the solution to Eq.~(\ref{ap}) in the plus-minus basis can be
formally written as
\begin{equation}
D(\mu^2)=T_{\mu^2}\left\{\exp{\int_{\mu_0^2}^{\mu^2}\frac{d\bar{\mu}^2}{\bar{\mu}^2}P(\bar{\mu}^2)}
\right\}D(\mu_0^2),
\label{gensol}
\end{equation}
where $T_{\mu^2}$ denotes the path ordering with respect to $\mu^2$ and
\begin{equation}
D=\left(\begin{array}{l} D^+ \\ D^-
\end{array}\right).
\end{equation}
As anticipated, we make the following ansatz to expand about the diagonal part
of the timelike-splitting-function matrix in the plus-minus basis: 
\begin{equation}
T_{\mu^2}\left\{\exp{\int_{\mu_0^2}^{\mu^2}\frac{d\bar{\mu}^2}{\bar{\mu}^2}P(\bar{\mu}^2)}
\right\}
=Z^{-1}(\mu^2)\exp\left[\int_{\mu_0^2}^{\mu^2}\frac{d\bar{\mu}^2}{\bar{\mu}^2}P^{D}
(\bar{\mu}^2)\right]Z(\mu_0^2),
\label{ansatz}
\end{equation}
where
\begin{equation}
P^D(\omega)=
\left(\begin{array}{ll} P_{++}(\omega) & 0 
\\ 0 & P_{--}(\omega)\end{array}\right)
\label{diagpart}
\end{equation}
is the diagonal part of Eq.~(\ref{sfdec}) and $Z$ is a matrix in the
plus-minus basis which has a perturbative expansion of the form
\begin{equation}
Z(\mu^2)=1+a_s(\mu^2)Z^{(1)}+\mathcal{O}(a_s^2).
\label{zpertexp}
\end{equation}

%Using Eq.~(\ref{running}) to perform a change of 
Changing integration variable in
Eq.~(\ref{ansatz}), we obtain
\begin{equation}
T_{a_s}\left\{\exp{\int_{a_s(\mu_0^2)}^{a_s(\mu^2)}\frac{d\bar{a}_s}{\beta(\bar{a}_s)}P(\bar{a}_s)
}
\right\}
=Z^{-1}(a_s(\mu^2))\exp\left[\int_{a_s(\mu_0^2)}^{a_s(\mu^2)}
\frac{d\bar{a}_s}{\beta(\bar{a}_s)}P^{D}
(\bar{a}_s)\right]Z(a_s(\mu_0^2)).
\label{ansatz2}
\end{equation}
Substituting then Eq.~(\ref{zpertexp}) into Eq.~(\ref{ansatz2}),
differentiating it with respect to $a_s$, and keeping only the first term in
the $a_s$ expansion, we obtain the following condition for the $Z^{(1)}$ matrix
(see Section 8.1 for the similar procedure in the spacelike case):
\begin{equation}
Z^{(1)}+\left[\frac{P^{(0)D}}{\beta_0},Z^{(1)}\right]=\frac{P^{(1)OD}}{\beta_0},~~
%\end{equation}
%where
%\begin{equation}
P^{(1)OD}(\omega)=
\left(\begin{array}{ll} 0 & P^{(1)}_{-+}(\omega) 
\\ P^{(1)}_{+-}(\omega) & 0\end{array}\right).
\end{equation}
Solving it, we find:
\begin{equation}
Z_{\pm\pm}^{(1)}(\omega)=0,\qquad
Z_{\pm\mp}^{(1)}(\omega)=\frac{P_{\pm\mp}^{(1)}(\omega)}{\beta_0+P_{\pm\pm}^{(0)}(\omega)
-P_{\mp\mp}^{(0)}(\omega)}.
\label{zmatrix}
\end{equation}

At this point, an important comment is in order.
In the conventional approach to solve Eq.(\ref{ap}), one expands about the
diagonal LO matrix given in Eq.~(\ref{pmbasis}), while here we expand about the
all-order diagonal part of the matrix given in Eq.~(\ref{sfdec}).
The motivation for us to do this arises from the fact that the functional
dependence of $P_{\pm\pm}(\omega,a_s)$ on $a_s$ is different after resummation.

Now reverting the change of basis specified in Eq.~(\ref{changebasisin}), we
find the gluon and quark-singlet fragmentation functions to be given by
%\begin{eqnarray}
\begin{equation}
D_g(\omega,\mu^2) = -\frac{\alpha_\omega}{\epsilon_\omega}D^+(\omega,\mu^2)+
\left(\frac{1-\alpha_\omega}{\epsilon_\omega}\right)D^-(\omega,\mu^2),~~~
%\nonumber\\
D_q(\omega,\mu^2) = D^+(\omega,\mu^2) + D^-(\omega,\mu^2).
\label{inverbasis}
\end{equation}
%\end{eqnarray}
As expected, this suggests to write the gluon and quark-singlet fragmentation
functions in the following way:
\begin{equation}
D_a(\omega,\mu^2)\equiv D_a^+(\omega,\mu^2)+D_a^-(\omega,\mu^2), \qquad a=g,q,
\label{decomp}
\end{equation} 
where $D_a^+(\omega,\mu^2)$ evolves like a plus component and
$D_a^-(\omega,\mu^2)$ like a minus component.

We now explicitly compute the functions $D_a^\pm(\omega,\mu^2)$ appearing in
Eq.~(\ref{decomp}).
To this end, we first substitute Eq.~(\ref{ansatz}) into Eq.~(\ref{gensol}).
Using Eqs.~(\ref{diagpart}) and (\ref{zmatrix}), we then obtain
\begin{eqnarray}
D^+(\omega,\mu^2)&=&\tilde{D}^+(\omega,\mu_0^2)\hat{T}_+(\omega,\mu^2,\mu_0^2)
-a_s(\mu^2)Z^{(1)}_{-+}(\omega)\tilde{D}^-(\omega,\mu_0^2)\hat{T}_-(\omega,\mu^2,\mu_0^2),
\nonumber\\
D^-(\omega,\mu^2)&=&\tilde{D}^-(\omega,\mu_0^2)\hat{T}_-(\omega,\mu^2,\mu_0^2)
-a_s(\mu^2)Z^{(1)}_{+-}(\omega)\tilde{D}^+(\omega,\mu_0^2)\hat{T}_+(\omega,\mu^2,\mu_0^2),
\label{result1}
\end{eqnarray}
where 
\begin{equation}
\tilde{D}^\pm(\omega,\mu_0^2) = D^\pm(\omega,\mu_0^2)
+a_s(\mu_0^2) Z_{\mp\pm}^{(1)}(\omega) D^\mp(\omega,\mu_0^2) ,
\label{renfact}
\end{equation}
and
\begin{equation}
\hat{T}_{\pm}(\omega,\mu^2,\mu_0^2)  
= \exp \left[\int^{a_s(\mu^2)}_{a_s(\mu_0^2)}
\frac{d\bar{a}_s}{\beta(\bar{a}_s)} \, 
P_{\pm\pm}(\omega,\bar{a}_s) \right] 
\label{rengroupexp}
\end{equation}
has a RG-type exponential form.
Finally, inserting Eq.~(\ref{result1}) into Eq.~(\ref{inverbasis}), we find
by comparison with Eq.~(\ref{decomp}) that
\begin{equation}
D_a^\pm(\omega,\mu^2)=\tilde{D}_a^\pm(\omega,\mu_0^2)
%\left[\frac{\alpha_s(\mu^2)}{\alpha_s(\mu_0^2)}\right]
%^{-\frac{P_{\pm\pm}^{(0)}}{2\beta_0}}
\hat{T}_{\pm}(\omega,\mu^2,\mu_0^2)
\, H_{a}^\pm(\omega,\mu^2),
\label{evolsol}
\end{equation}
where
\begin{eqnarray}
%\begin{equation}
\tilde{D}_g^+(\omega,\mu_0^2) = -\frac{\alpha_\omega}{\epsilon_\omega}
\tilde{D}_q^+(\omega,\mu_0^2),~~
\tilde{D}_g^-(\omega,\mu_0^2)=\frac{1-\alpha_\omega}{\epsilon_\omega}
\tilde{D}_q^-(\omega,\mu_0^2),~~
\nonumber\\
\tilde{D}_q^+(\omega,\mu_0^2) = \tilde{D}^+(\omega,\mu_0^2),~~
\tilde{D}_q^-(\omega,\mu_0^2)=\tilde{D}^-(\omega,\mu_0^2),
\label{rlo}
%\end{equation}
\end{eqnarray}
and $H_a^\pm(\omega,\mu^2)$ are perturbative functions given by
\begin{equation}
H_a^\pm(\omega,\mu^2) =1- a_s(\mu^2)
Z_{\pm\mp,a}^{(1)}(\omega)+\mathcal{O}(a_s^2).
\label{pertfun}
\end{equation}
At $\mathcal{O}(\alpha_s)$, we have
\begin{equation}
Z_{\pm\mp,g}^{(1)}(\omega)=-Z_{\pm\mp}^{(1)}(\omega)
{ \left(\frac{1-\alpha_\omega}{\alpha_\omega}\right)}^{\pm 1},\qquad
Z_{\pm\mp,q}^{(1)}(\omega)=Z_{\pm\mp}^{(1)}(\omega),
\end{equation}
where $Z_{\pm\mp}^{(1)}(\omega)$ is given by Eq.~(\ref{zmatrix}).

\section{
Generalized DAS
approach} \indent

The flat initial condition (\ref{1}) corresponds to the case when parton density
%distributions
tend  to some constant value at $x \to 0$ and at some initial value $Q^2_0$.
%(\ref{1}).
The main ingredients of the results \cite{Q2evo,HT}, are:
%\begin{itemize}
%\item

{{\color{red} A.} ~
Both, the gluon and quark singlet densities are presented in terms of two
components ($"+"$ and $"-"$) which are obtained from the analytic 
$Q^2$-dependent expressions of the corresponding ($"+"$ and $"-"$) PDF
%parton distributions 
moments.
%\footnote{Such an approach has been developed  \cite{Albino:2011si}
%recently also for the fragmentation function, 
%whose first moments (ie mean multiplicities of quarks and gluons) were analyzed
%\cite{Bolzoni:2012ii}. The results are 
%in good agreement with the experimental data
%(see contribution \cite{Bolzoni:2012cv} by Paolo Bolzoni to this Proceedings).}

{{\color{red} B.} ~
The twist-two part of the $"-"$ component is constant at small $x$ at any 
values of $Q^2$,
whereas the one of the $"+"$ component grows at $Q^2 \geq Q^2_0$ as
\begin{equation}
\sim e^{\sigma_{\rm NLO}},~~~
%\exp{\sigma},~~~
\sigma_{\rm NLO} = 2\sqrt{\left[ \left|\hat{d}_+\right| s_{\rm NLO}
%\ln \left( \frac{a_s(Q^2_0)}{a_s(Q^2)} \right) 
- \left( \hat{d}_{++}
%\hat{D}_+ 
+  \left|\hat{d}_+\right| b_1
%\hat{d}_+ b_1
%\frac{\beta_1}{\beta_0} 
\right) p_{\rm NLO}
% \Bigl( a_s(Q^2_0) - a_s(Q^2) \Bigr)
\right] \ln \left( \frac{1}{x} \right)}  \ ,~~~ 
\rho_{\rm NLO}=\frac{\sigma_{\rm NLO}}{2\ln(1/x)} \ ,
\label{intro:1}
\end{equation}
where $\sigma$ and $\rho$
%$=\sigma/(2\ln(1/x))$ 
are the generalized Ball--Forte
variables,
\begin{equation}
s_{\rm NLO}=\ln \left( 
\frac{a^{\rm NLO}_s(Q^2_0)}{a^{\rm NLO}_s(Q^2)} \right),~~
p_{\rm NLO}= a^{\rm NLO}_s(Q^2_0) - a^{\rm NLO}_s(Q^2),~~~
\hat{d}_+ = - \frac{12}{\beta_0},~~~
%\hat{D}_+ 
\hat{d}_{++} =  \frac{412}{27\beta_0}.
\label{intro:1a}
\end{equation}
and $\beta_0$ and $\beta_1$ are given in Eq. (\ref{alpha1})
%\end{itemize}

\subsection{Parton distributions and the structure function $F_2$
%$Q^2$ dependence of the slope $d \ln F_2/d\ln (1/x)$ in g
%Generalized DAS approach
%double-logarithmic approximation
} 

%Here, for simplicity we consider only  the LO
%%leading order (LO) 
%approximation\footnote{
%The NLO results may be found in  \cite{Q2evo,HT}.}.
The results for parton densities and $F_2$
%of Refs. \cite{Q2evo,HT} 
are following:
\begin{itemize}
\item
The structure function $F_2$ has the form:
%Both, the gluon and quark singlet densities are presented in terms of two
%components ($'+'$ and $'-'$) 
\begin{eqnarray}
	F_{2}^{\rm LO}(x,Q^2) &=& e \, f_{q,,{\rm LO}}(x,Q^2),~~~
%\label{r10} 
%\nonumber \\
	f_{a,{\rm LO}}(x,Q^2) 
%&=& 
~=~ f_{a,{\rm LO}}^{+}(x,Q^2) + f_{a,{\rm LO}}^{-}(x,Q^2)
\label{8a}
\end{eqnarray}
at the LO approximation, where $e$
%which are obtained from the analytic 
%$Q^2$-dependent expressions of the corresponding ($'+'$ and $'-'$) PDF
%%parton distributions 
%moments. Here, 
%\be e=(\sum_1^f e_i^2)/f  \label{8aa}
%\ee
is the average charge square (\ref{RNS}), and \\
%and $f$ is the number of active quark flavors.\\
\begin{eqnarray}
	F_2^{\rm NLO}(x,Q^2) &=& e \, \left( f_{q,{\rm NLO}}(x,Q^2) 
+ \frac{2}{3} f a^{\rm NLO}_s(Q^2)
f_{g,{\rm NLO}}(x,Q^2)\right),
%\label{r10} 
\nonumber \\
	f_{a,{\rm NLO}}(x,Q^2) &=& f_{a,{\rm NLO}}^{+}(x,Q^2) + 
f_{a,{\rm NLO}}^{-}(x,Q^2)
\label{8ab}
\end{eqnarray}
at the NLO approximation.

\item
The small-$x$ asymptotic results for the LO parton densities $f^{\pm}_{a,{\rm LO}}$ are
\begin{eqnarray}
	f^{+}_{g,{\rm LO}}(x,Q^2) &=& \biggl(A_g + \frac{4}{9} A_q \biggl)
		\tilde I_0(\sigma_{\rm LO}) \; e^{-\overline d_{+} s_{\rm LO}} 
~+~ O(\rho_{\rm LO}),
	\label{8.0} \\
	f^{+}_{q,{\rm LO}}(x,Q^2) &=& 
\frac{f}{9}\biggl(A_g + \frac{4}{9} A_q \biggl)
		\rho_{\rm LO} \tilde I_1(\sigma_{\rm LO}) \; e^{-\overline d_{+} s_{\rm LO}} 
~+~ O(\rho_{\rm LO}),
	\label{8.01} \\
	f^{-}_{g,{\rm LO}}(x,Q^2) &=& -\frac{4}{9} A_q e^{- d_{-} s_{\rm LO}} ~+~ O(x),	~~
%\label{8.00} \\
	f^{-}_{q,{\rm LO}}(x,Q^2) ~=~
%&=&  
A_q e^{-d_{-} s_{\rm LO}} ~+~ O(x),
	\label{8.02}
\end{eqnarray}
where
\footnote{The dependence on the colour factors $C_A=3$, $C_F=4/3$ in Eqs. (\ref{intro:1a}), 
(\ref{8.02a1}) and (\ref{8.02b}) can be found in \cite{HT}.}
\be \overline d_{+} = 1 + \frac{20f}{27\beta_0},~~
%$ and          
d_{-} = \frac{16f}{27\beta_0} \label{8.02a1} 
\ee
are the regular parts of the anomalous dimensions $d_{+}(n)$ and $d_{-}(n)$, 
respectively, in the limit $n\to1$\footnote{
We denote the singular and regular parts of a given quantity $k(n)$ in the
limit $n\to1$ by $\hat k/(n-1)$ and $\overline k$, respectively.}.
%Here $n$ is the variable of the PD Mellin transform.
Here $n$ is
the variable in Mellin space.
%
%We define the variable
% \begin{eqnarray}
%s=ln\left(\frac{a_s(Q^2_0)}{a_s(Q^2)}\right)
%\label{2.4}
% \end{eqnarray}
%
The functions $\tilde I_{\nu}$ ($\nu=0,1$) 
%in Eqs. (\ref{8.0,8.01}) 
are related to the modified Bessel
function $I_{\nu}$
and to the Bessel function $J_{\nu}$ by:
\begin{equation}
\tilde I_{\nu}(\sigma) =
\left\{
\begin{array}{ll}
I_{\nu}(\sigma), & \mbox{ if } s \geq 0 \\
i^{-\nu} J_{\nu}(i\sigma), \ i^2=-1, \ & \mbox{ if } s \leq 0 
\end{array}
\right. .
\label{4}
\end{equation}
At the LO, 
the variables 
%$s$, 
$\sigma_{\rm LO}$ and $\rho_{\rm LO}$ are
%argument $\sigma$ is 
given by Eq. (\ref{intro:1}) when $p=0$, i.e.
\begin{equation}
\sigma_{\rm LO} = 2\sqrt{\left|\hat{d}_+\right| s_{\rm LO}
 \ln \left( \frac{1}{x} \right)}  \ ,~~~ \rho_{\rm LO}=\frac{\sigma_{\rm LO}}{2\ln(1/x)} \ ,
\label{intro:1b}
\end{equation}
and the variable $s_{\rm LO}$ is given by Eq. (\ref{intro:1a}) with $a_s^{\rm LO}(Q^2)$
 as in Eq. (\ref{as:LO}).

\item
The small-$x$ asymptotic results for the NLO parton densities $f^{\pm}_a$ are
\begin{eqnarray}
	f^{+}_{g,{\rm NLO}}(x,Q^2) &=& A_{g,{\rm NLO}}^+(Q^2,Q_0^2)
%\biggl(A_g + \frac{4}{9} A_q \biggl)
		\tilde I_0(\sigma_{\rm NLO}) \; e^{-\overline d_{+} s_{\rm NLO}
 -\overline D_{+} p_{\rm NLO}} ~+~ O(\rho_{\rm NLO}),
	\label{8.0A} \\
	f^{+}_{q,{\rm NLO}}(x,Q^2) &=& A_{q,{\rm NLO}}^+
%\frac{f}{9}\biggl(A_g + \frac{4}{9} A_q \biggl) 
\Biggl[\left(1-\overline d_{+-}^q a^{\rm NLO}_s(Q^2)\right)
		\rho_{\rm NLO} \tilde I_1(\sigma_{\rm NLO}) 
\nonumber \\ &&+ 
20 a^{\rm NLO}_s(Q^2) I_0(\sigma_{\rm NLO}) \Biggr]
\; e^{-\overline d_{+}(1) s_{\rm NLO} -\overline D_{+} p_{\rm NLO}} 
~+~ O(\rho_{\rm NLO}),~~~~~~~~
	\label{8.01A} \\
	f^{-}_{g,{\rm NLO}}(x,Q^2) &=& A_{g,{\rm NLO}}^-(Q^2,Q_0^2)
%-\frac{4}{9} A_q 
e^{- d_{-}(1) s_{\rm NLO}-D_{-} p_{\rm NLO}} ~+~ O(x),
	\label{8.00A} \\
	f^{-}_{q,{\rm NLO}}(x,Q^2) &=& A_{q,{\rm NLO}}^-
%A_q 
e^{-d_{-}(1) s_{\rm NLO}-D_{-} p_{\rm NLO}} ~+~ O(x),
	\label{8.02NLO}
\end{eqnarray}
where $(b_1=\beta_1/\beta_0)$
\be
 D_{\pm}=d_{\pm\pm}-
%\frac{\beta_1}{\beta_0} 
d_{\pm} b_1
   \label{8.02aa}
\ee
and similar for $\hat D_{+}$ and $\overline D_{+}$,
%\begin{eqnarray}
\bea
A_{g,{\rm NLO}}^{+}(Q^2,Q_0^2) &=& \left(1-
\frac{80f}{81}a^{\rm NLO}_s(Q)\right)A_g 
\nonumber \\&&
+ \frac{4}{9}
\left(1+\Bigl(3+\frac{f}{27}\Bigr)a_s^{\rm NLO}(Q_0) 
-\frac{80f}{81}a_s^{\rm NLO}(Q)\right)A_q,
\nonumber \\
\qquad A_{g,{\rm NLO}}^{-}(Q^2,Q_0^2) &=& A_g 
- A_{g,{\rm NLO}}^{+}(Q^2_0,Q^2) \, .
\label{8.02a}
\eea
%\end{eqnarray}
%and

The coupling constant $a_s(Q^2)$ 
is introduced in Eq. (\ref{as:NLO}). The variables $\hat d_{+}$, $\hat d_{++}$
$\overline d_{+}$ and $d_{-}$ are diven in Eqs. (\ref{intro:1a}) and (\ref{8.02a1}), respectively.
The nonzero variables $\overline d_{++}$, $d_{--}$ and $d_{+-}^a$ $(a=q,g)$ have the form
\begin{eqnarray}
%a_s(\mu)&=&\frac{\alpha_s(\mu)}{4\pi},\qquad
%\hat d_{+} = - \frac{12}{\beta_0},\qquad
%\overline d_{+}
%%(1) 
%= 1 + \frac{20f}{27\beta_0},\qquad
%d_{-}
%%(1) 
%= \frac{16f}{27\beta_0},\nonumber \\
%\hat d_{++}&=& \frac{412f}{27\beta_0},\qquad
\overline d_{++} &=&
%(1) = 
\frac{8}{\beta_0} \left(
36\zeta_3+33\zeta_2-\frac{1643}{12} + \frac{2f}{9}
\left[\frac{68}{9}-4\zeta_2-\frac{13f}{243}\right]\right),~ \overline d_{+-}^g = \frac{80f}{61}, ~
\overline d_{-+}^g = -3-\frac{f}{27},
\nonumber \\
d_{--}
%(1) 
&=& \frac{16}{9\beta_0} \left(
2\zeta_3-3\zeta_2+\frac{13}{4} + f
\left[4\zeta_2-\frac{23}{18}+\frac{13f}{243}\right]\right), ~~
%~~~ \overline d_{-+}^g = -3-\frac{f}{27},~~ \nonumber \\
d_{+-}^q = 23 -12\zeta_2 - \frac{13f}{81} \, ,
\label{8.02b}
\end{eqnarray}
with $\zeta_3$ and $\zeta_2$ are Eller functions.
%$\beta_0$ and $\beta_1$ are first two coefficients of QCD $\beta$-function.
%\footnote{Note that evaluation of the results (\ref{8.0})-(\ref{8.02b})
%need the knowledge of the analytic continuation of the anomalous dimansions
%and coefficient functions. The analytic continuation can be found in Refs.
%\cite{Kazakov:1987jk}. It was used also for the fits 
%\cite{Ourfits,Kotikov:2010bm}.}
\end{itemize}

\section{Resummation in FFs}

As already mentioned in Introduction,
%Section~\ref{sec:intro}, 
reliable computations of AJMs
%average jet multiplicities 
require resummed analytic expressions for the
splitting functions because one has to evaluate the first Mellin moment
(corresponding to $\omega=0$), which is a divergent quantity in the
fixed-order perturbative approach.
As is well known, resummation overcomes this problem, as demonstrated in the
pioneering works by Mueller \cite{Mueller:1981ex} and others 
\cite{Ermolaev:1981cm}.
%,Dokshitzer:1982xr,Dokshitzer:1982fh,Dokshitzer:1982ia}.

In particular, as we shall see in previous subsection,
%Section~\ref{multiplicities}, 
resummed
expressions for the first Mellin moments of the timelike splitting functions
in the plus-minus basis appearing in Eq.~(\ref{pmbasis}) are required in our
approach.
Up to the NNLL level in the $\overline{\mathrm{MS}}$ scheme, these may be
extracted from the available literature
\cite{Mueller:1981ex,Vogt:2011jv,Albino:2011cm,Kom:2012hd} in closed analytic
form using the relations in Eq.~(\ref{changebasis}).
%Note that the expressions are generally simpler in the plus-minus basis (see Ref. 
%\cite{Bolzoni:2013rsa}),%
%\footnote[1]{In fact, one can see from Eq.~(3.3) of Ref.~\cite{Kom:2012hd} that
%the resummation of the combination $P_{gg}(\omega,a_s)+P_{qq}(\omega,a_s)$, which
%according to Eq.~(\ref{changebasisin}) gives $P_{++}(\omega,a_s)$ because
%$P_{--}(\omega,a_s)$ does not need resummation, is much simpler than that of
%$P_{gg}(\omega,a_s)$ alone.}
%while the corresponding results for the resummation of $P_{gg}(\omega,a_s)$ and
%$P_{gq}(\omega,a_s)$ can be highly nontrivial and complicated in higher orders
%of resummation.
%%An analogous observation was made for the double-logarithm aymptotics in the
%%Kirschner-Lipatov approach \cite{Kirschner:1982xw}, where 
%%the corresponding amplitudes obey nontrivial equations, whose solutions are
%%rather complicated special functions.

For future considerations, we remind the reader of an assumpion already
made in Ref.~\cite{Albino:2011cm} according to which the splitting functions
$P^{(k)}_{--}(\omega)$ and $P^{(k)}_{+-}(\omega)$ are supposed to be free of
singularities in the limit $\omega \to 0$.
%Hence, neglecting all non-singular terms we have that,
%\be
%P_{--}(\omega)=P_{+-}(\omega)=0.
%\label{appassum}
%\ee
In fact, this is expected to be true to all orders.
This is certainly true at the LL and NLL levels for
%for the leading DL (LL) contributions to 
the timelike splitting functions,
%, for the next-to-leading DL (NLL) contributions in the MG scheme 
%given in Eqs.\ (\ref{mgsplittingfunctions1})--(\ref{mgsplittingfunctions3}), 
%and through NNLO \cite{Moch:2007tx,Almasy:2011eq},
as was verified in \cite{Albino:2011cm}.
This is also true at the NNLL level, as may be explicitly checked by inserting
the results of Ref.~\cite{Kom:2012hd} in Eq.~(\ref{changebasis}). 
Moreover, this is true through NLO in the spacelike case \cite{Q2evo}
and holds for the LO and NLO singularities \cite{FaLi,KoLi,Kotikov:2002ab}
to all orders in the framework of the BFKL
%Balitski-Fadin-Kuraev-Lipatov (BFKL)
dynamics \cite{BFKL},
%,Kuraev:1976ge,Kuraev:1977fs,Balitsky:1978ic},
a fact that was exploited in various approaches (see, e.g., 
Refs.~\cite{Ciafaloni:2007gf} and references cited therein).
We also note that the timelike splitting functions share a number of simple 
properties with their spacelike counterparts.
In particular, the LO splitting functions are the same, and the diagonal
splitting functions grow like $\ln \omega$ for $\omega \to \infty$
at all orders.
This suggests the conjecture that the double-logarithm resummation in the
timelike case and the BFKL resummation in the spacelike case are only related
via the plus components. 
The minus components are devoid of singularities as $\omega \to 0$ and thus 
are not resummed.
Now that this is known to be true for the first three orders of resummation,
one has reason to expect this to remain true for all orders.

Using the relationships between the components of the splitting functions in
the two bases given in Eq.~(\ref{changebasis}), we find that the absence of
singularities for $\omega=0$ in $P_{--}(\omega,a_s)$ and $P_{+-}(\omega,a_s)$
implies that the singular terms are related as
%\begin{eqnarray}
\begin{equation}
P_{gq}^{\rm sing}(\omega,a_s) = 
-\frac{\epsilon_\omega}{\alpha_\omega}P_{g g}^{\rm sing}(\omega,a_s),~~~
%\label{tolja3}\\
P_{qg}^{\rm sing}(\omega,a_s) = 
-\frac{\alpha_\omega}{\epsilon_\omega}P_{qq}^{\rm sing}(\omega,a_s),
\label{tolja3.1}
\end{equation}
%\end{eqnarray}
where, through the NLL level,
\footnote{To have a possibility to compare different approximations, it is
convenient to keep the general forms of the colour factors $C_A=3$, $C_F=4/3$
in the present and the next sections.}

\begin{equation}
-\frac{\alpha_\omega}{\epsilon_\omega}=\frac{C_A}{C_F}\left[1-\frac{\omega}{6}\left(
1+2\frac{T_F}{C_A}
-4\,\frac{C_F T_F}{C_A^2}\right)\right]+\mathcal{O}(\omega^2).
\label{motivation}
\end{equation}
An explicit check of the applicability of the relationships in Eqs.
%~(\ref{tolja3}) and 
(\ref{tolja3.1}) for $P_{ij}(\omega,a_s)$ with
$i,j=g,g$ themselves is performed in the Appendix of Ref. \cite{Bolzoni:2013rsa}.
Of course, the relationships in Eqs.
%~(\ref{tolja3}) and 
(\ref{tolja3.1}) may be
used to fix the singular terms of the off-diagonal timelike splitting functions
$P_{qg}(\omega,a_s)$ and $P_{gq}(\omega,a_s)$ using known results for the
diagonal timelike splitting functions $P_{qq}(\omega,a_s)$ and
$P_{gg}(\omega,a_s)$.
Since Refs.~\cite{Vogt:2011jv,Almasy:2011eq} became available during the
preparation of Ref.~\cite{Albino:2011cm}, the relations in Eqs.
%~(\ref{tolja3}) and 
(\ref{tolja3.1}) provided an important independent check rather than a
prediction.

We take here the opportunity to point out that Eqs.~(\ref{evolsol}) and
(\ref{rlo}) together with Eq.~(\ref{motivation}) support the motivations for
the numerical effective approach that we used in Ref.~\cite{Bolzoni:2012ed,Bolzoni:2013rsa} to
study the 
%average 
gluon-to-quark AJM
%jet multiplicity 
ratio. 
In fact, according to the findings of Ref.~\cite{Bolzoni:2012ed,Bolzoni:2013rsa}, 
substituting $\omega=\omega_\mathrm{eff}$, where
\begin{equation}
\omega_\mathrm{eff}=2\sqrt{2C_A a_s},
\label{replacement}
\end{equation}
into Eq.~(\ref{motivation}) exactly reproduces the result for the average
gluon-to-quark jet multiplicity ratio $r(Q^2)$ obtained in
Ref.~\cite{Mueller:1983cq}.
In the next section, we shall obtain improved analytic formulae for the 
ratio $r(Q^2)$ and also for the average gluon and quark jet multiplicities.

Here we would also like to note that, at first sight, the substitution
$\omega=\omega_{\rm eff}$ should induce a $Q^2$ dependence 
%in Eq.~(\ref{elements}), which should contribute 
to the diagonalization matrix.
This is not the case, however, because to double-logarithmic accuracy the $Q^2$
dependence of $a_s(Q^2)$ can be neglected, so that the factor
$\alpha_\omega/\epsilon_\omega$ does not recieve any $Q^2$ dependence upon the
substitution $\omega=\omega_{\rm eff}$.
This supports the possibility to use this substitution in our analysis and
gives an explanation of the good agreement with other approaches, e.g.\ that of
Ref.~\cite{Mueller:1983cq}.
Nevertheless, this substitution only carries a phenomenological meaning.
It should only be done in the factor $\alpha_\omega/\epsilon_\omega$, but not
in the RG exponents of Eq.~(\ref{rengroupexp}), where it
would lead to a double-counting problem.
In fact, the dangerous terms are already resummed in Eq.~(\ref{rengroupexp}).

In order to be able to obtain the AJMs,
%average jet multiplicities, 
we have to first 
evaluate the first Mellin momoments of the timelike splitting functions in the
plus-minus basis. 
According to Eq.~(\ref{changebasis}) together with the results given in 
Refs.~\cite{Mueller:1981ex,Kom:2012hd}, we have
\begin{equation}
P_{++}^\mathrm{NNLL}(\omega=0)=\gamma_0(1 - K_1 \gamma_0 + K_2  \gamma_0^2),
\label{nllfirst}
\end{equation}
where
\begin{eqnarray}
\gamma_0&=&P_{++}^\mathrm{LL}(\omega=0)=\sqrt{2 C_A a_s},~~~~~
%\label{llgamma0}\\
K_1 ~=~ \frac{1}{12} 
\left[11 +4\frac{T_F}{C_A} \left(1-\frac{2C_F}{C_A}\right)\right], \label{llgamma0}\\
K_2&=&\frac{1}{288} 
\left[1193-576\zeta_2 -56\frac{T_F}{C_A} 
\left(5+2\frac{C_F}{C_A}\right)\right]
+ 16 \frac{T^2_F}{C^2_A}
\left(1+4\frac{C_F}{C_A}-12\frac{C^2_F}{C^2_A}\right),\quad
\end{eqnarray}
and
\begin{equation}
P_{-+}^\mathrm{NNLL}(\omega=0)=-\frac{C_F}{C_A}\,P_{qg}^{NNLL}(\omega=0),
\label{evolsolaa}
\end{equation}
where
\begin{equation}
P_{qg}^\mathrm{NNLL}(\omega=0) = \frac{16}{3} T_F a_s
-\frac{2}{3} T_F
\left[17-4\,\frac{T_F}{C_A} \left(1-\frac{2C_F}{C_A}\right)\right]
{\left(2C_A a_s^3\right)}^{1/2}.
\label{nllsecondA}
\end{equation}
For the $P_{+-}$ component, we obtain
\begin{equation}
P_{+-}^\mathrm{NNLL}(\omega=0)= \mathcal{O}(a_s^2) .
\end{equation}
Finally, as for the $P_{--}$ component, we note that its LO expression produces
a finite, nonvanishing term for $\omega=0$ that is of the same order in $a_s$
as the NLL-resummed results in Eq.~(\ref{nllfirst}), which leads us to use the
following expression for the $P_{--}$ component: 
\begin{equation}
P_{--}^\mathrm{NNLL}(\omega=0)=-\frac{8T_F C_F}{3 C_A}\,a_s
+ \mathcal{O}(a_s^2),
\label{nllsecond}
\end{equation}
at NNLL accuracy.

We can now perform the integration in Eq.~(\ref{rengroupexp}) through the NNLL
level, which yields
\begin{eqnarray}
\hat{T}_{\pm}^\mathrm{NNLL}(0,Q^2,Q^2_0)&=& 
\frac{T_{\pm}^\mathrm{NNLL}(Q^2)}{T_{\pm}^\mathrm{NNLL}(Q^2_0)},~~
T_{-}^\mathrm{NNLL}(Q^2) =
%&=&
T_{-}^\mathrm{NLL}(Q^2) =
\left(a_s(Q^2)\right)^{d_-},
\label{nnllresult}\\
T_{+}^\mathrm{NNLL}(Q^2) &=& \exp\left\{\frac{4C_A}{\beta_0 \gamma^0(Q^2)}
\left[1+\left(b_1-2C_AK_2\right)a_s(Q^2)\right]\right\}
\left(a_s(Q^2)\right)^{d_+},
%\\
%T_{-}^\mathrm{NNLL}(Q^2)&=&T_{-}^\mathrm{NLL}(Q^2) =
%\left(a_s(Q^2)\right)^{d_-},
%\label{nnllresultm}
%D_a^-(Q^2) =0,  
%\end{equation}
\label{nnllresultm}
\end{eqnarray}
where
\begin{equation}
b_1=\frac{\beta_1}{\beta_0},\qquad
d_{-} = \frac{8 T_F C_F}{3 C_A \beta_0},\qquad
d_{+} = \frac{2 C_A K_1}{\beta_0}.
\label{anomdim}
\end{equation}

\section{Multiplicities}
\label{multiplicities}

According to Eqs.~(\ref{rengroupexp}) and (\ref{evolsol}), the $\pm\mp$
components are not involved in the AJM $Q^2$ evolution,
%of average jet multiplicities, 
which is performed at $\omega =0$ using the resummed
expressions for the plus and minus components given in Eq.~(\ref{nllfirst})
and (\ref{nllsecond}), respectively.
We are now ready to define the average gluon and quark jet multiplicities in
our formalism, namely
\begin{equation}
\langle n_h(Q^2)\rangle_a\equiv D_a(0,Q^2)= D_a^+(0,Q^2) + D_a^-(0,Q^2),~~~(a=g,q) \, .
\label{multdef}
\end{equation}
%with $a=g,q$, respectively.

On the other hand, from Eqs.~(\ref{evolsol}) and (\ref{rlo}), it follows that
%\begin{eqnarray}
\begin{equation}
r_+(Q^2) \equiv 
\frac{D_g^+(0,Q^2)}{D_q^+(0,Q^2)}=
-\lim_{\omega\rightarrow 0}\frac{\alpha_\omega}{\epsilon_\omega}\,
\frac{H^+_g(\omega,Q^2)}{H^+_q(\omega,Q^2)},~~
%\label{evolsola}\\
r_-(Q^2) \equiv \frac{D_g^-(0,Q^2)}{D_q^-(0,Q^2)}=\lim_{\omega\rightarrow 0}
\frac{1-\alpha_\omega}{\epsilon_\omega}\,
\frac{H^-_g(\omega,Q^2)}{H^-_q(\omega,Q^2)}.
\label{rmin}
\end{equation}
%\end{eqnarray}
Using these definitions and again Eq.~(\ref{evolsol}), we may write general
expressions for the 
%average 
gluon and quark AJMs:
%jet multiplicities:
\begin{eqnarray}
\langle n_h(Q^2)\rangle_g&=&\tilde{D}_g^+(0,Q_0^2)\hat{T}_+^\mathrm{res}(0,Q^2,Q_0^2)
H^+_g(0,Q^2)
%\nonumber\\&&{}
+\tilde{D}_q^-(0,Q_0^2)r_-(Q^2)\hat{T}_-^\mathrm{res}(0,Q^2,Q_0^2)
H^-_q(0,Q^2),\nonumber\\
\langle n_h(Q^2)\rangle_s&=&\frac{\tilde{D}_g^+(0,Q_0^2)}{r_+(Q^2)}\hat{T}_+^\mathrm{res}(0,Q^2,Q_0^2)
H^+_g(0,Q^2)
%\nonumber\\&&{}
+\tilde{D}_q^-(0,Q_0^2)\hat{T}_-^\mathrm{res}(0,Q^2,Q_0^2)
H^-_q(0,Q^2).
\end{eqnarray}
At the LO in $a_s$, the coefficients of the RG exponents are given by
%\begin{eqnarray}
\begin{equation}
r_+(Q^2) = \frac{C_A}{C_F},\qquad r_-(Q^2)=0, \qquad
%\nonumber\\ \qquad 
H^{\pm}_s(0,Q^2) = 1,\qquad 
\tilde{D}_a^\pm(0,Q_0^2)=D_a^\pm(0,Q_0^2).
\label{lonnll}
\end{equation}
%\end{eqnarray}
%for $a=g,q$.
%It is correct because
%\begin{equation}
%\label{evolsolaA}
%Z^{(1)}_{\pm\pm,s} \sim O(\omega),~~ Z^{(1)}_{+-,g} \sim O(\omega^2), 
%~~ Z^{(1)}_{-+,g}(\omega=0) = +\frac{3}{8n_f} P_{qg}(\omega=0)= a_s +
%O(\omega^{3/2})
%\end{equation}
%Hence in this approximation that we call $LO+NNLL$ 
%using also Eq.(\ref{changebasisin}) we have 
%\bea
%D_g^{+}(0,Q^2_0)U=& D_g(0,Q^2_0),~~ D_g^{-}(0,Q^2_0)= 0,\nonumber \\
%D_s^{+}(0,Q^2_0)&=& \frac{C_F}{C_A} D_g^{+}(0,Q^2_0) = 
%\frac{C_F}{C_A} D_g(0,Q^2_0),~~ D_s^{-}(0,Q^2_0)= D_s(0,Q^2_0) -
%\frac{C_F}{C_A} D_g(0,Q^2_0).
%\label{onlytwo}
%\eea

It would, of course, be desirable to include higher-order corrections in
Eqs.~(\ref{lonnll}).
However, this is highly nontrivial because the general perturbative structures
of the functions $H^{\pm}_a(\omega,\mu^2)$ and $Z_{\pm\mp,a}(\omega,a_s)$, which
would allow us to resum those higher-order corrections, are presently unknown.
Fortunatly, some approximations can be made.
On the one hand, it is well-known that the plus components by themselves
represent the dominant contributions to both the 
%average 
gluon and quark AJMs
%jet multiplicities 
(see, e.g., Ref.~\cite{Schmelling:1994py} for the gluon case and
Ref.~\cite{Dremin:2000ep} for the quark case).
On the other hand, Eq.~(\ref{rmin}) tells us that $D^-_g(0,Q^2)$ is suppressed
with respect to $D^-_q(0,Q^2)$ because $\alpha_\omega\sim 1+\mathcal{O}(\omega)$.
These two observations suggest that keeping $r_-(Q^2)=0$ also beyond LO should
represent a good approximation.
Nevertheless, we shall explain below how to obtain the first nonvanishing
contribution to $r_-(Q^2)$.
Furthermore, we notice that higher-order corrections to $H^{\pm}_a(0,Q^2)$ and 
$\tilde{D}^\pm_a(0,Q_0^2)$ just represent redefinitions of $D^\pm_a(0,Q_0^2)$ by
constant factors apart from running-coupling effects.
Therefore, we assume that these corrections can be neglected.

Note that the resummation of the $\pm\pm$ components was performed similarly
to Eq.~(\ref{rengroupexp}) for the case of parton distribution functions in
Ref.~\cite{Q2evo}.
Such resummations are very important because they reduce the $Q^2$ dependences
of the considered results at fixed order in perturbation theory by properly
taking into account terms that are potentially large in the limit
$\omega \to 0$ \cite{Illarionov:2004nw,Cvetic:2009kw}.
We anticipate similar properties in the considered case, too, which is in line
with our approximations.
Some additional support for this may be obtained from $\mathcal{N}=4$ super
Yang-Mills theory, where the diagonalization can be performed exactly in any
order of perturbation theory because the coupling constant and the
corresponding martices for the diagonalization do not depended on $Q^2$.
Consequently, there are no $Z_{\pm\mp,a}^{(k)}(\omega)$ terms, and only
$P_{\pm\pm}^{(k)}(\omega)$ terms contribute to the integrand of the RG exponent.
Looking at the r.h.s.\ of Eqs.~(\ref{renfact}) and (\ref{pertfun}), we indeed
observe that the corrections of $\mathcal{O}(a_s)$ would cancel each other if
the  coupling constant were scale independent.

We now discuss higher-order corrections to $r_+(Q^2)$.
As already mentioned above, we introduced in Ref.~\cite{Bolzoni:2012ed} an
effective approach to perform the resummation of the first Mellin moment of the
plus component of the anomalous dimension.
In that approach, resummation is performed by taking the fixed-order plus
component and substituting $\omega=\omega_\mathrm{eff}$, where
$\omega_\mathrm{eff}$ is given in Eq.~(\ref{replacement}).
We now show that this approach is exact to $\mathcal{O}(\sqrt{a_s})$.
We indeed recover Eq.~(\ref{llgamma0}) by substituting
$\omega=\omega_\mathrm{eff}$ in the leading singular term of the LO splitting
function $P_{++}(\omega,a_s)$,
\begin{equation}
P^\mathrm{LO}_{++}(\omega)=\frac{4C_Aa_s}{\omega}+\mathcal{O}(\omega^0).
\end{equation}
We may then also substitute $\omega=\omega_\mathrm{eff}$ in
Eq.~(\ref{rmin}) before taking the limit in $\omega=0$.
Using also Eq.~(\ref{motivation}), we thus find
\begin{equation}
r_+(Q^2)=\frac{C_A}{C_F}\left[1-\frac{\sqrt{2a_s(Q^2) C_A}}{3}\left(
1+2\frac{T_F}{C_A}
-4\frac{C_F T_F}{C_A^2}\right)\right]+\mathcal{O}(a_s),
\label{rplusll}
\end{equation}
which coincides with the result obtained by Mueller in
Ref.~\cite{Mueller:1983cq}.
For this reason and because, in Ref.~\cite{Dremin:1999ji}, the 
%average 
gluon and quark AJMs
%jet multiplicities 
evolve with only one RG exponent, we inteprete
the result in Eq.~(5) of Ref.~\cite{Capella:1999ms} as higher-order
corrections to Eq.~(\ref{rplusll}).
Complete analytic expressions for all the coefficients of the expansion through
$\mathcal{O}(a_s^{3/2})$ may be found in Appendix~1 of
Ref.~\cite{Capella:1999ms}.
This interpretation is also explicitely confirmed in Chapter 7 of
Ref.~\cite{Dokshitzer:1991wu} through $\mathcal{O}(a_s)$.

Since we showed that our approach reproduces exact analytic results at
$\mathcal{O}(\sqrt{a_s})$, we may safely apply it to predict the first
non-vanishing correction to $r_-(Q^2)$ defined in Eq.~(\ref{rmin}), which
yields
\begin{equation}
r_-(Q^2)=-\frac{4 T_F}{3}\sqrt{\frac{2 a_s(Q^2)}{C_A}} +\mathcal{O}(a_s).
\label{frminus}
\end{equation} 
However, contributions beyond $\mathcal{O}(\sqrt{\alpha_s})$ obtained in this
way cannot be trusted, and further investigation is required.
Therefore, we refrain from considering such contributions here.

For the reader's convenience, we list here expressions with numerical
coefficients for $r_+(Q^2)$ through $\mathcal{O}(a_s^{3/2})$ and for $r_-(Q^2)$
through $\mathcal{O}(\sqrt{a_s})$ in QCD with $n_f=5$:
\begin{eqnarray}
r_{+}(Q^2)&=&2.25-2.18249\,\sqrt{a_s(Q^2)}
-27.54\,a_s(Q^2)
+10.8462\,a_s^{3/2}(Q^2)+\mathcal{O}(a_s^2),\label{dreminscaleplus}\\
r_{-}(Q^2)&=&-2.72166\,\sqrt{a_s(Q^2)}+\mathcal{O}(a_s).
\label{dreminscaleminus}
\end{eqnarray}

We denote the approximation in which
Eqs.~(\ref{nnllresult})--(\ref{nnllresultm}) and (\ref{lonnll}) are used as
$\mathrm{LO}+\mathrm{NNLL}$, the improved approximation in which the
expression for $r_+(Q^2)$ in Eq.~(\ref{lonnll}) is replaced by
Eq.~(\ref{dreminscaleplus}), i.e.\ Eq.~(5) in Ref.~\cite{Capella:1999ms}, as
$\mathrm{N}^3\mathrm{LO}_\mathrm{approx}+\mathrm{NNLL}$, and our best
approximation in which, on top of that, the expression for $r_-(Q^2)$ in
Eq.~(\ref{lonnll}) is replaced by Eq.~(\ref{dreminscaleminus}) as
$\mathrm{N}^3\mathrm{LO}_\mathrm{approx}+\mathrm{NLO}+\mathrm{NNLL}$. 
We shall see in the next Section,
%~\ref{analysis}, 
where we compare with the
experimental data and extract the strong-coupling constant, that the latter
two approximations are actually very good and that the last one yields the
best results, as expected.

In all the approximations considered here, we may summarize our main
theoretical results for the 
%average 
gluon and quark AJMs
%jet multiplicities 
in the following way:
\begin{eqnarray}
\langle n_h(Q^2)\rangle_g&=&n_1(Q_0^2)\hat{T}_+^\mathrm{res}(0,Q^2,Q_0^2)
+n_2(Q_0^2)\,r_-(Q^2)\hat{T}_-^\mathrm{res}(0,Q^2,Q_0^2),
\nonumber\\
\langle n_h(Q^2)\rangle_s&=&n_1(Q_0^2)\frac{\hat{T}_+^\mathrm{res}(0,Q^2,Q_0^2)}
{r_+(Q^2)}+n_2(Q_0^2)\,\hat{T}_-^\mathrm{res}(0,Q^2,Q_0^2),
\label{quarkgen}
\end{eqnarray}
where 
%\begin{eqnarray}
\begin{equation}
n_1(Q_0^2) = r_+(Q_0^2)
\frac{D_g(0,Q_0^2)-r_-(Q_0^2)D_s(0,Q_0^2)}{r_+(Q_0^2)-r_-(Q_0^2)},~~
%\nonumber\\
n_2(Q_0^2) = \frac{r_+(Q_0^2)D_s(0,Q_0^2)-D_g(0,Q_0^2)}{r_+(Q_0^2)-r_-(Q_0^2)}.
\label{n2}
\end{equation}
%\end{eqnarray}
The 
%average 
gluon-to-quark AJM
%jet multiplicity 
ratio may thus be written as
\begin{equation}
r(Q^2)\equiv\frac{\langle n_h(Q^2)\rangle_g}{\langle n_h(Q^2)\rangle_s}
=r_+(Q^2)\left[\frac{1+r_-(Q^2)R(Q_0^2)
\hat{T}_{-}^\mathrm{res}(0,Q^2,Q^2_0)/\hat{T}_{+}^\mathrm{res}(0,Q^2,Q^2_0)}
{1+r_+(Q^2)R(Q_0^2)\hat{T}_{-}^\mathrm{res}(0,Q^2,Q^2_0)/
\hat{T}_{+}^\mathrm{res}(0,Q^2,Q^2_0)}\right],
\label{ratiogen}
\end{equation}
where 
\begin{equation}
R(Q_0^2)=\frac{n_2(Q_0^2)}{n_1(Q_0^2)}.
\end{equation}
It follows from the definition of $\hat{T}^\mathrm{res}_\pm(0,Q^2,Q_0^2)$ in
Eq.~(\ref{nnllresult}) and from Eq.~(\ref{n2}) that, for $Q^2=Q_0^2$,
Eqs.~(\ref{quarkgen}) and (\ref{ratiogen}) become
\begin{equation}
\langle n_h(Q_0^2)\rangle_g=D_g(0,Q_0^2),\qquad
\langle n_h(Q_0^2)\rangle_q=D_s(0,Q_0^2),\qquad
r(Q_0^2)=\frac{D_g(0,Q_0^2)}{D_s(0,Q_0^2)}.
\label{incond}
\end{equation}
These represent the initial conditions for the $Q^2$ evolution at an arbitrary
initial scale $Q_0$.
In fact, Eq.~(\ref{quarkgen}) is independ of $Q_0^2$, as
may be observed by noticing that
\begin{equation}
\hat{T}_\pm^\mathrm{res}(0,Q^2,Q_0^2)=\hat{T}_\pm^\mathrm{res}(0,Q^2,Q_1^2)
\hat{T}_\pm^\mathrm{res}(0,Q_1^2,Q_0^2),
\end{equation}
for an arbitrary scale $Q_1$ (see also Ref.~\cite{Bolzoni:2012cv} for a
detailed discussion of this point).

In the approximations with $r_-(Q^2)=0$ \cite{Bolzoni:2012ii}, i.e.\ the
$\mathrm{LO}+\mathrm{NNLL}$ and
$\mathrm{N}^3\mathrm{LO}_\mathrm{approx}+\mathrm{NNLL}$ ones, our general results
in Eqs.~(\ref{quarkgen}), and (\ref{ratiogen}) collapse to
\begin{eqnarray}
\langle n_h(Q^2)\rangle_g&=&D_g(0,Q_0^2)\hat{T}_+^\mathrm{res}(0,Q^2,Q_0^2),
\nonumber\\
\langle n_h(Q^2)\rangle_s&=&D_g(0,Q_0^2)
\frac{\hat{T}_+^\mathrm{res}(0,Q^2,Q_0^2)}{r_+(Q^2)}
+\left[D_s(0,Q_0^2)-\frac{D_g(0,Q_0^2)}{{r_+(Q_0^2)}}\right]
\hat{T}_-^\mathrm{res}(0,Q^2,Q_0^2),
\nonumber\\
r(Q^2)  &=& 
\frac{r_{+}(Q^2)}{\left[1 + \frac{r_{+}(Q^2)}{r_{+}(Q_0^2)}\left(
\frac{D_s(0,Q^2_0)r_{+}(Q_0^2)}{D_g(0,Q^2_0)}
-1 \right)
\frac{\hat{T}_{-}^\mathrm{res}(0,Q^2,Q^2_0)}{\hat{T}_{+}^\mathrm{res}(0,Q^2,Q^2_0)}\right]} .
\end{eqnarray}

%We note that the $Q^2$ dependence of $\gamma_0$ in Eq.~(\ref{llgamma0}) is
%entirely generated via $a_s$ according to Eq.~(\ref{running}).
The NNLL-resummed expressions for the 
%average 
gluon and quark AJMs
%jet multiplicites
given by Eq.~(\ref{quarkgen}) only depend on two nonperturbative constants,
namely $D_g(0,Q^2_0)$ and $D_s(0,Q^2_0)$.
These allow for a simple physical interpretation.
In fact, according to Eq.~(\ref{incond}), they are the average gluon and quark
jet multiplicities at the arbitrary scale $Q_0$. 
We should also mention that identifying the quantity $r_+(Q^2)$ with the one
computed in Ref.~\cite{Capella:1999ms}, we assume the scheme dependence to be
negligible.
This should be justified because of the scheme independence through NLL  
established in Ref.~\cite{Albino:2011cm}.

We note that the $Q^2$ dependence of our results is always generated via
$a_s(Q^2)$ according to Eq.~(\ref{alpha}).
This allows us to express Eq.~(\ref{nnllresult}) entirely in terms of
$\alpha_s(Q^2)$.
In fact, substituting the QCD values for the color factors and choosing
$n_f=5$ in the formulae given in Refs.~\cite{Bolzoni:2012ii,Bolzoni:2013rsa}, we may write at
NNLL
\be
\hat{T}_{\pm}^\mathrm{res}(Q^2,Q_0^2)=
\frac{T_{\pm}^\mathrm{res}(Q^2)}{T_{\pm}^\mathrm{res}(Q^2_0)},~~
T_{-}^\mathrm{res}(Q^2)= \alpha^{d_1}_s(Q^2),~~
T_{+}^\mathrm{res}(Q^2)= 
\exp\left[\frac{d_2+d_3\alpha_s(Q^2)}{\sqrt{\alpha_s(Q^2)}}\right]
\alpha^{d_4}_s(Q^2),
\ee
where
\begin{equation}
d_1=0.38647,\qquad
d_2=2.65187,\qquad
d_3=-3.87674,\qquad
d_4=0.97771. 
\end{equation}

\begin{figure}[t]
\includegraphics[height=0.55\textheight,width=0.95\textwidth]{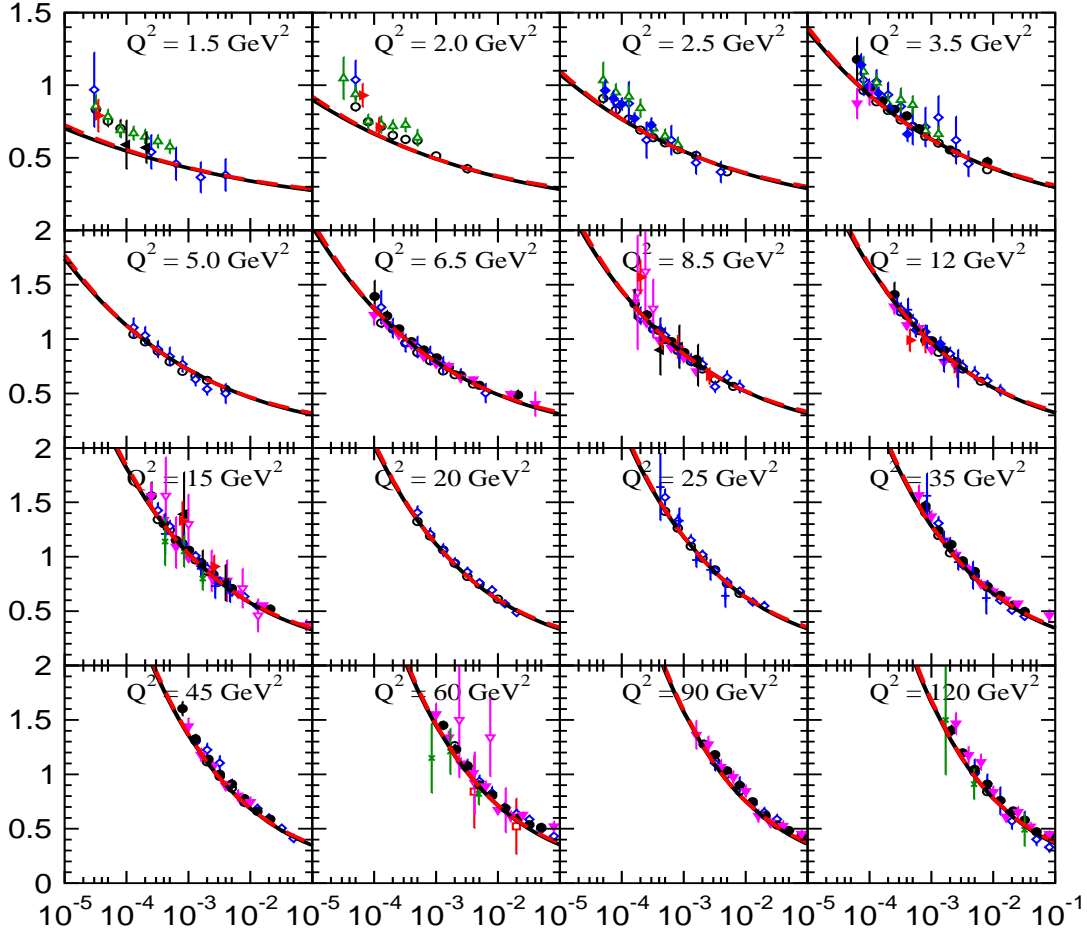}
\vskip 1.3cm
\caption{$F_2(x,Q^2)$ as a function of $x$ for different $Q^2$ bins. 
The experimental points are from H1 \cite{H197} (open points) and ZEUS 
\cite{ZEUS01} (solid points) at 
$Q^2 \geq 1.5$ GeV$^2$.
The solid curve represents the NLO fit. The dashed curve (hardly 
distinguishable 
from the solid one) represents the LO fit.}
\label{fig1}
\end{figure}

\section{Comparison with experimental data 
%for SF $F_2$.
% and the slope $\lambda_{\rm F_2}$
%Results of the fits
} \indent

Here we compare our formulae with experimental data for DIS SF $F_2(x,Q^2)$ and for the AJMs.
%average jet multiplicities. 
In the DIS case, we limite ourselves by consideration only
the SF $F_2(x,Q^2)$. The comparison of the generalized DAS approach predictions with the data
for the slope $\partial \ln F_2/\partial \ln(1/x)$  \cite{Surrow} and for the heavy parts of 
$F_2$ \cite{Collaboration:2009jy} can be found in Refs. \cite{KoPa02,Cvetic:2009kw} and 
\cite{Illarionov:2011km}, respectively (see also the review \cite{Kotikov:2012ad}). An estimation
of the cross-sections of very high-energy neutrino and nucleon scattering has been found in
\cite{Illarionov:2011wc}.

\subsection{ DIS SF $F_2$}

Using the results of section 3 we have
analyzed  HERA data for $F_2$ 
%and the slope $\partial \ln F_2/\partial \ln (1/x)$
at small $x$ from the H1 and ZEUS Collaborations \cite{H197,ZEUS01,Aaron:2009aa}.
%\cite{H197,ZEUS01,Surrow,H1slo,DIS02}.

In order to keep the analysis as simple as possible,
we fix $f=4$ and $\alpha_s(M^2_Z)=0.1166 $ (i.e., $\Lambda^{(4)} = 284$ MeV) in agreement
with the more recent ZEUS results \cite{ZEUS01}.

\begin{figure}[t]
\includegraphics[height=0.55\textheight,width=0.95\textwidth]{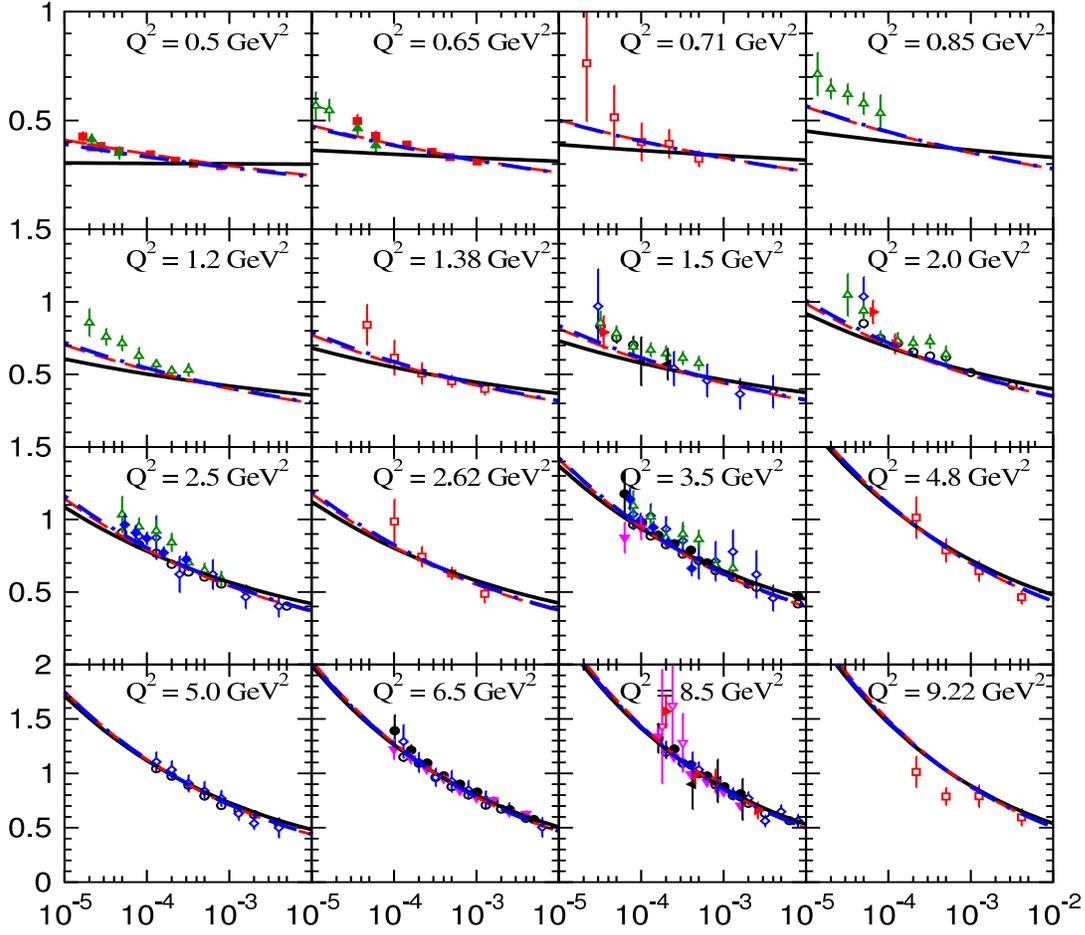}
\vskip 1.5cm
\caption{ $x$ dependence of $F_2(x, Q^2)$ in bins of $Q^2$. The experimental data from H1 (open points) 
and ZEUS (solid points) are compared with the NLO fits for $Q^2\geq 0.5$ GeV$^2$
implemented with the canonical (solid lines), frozen (dot-dashed lines), and analytic (dashed lines) 
versions of the strong-coupling constant. For comparison, also the results
obtained in Ref. \cite{HT} through a fit based on the renormalon model of higher-twist terms are shown 
(dotted lines).}
\label{fig3}
\end{figure}

%\begin{\Large}
\begin{table}
\caption{
%\label{Tab:H1+ZEUS:96/97}\sffamily
The result of the LO and NLO fits to H1 
%(1996/97) \protect\cite{Adloff:1999}
and ZEUS 
%(1996/97) \protect\cite{Chekanov:2001} 
data  for different low
$Q^2$ cuts.  In the fits $f$ is fixed to 4 flavors.
}
\centering
\footnotesize
%\small
\large
\vspace{0.3cm}
%\begin{ruledtabular}
\begin{tabular}{|l||c|c|c||r|} \hline \hline
& $A_g$ & $A_q$ & $Q_0^2~[{\rm GeV}^2]$ &
 $\chi^2 / n.o.p.$~ \\
\hline\hline
~$Q^2 \geq 1.5 {\rm GeV}^2 $  &&&& \\
 LO & 0.784$\pm$.016 & 0.801$\pm$.019 & 0.304$\pm$.003 & 754/609 \\
 LO$\&$an. & 0.932$\pm$.017 & 0.707$\pm$.020 & 0.339$\pm$.003 & 632/609  \\
  LO$\&$fr. & 1.022$\pm$.018 & 0.650$\pm$.020 & 0.356$\pm$.003 & 547/609   \\
\hline
 NLO & -0.200$\pm$.011 & 0.903$\pm$.021 & 0.495$\pm$.006 & 798/609 \\
 NLO$\&$an. & 0.310$\pm$.013 & 0.640$\pm$.022 & 0.702$\pm$.008 & 655/609  \\
  NLO$\&$fr. & 0.180$\pm$.012 & 0.780$\pm$.022 & 0.661$\pm$.007 & 669/609   \\
\hline\hline
~$Q^2 \geq 0.5 {\rm GeV}^2 $  &&&& \\
 LO & 0.641$\pm$.010 & 0.937$\pm$.012 & 0.295$\pm$.003 & 1090/662 \\
 LO$\&$an. & 0.846$\pm$.010 & 0.771$\pm$.013 & 0.328$\pm$.003 & 803/662  \\
  LO$\&$fr. & 1.127$\pm$.011 & 0.534$\pm$.015 & 0.358$\pm$.003 & 679/662   \\
\hline
 NLO & -0.192$\pm$.006 & 1.087$\pm$.012 & 0.478$\pm$.006 & 
{\color{red} 1229/662} \\
 NLO$\&$an. & 0.281$\pm$.008 & 0.634$\pm$.016 & 0.680$\pm$.007 & 
{\color{red} 633/662}  \\
  NLO$\&$fr. & 0.205$\pm$.007 & 0.650$\pm$.016 & 0.589$\pm$.006 & 
{\color{red} 670/662}   \\
\hline \hline
%\normale
\end{tabular}
%\end{ruledtabular}
\end{table}

As it is possible to see in Fig. 1 (see also \cite{Q2evo,HT}), the twist-two
approximation is reasonable at $Q^2 \geq 2$ GeV$^2$. At smaller $Q^2$, some
modification of the approximation should be considered. In Ref.
%the recent article
\cite{HT} we have added the higher twist corrections.
For renormalon model of higher twists, we
have found a good
agreement with experimental data at essentially lower $Q^2$ values:
$Q^2 \geq 0.5$ GeV$^2$ (see Figs. 4 and 5 in Ref. \cite{HT}), but we have 
added 4 additional parameters:
amplitudes of twist-4 and twist-6 corrections to quark and gluon densities.

Moreover, 
 the results of fits in \cite{HT} have an important property: they are
very similar in LO and NLO approximations of perturbation theory.
The similarity is related to the fact that the small-$x$ asymptotics of 
the NLO corrections
are usually large and negative (see, for example, $\alpha_s$-corrections 
\cite{FaLi,KoLi} to
BFKL kernel \cite{BFKL}\footnote{It seems that it is a property of 
any processes in which gluons,
but not quarks play a basic role.}).
% and 
Then, the LO form $\sim \alpha_s(Q^2)$ for
some observable and the NLO one 
$\sim \alpha_s(Q^2) (1-K\alpha_s(Q^2)) $
with a large value of $K$ are similar, because 
%usually 
$\Lambda_{\rm NLO} \gg
\Lambda_{\rm LO}$\footnote{The equality of
%similarity between 
$\alpha_s(M_Z^2)$ at LO and NLO approximations,
%and $\alpha^{\rm LO}_s(M_Z^2)$,
where $M_Z$ is the $Z$-boson mass, relates $\Lambda_{\rm NLO}$ and $\Lambda_{\rm LO}$:
$\Lambda^{(4)}_{\rm NLO} = 284$ MeV (as in \cite{ZEUS01}) corresponds to 
$\Lambda_{\rm LO} = 112$ MeV (see \cite{HT}).}
and, thus, $\alpha_s(Q^2)$ at LO is considerably smaller  then 
$\alpha_s(Q^2)$ at NLO  for HERA $Q^2$ values.

In other words, performing some resummation procedure (such as Grunberg's 
effective-charge method \cite{Grunberg}), one can see that the
%the NLO form 
results up to NLO approximation may
%can 
be represented as $\sim \alpha_s(Q^2_{\rm eff})$,
where $Q^2_{\rm eff} \gg Q^2$. 
Indeed, from 
different studies
\cite{DoShi,bfklp,Andersson},
it is well known that at small-$x$ values the effective
argument of the coupling constant is higher then $Q^2$.
As it was shown in \cite{Kotikov:2014faa},
the usage of the effective scale in the generalized DAS approach improves
the agreement with data for SF $F_2(x,Q^2)$.

Here, to improve the agreement at small $Q^2$ values without additional parameters,
we modify the QCD coupling constant.
We consider two modifications, 
which effectively increase the argument of the coupling constant 
at small $Q^2$ values (in agreement with \cite{DoShi,bfklp,Andersson}).

In one case, which is more phenomenological, we introduce freezing
of the coupling constant by changing its argument $Q^2 \to Q^2 + M^2_{\rho}$,
where $M_{\rho}$ is the $\rho $-meson mass (see \cite{Greco}). Thus, in the 
formulae of the
Section 2 we should do the following replacement:
\begin{equation}
 a_s(Q^2) \to a_{\rm fr}(Q^2) \equiv a_s(Q^2 + M^2_{\rho})
\label{Intro:2}
\end{equation}

The second possibility incorporates the Shirkov--Solovtsov idea 
\cite{ShiSo}-\cite{Cvetic}
about analyticity of the coupling constant that leads to the additional its
power dependence. Then, in the formulae of the previous section
%and \ref{Sec:3}
the coupling constant $a_s(Q^2)$ should be replaced as follows:
\begin{eqnarray}
 a^{\rm LO}_{\rm an}(Q^2) \, = \, a_s(Q^2) - \frac{1}{\beta_0}
 \frac{\Lambda^2_{\rm LO}}{Q^2 - \Lambda^2_{\rm LO}}
\label{an:LO} 
\end{eqnarray}
at the LO
%leading (LO) 
approximation and
\begin{eqnarray}
 a_{\rm an}(Q^2) \, = \, a_s(Q^2) - \frac{1}{2\beta_0}
 \frac{\Lambda^2}{Q^2 - \Lambda^2} 
+ \ldots \, ,
%- \frac{1}{\beta_0}
% \sum_{k=1}^\infty \left(\frac{\Lambda^2}{Q^2}\right)^k \, C_k[f]
\label{an:NLO}
\end{eqnarray}
at the NLO approximation,
where the symbol $\ldots$ stands for terms which have negligible
contributions
at $Q \geq 1$ GeV \cite{ShiSo}\footnote{Note that in \cite{Nesterenko,Cvetic} 
more accurate, but essentially more
cumbersome approximations of $a_{an}(Q^2)$ have been proposed.
We limit ourselves by above simple form (\ref{an:LO}), (\ref{an:NLO})
and plan to add the other modifications in our future investigations.}.

Figure~2 and Table 1 show  a strong improvement of the agreement with experimental data
for $F_2$ ($\chi^2$ values decreased almost 2 times!). 
%Similar results can be seen also in Figs. 5 and 6 for
%%shows 
%the experimental data for $\lambda_{F_2}^{\rm eff}(x,Q^2)$
%at $x\sim 10^{-3}$, which represents an average of the $x$-values of HERA experimental 
%data. The top dashed line represents the aforementioned linear rise of
%$\lambda(Q^2)$ with $\ln(Q^2)$.

\subsubsection{H1$\&$ZEUS data}

\begin{figure}[t]
\centering
\vskip 0.5cm
\includegraphics[height=0.7\textheight,width=1.0\hsize]{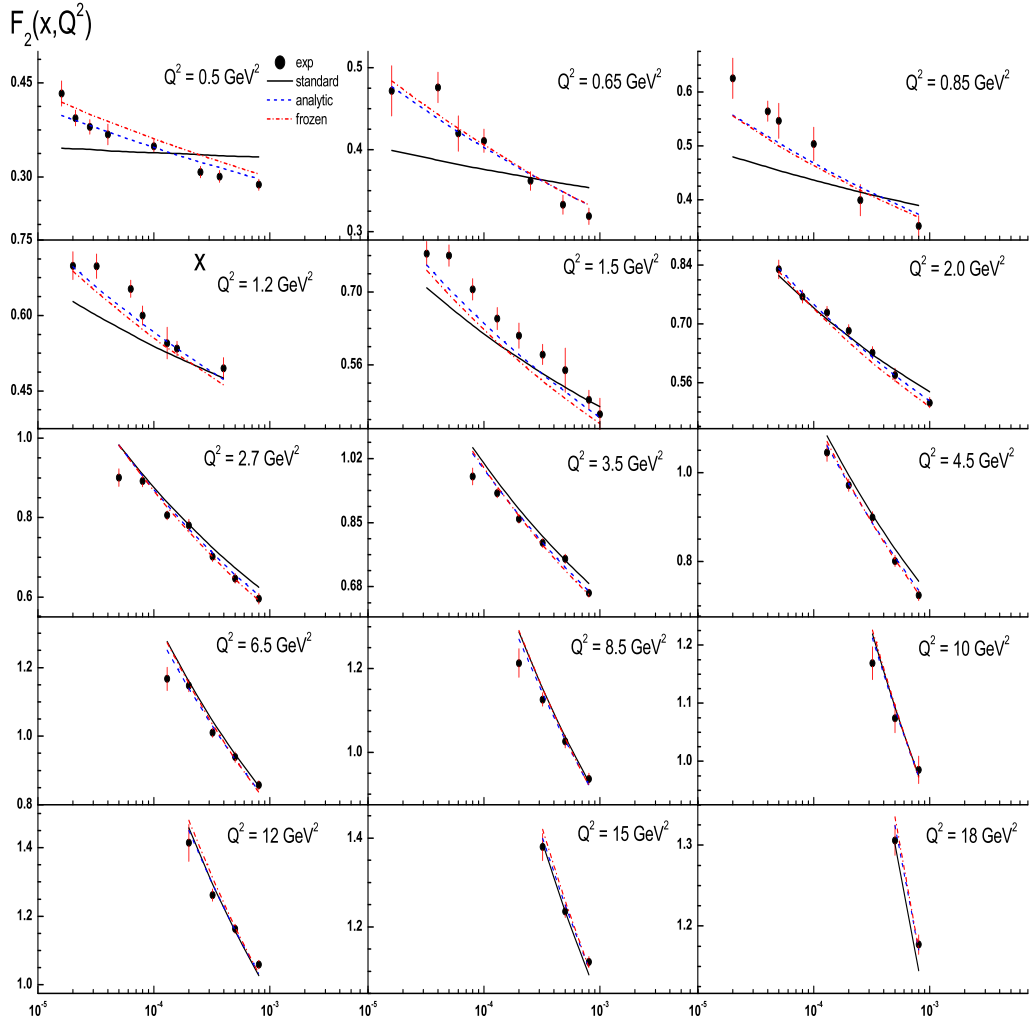}
\vskip -0.3cm
\caption{$x$ dependence of $F_2(x,Q^2)$ in bins of $Q^2$.
The combined experimental data from H1 and ZEUS Collaborations
\cite{Aaron:2009aa} are
compared with the NLO fits for $Q^2\geq0.5$~GeV$^2$ implemented with the
standard (solid lines), frozen (dot-dashed lines), and analytic (dashed lines)
versions of the strong coupling constant.}
\label{fig:F1}
\end{figure}

%By using the results of the previous section 
Here we have
analyzed the very precise H1$\&$ZEUS data for $F_2$ \cite{Aaron:2009aa}.
%% and the slope $\partial \ln F_2/\partial \ln (1/x)$ \cite{H1slo,DIS02}
%%at small $x$ from the H1 and ZEUS Collaborations.
%%
%In order to keep the analysis as simple as possible,
%we fix $f=4$ and $\alpha_s(M^2_Z)=0.1168$ (i.e., $\Lambda^{(4)} = 284$ MeV) in agreement
%with more recent ZEUS results given in~\cite{H1ZEUS}.
As can be seen from Fig.~3 and Table~2,
%(see also \cite{Q2evo,HT}),
the twist-two approximation is reasonable for $Q^2 \geq 4$ GeV$^2$.
At lower $Q^2$ we observe that the fits in the cases with ``frozen'' and
analytic strong coupling constants are very similar
(see also \cite{KoLiZo}) and describe the data in
the low $Q^2$ region significantly better than the standard fit
($\chi^2$ values decreased 2$\div$ 3 times!)
Nevertheless, for $Q^2 \leq 1.5$~GeV$^2$
%in the case of $\lambda^{\rm eff}_{F_2}(x,Q^2_0)$,
there is still some disagreement with
the data, which needs to be additionally studied.
In particular,  the BFKL
%Balitsky--Fadin--Kuraev--Lipatov (BFKL)
resummation \cite{BFKL} may be important here \cite{Kowalski:2012ur}.
It can be added in the generalized DAS approach according to the discussion
in Ref. \cite{Kotikov:2012ad}.

%\begin{\Large}
\begin{table}
\caption{
%\label{Tab:H1+ZEUS:96/97}\sffamily
The results of LO and NLO fits to  H1 $\&$ ZEUS data
%for $F_2$
\cite{Aaron:2009aa},
%H1
%%(1996/97) \protect\cite{Adloff:1999}
%and ZEUS
%%(1996/97) \protect\cite{Chekanov:2001}
with various lower cuts on $Q^2$; in the fits
the number of flavors $f$ is fixed to 4.
}
\centering
\footnotesize
%\small
%\large
\vspace{0.3cm}
%\begin{ruledtabular}
\begin{tabular}{|l||c|c|c||r|} \hline \hline
& $A_g$ & $A_q$ & $Q_0^2~[{\rm GeV}^2]$ &
 $\chi^2 / n.d.f.$~ \\
\hline\hline
~$Q^2 \geq 5 {\rm GeV}^2 $  &&&& \\
 LO & 0.623$\pm$0.055 & 1.204$\pm$0.093 & 0.437$\pm$0.022 & 1.00 \\
 LO$\&$an. & 0.796$\pm$0.059 & 1.103$\pm$0.095 & 0.494$\pm$0.024 & 0.85  \\
  LO$\&$fr. & 0.782$\pm$0.058 & 1.110$\pm$0.094 & 0.485$\pm$0.024 & 0.82   \\
\hline
 NLO & -0.252$\pm$0.041 & 1.335$\pm$0.100 & 0.700$\pm$0.044 & 1.05 \\
 NLO$\&$an. & 0.102$\pm$0.046 & 1.029$\pm$0.106 & 1.017$\pm$0.060 & 0.74  \\
  NLO$\&$fr. & -0.132$\pm$0.043 & 1.219$\pm$0.102 & 0.793$\pm$0.049 & 0.86   \\
\hline\hline
~$Q^2 \geq 3.5 {\rm GeV}^2 $  &&&& \\
 LO & 0.542$\pm$0.028 & 1.089$\pm$0.055 & 0.369$\pm$0.011 & 1.73 \\
 LO$\&$an. & 0.758$\pm$0.031 & 0.962$\pm$0.056 & 0.433$\pm$0.013 & 1.32  \\
  LO$\&$fr. & 0.775$\pm$0.031 & 0.950$\pm$0.056 & 0.432$\pm$0.013 & 1.23   \\
\hline
 NLO & -0.310$\pm$0.021 & 1.246$\pm$0.058 & 0.556$\pm$0.023 & 1.82 \\
 NLO$\&$an. & 0.116$\pm$0.024 & 0.867$\pm$0.064 & 0.909$\pm$0.330 & 1.04  \\
  NLO$\&$fr. & -0.135$\pm$0.022 & 1.067$\pm$0.061 & 0.678$\pm$0.026 & 1.27 \\
\hline \hline
~$Q^2 \geq 2.5 {\rm GeV}^2 $  &&&& \\
 LO & 0.526$\pm$0.023 & 1.049$\pm$0.045 & 0.352$\pm$0.009 & 1.87 \\
 LO$\&$an. & 0.761$\pm$0.025 & 0.919$\pm$0.046 & 0.422$\pm$0.010 & 1.38  \\
  LO$\&$fr. & 0.794$\pm$0.025 & 0.900$\pm$0.047 & 0.425$\pm$0.010 & 1.30   \\
\hline
 NLO & -0.322$\pm$0.017 & 1.212$\pm$0.048 & 0.517$\pm$0.018 & 2.00 \\
 NLO$\&$an. & 0.132$\pm$0.020 & 0.825$\pm$0.053 & 0.898$\pm$0.026 & 1.09  \\
  NLO$\&$fr. & -0.123$\pm$0.018 & 1.016$\pm$0.051 & 0.658$\pm$0.021 & 1.31   \\
\hline\hline
~$Q^2 \geq 0.5 {\rm GeV}^2 $  &&&& \\
 LO & 0.366$\pm$0.011 & 1.052$\pm$0.016 & 0.295$\pm$0.005 & 5.74 \\
 LO$\&$an. & 0.665$\pm$0.012 & 0.804$\pm$0.019 & 0.356$\pm$0.006 & 3.13  \\
  LO$\&$fr. & 0.874$\pm$0.012 & 0.575$\pm$0.021 & 0.368$\pm$0.006 & 2.96   \\
\hline
 NLO & -0.443$\pm$0.008 & 1.260$\pm$0.012 & 0.387$\pm$0.010 & 
{\color{red} 6.62} \\
 NLO$\&$an. & 0.121$\pm$0.008 & 0.656$\pm$0.024 & 0.764$\pm$0.015 & 
{\color{red} 1.84}  \\
  NLO$\&$fr. & -0.071$\pm$0.007 & 0.712$\pm$0.023 & 0.529$\pm$0.011 & 
{\color{red} 2.79} \\
\hline \hline
%\hline
%\normale
\end{tabular}
%\end{ruledtabular}
\end{table}

\subsection{%Analysis
Average multiplicity and experimendal data
}
\label{analysis}

Now we show the results in \cite{Bolzoni:2013rsa} obtained from
%We are now in a position to perform 
a global fit to the available experimental
data of our formulas in Eq.~(\ref{quarkgen}) in the
$\mathrm{LO}+\mathrm{NNLL}$,
$\mathrm{N}^3\mathrm{LO}_\mathrm{approx}+\mathrm{NNLL}$, and 
$\mathrm{N}^3\mathrm{LO}_\mathrm{approx}+\mathrm{NLO}+\mathrm{NNLL}$
approximations, so as to extract the nonperturbative constants $D_g(0,Q^2_0)$
and $D_s(0,Q^2_0)$. 
We have to make a choice for the scale $Q_0$, which, in principle, is
arbitrary.
In \cite{Bolzoni:2013rsa}, we adopted $Q_0=50$~GeV.

\begin{table}
\centering
\begin{tabular}{|c|c|c|c|}
\hline
 & $\mathrm{LO}+\mathrm{NNLL}$ &
$\mathrm{N}^3\mathrm{LO}_\mathrm{approx}+\mathrm{NNLL}$ &
$\mathrm{N}^3\mathrm{LO}_\mathrm{approx}+\mathrm{NLO}+\mathrm{NNLL}$ \\
\hline
$\langle n_h(Q_0^2)\rangle_g$ & $24.31\pm0.85$ & $24.02\pm0.36$ &
$24.17\pm 0.36$ \\
$\langle n_h(Q_0^2)\rangle_q$ & $15.49\pm0.90$ & $15.83\pm0.37$ &
$15.89\pm 0.33$ \\
$\chi_\mathrm{dof}^2$ & {\color{red} 18.09} & {\color{red} 3.71} & 
{\color{red} 2.92} \\
\hline
\end{tabular}
\caption{\footnotesize%
Fit results for $\langle n_h(Q_0^2)\rangle_g$ and $\langle n_h(Q_0^2)\rangle_q$
at $Q_0=50$~GeV with 90\% CL errors and minimum values of
$\chi_\mathrm{dof}^2$ achieved in the $\mathrm{LO}+\mathrm{NNLL}$,
$\mathrm{N}^3\mathrm{LO}_\mathrm{approx}+\mathrm{NNLL}$, and 
$\mathrm{N}^3\mathrm{LO}_\mathrm{approx}+\mathrm{NLO}+\mathrm{NNLL}$
approximations.}
\label{tab:fit}
\end{table}

The 
%average 
gluon and quark AJMs
%jet multiplicities 
extracted from experimental
data strongly depend on the choice of jet algorithm.
We adopt the selection of experimental data from Ref.~\cite{Abdallah:2005cy}
performed in such a way that they correspond to compatible jet algorithms.
Specifically, these include the gluon AJM measurements 
%of average gluon jet multiplicities 
in Refs.~\cite{Abdallah:2005cy}-\cite{Siebel:2003zz}
%\cite{Abdallah:2005cy,Nakabayashi:1997hr,Abbiendi:1999pi,Abbiendi:2004pr,Siebel:2003zz}
and quark ones
%those of average quark jet multiplicities 
in Refs.~\cite{Nakabayashi:1997hr,Kluth:2003uq},
%-\cite{Abbiendi:1999sx},
%Althoff:1983ew,Braunschweig:1989bp,Aihara:1986mv,%
%Rowson:1985xh,Derrick:1986jx,Zheng:1990iq,Abrams:1989rz,%
%Decamp:1989tf,Decamp:1991uz,Buskulic:1995xz,Barate:1996fi,Abreu:1990cc,%
%Abreu:1991yc,Adeva:1991it,Adeva:1992gv,Akrawy:1990yx,Acton:1992ry,%
%Acton:1991aa,Abreu:1998vq,Ackerstaff:1998hz,Acciarri:1995ia,Abreu:1996va,%
%Alexander:1996kh,Buskulic:1996tt,Ackerstaff:1997kk,Abreu:1997dm,%
%Abbiendi:1999sx,Abreu:2000gw}, 
which include 27 and 51 experimental data points, respectively.
The results for $\langle n_h(Q_0^2)\rangle_g$ and
$\langle n_h(Q_0^2)\rangle_q$ at $Q_0=50$~GeV together with the
$\chi_\mathrm{dof}^2$ values obtained in our $\mathrm{LO}+\mathrm{NNLL}$,
$\mathrm{N}^3\mathrm{LO}_\mathrm{approx}+\mathrm{NNLL}$, and 
$\mathrm{N}^3\mathrm{LO}_\mathrm{approx}+\mathrm{NLO}+\mathrm{NNLL}$ fits are
listed in Table~\ref{tab:fit}.
The errors correspond to 90\% CL as explained above.
All these fit results are in agreement with the experimental data.
Looking at the $\chi_\mathrm{dof}^2$ values, we observe that the qualities of
the fits improve as we go to higher orders, as they should.
The improvement is most dramatic in the step from
$\mathrm{LO}+\mathrm{NNLL}$ to
$\mathrm{N}^3\mathrm{LO}_\mathrm{approx}+\mathrm{NNLL}$, where the errors on
$\langle n_h(Q_0^2)\rangle_g$ and $\langle n_h(Q_0^2)\rangle_q$ are more than
halved.
The improvement in the step from
$\mathrm{N}^3\mathrm{LO}_\mathrm{approx}+\mathrm{NNLL}$ to 
$\mathrm{N}^3\mathrm{LO}_\mathrm{approx}+\mathrm{NLO}+\mathrm{NNLL}$, albeit
less pronounced, indicates that the inclusion of the first correction to
$r_-(Q^2)$ as given in Eq.~(\ref{frminus}) is favored by the experimental data.
We have verified that the values of $\chi_\mathrm{dof}^2$ are insensitive to
the choice of $Q_0$, as they should. 
Furthermore, the central values converge in the sense that the shifts in the
step from $\mathrm{N}^3\mathrm{LO}_\mathrm{approx}+\mathrm{NNLL}$ to 
$\mathrm{N}^3\mathrm{LO}_\mathrm{approx}+\mathrm{NLO}+\mathrm{NNLL}$ are
considerably smaller than those in the step from $\mathrm{LO}+\mathrm{NNLL}$ to
$\mathrm{N}^3\mathrm{LO}_\mathrm{approx}+\mathrm{NNLL}$ and that, at the same
time, the central values after each step are contained within error bars before
that step.
In the fits presented so far, the strong-coupling constant was taken to be the
central value of the world avarage, $\alpha_s^{(5)}(m_Z^2)=0.1184$
\cite{Beringer:1900zz}.
In the next Section,
%~\ref{coupling}, 
we shall include $\alpha_s^{(5)}(m_Z^2)$ among the
fit parameters.

\begin{figure}
\centering
\includegraphics[width=0.85\textwidth]{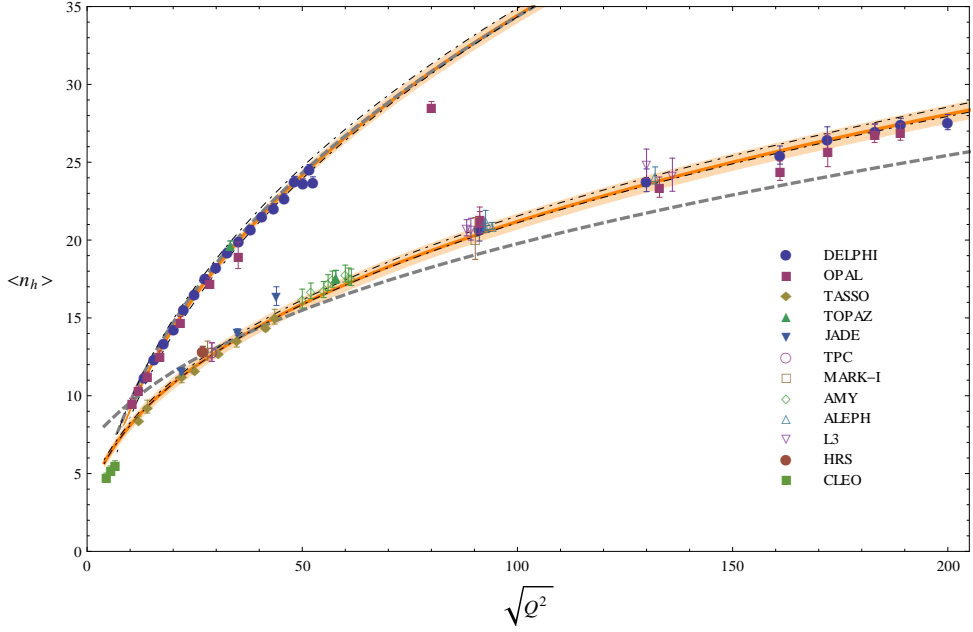}
\caption{
\footnotesize{
The average gluon (upper curves) and quark (lower curves) jet multiplicities
evaluated from Eq. (5.10),
%(\ref{quarkgen}), 
respectively, 
in the $\mathrm{LO}+\mathrm{NNLL}$ (dashed/gray lines) and
$\mathrm{N}^3\mathrm{LO}_\mathrm{approx}+\mathrm{NLO}+\mathrm{NNLL}$ 
(solid/orange lines) approximations using the corresponding fit results for
$\langle n_h(Q_0^2)\rangle_g$ and $\langle n_h(Q_0^2)\rangle_q$ from
%%Table~\ref{tab:fit} 
Table 3 are compared with the experimental data included in the
fits.
The experimental and theoretical uncertainties in the
$\mathrm{N}^3\mathrm{LO}_\mathrm{approx}+\mathrm{NLO}+\mathrm{NNLL}$ results
are indicated by the shaded/orange bands and the bands enclosed between the
dot-dashed curves, respectively.}
}
\label{Fig:plotmult}
\end{figure}

In Fig.~\ref{Fig:plotmult}, we show as functions of $Q$ the 
%average 
gluon and quark AJMs 
%jet multiplicities 
evaluated from Eq.~(\ref{quarkgen}) at
$\mathrm{LO}+\mathrm{NNLL}$ and 
$\mathrm{N}^3\mathrm{LO}_\mathrm{approx}+\mathrm{NLO}+\mathrm{NNLL}$ using the
corresponding fit results for $\langle n_h(Q_0^2)\rangle_g$ and
$\langle n_h(Q_0^2)\rangle_q$ at $Q_0=50$~GeV from Table~\ref{tab:fit}.
For clarity, we refrain from including in Fig.~\ref{Fig:plotmult} the
$\mathrm{N}^3\mathrm{LO}_\mathrm{approx}+\mathrm{NNLL}$ results, which are very
similar to the 
$\mathrm{N}^3\mathrm{LO}_\mathrm{approx}+\mathrm{NLO}+\mathrm{NNLL}$ ones
already presented in Ref.~\cite{Bolzoni:2012ii}.
In the $\mathrm{N}^3\mathrm{LO}_\mathrm{approx}+\mathrm{NLO}+\mathrm{NNLL}$
case, Fig.~\ref{Fig:plotmult} also displays two error bands, namely the
experimental one induced by the 90\% CL errors on the respective fit parameters
in Table~\ref{tab:fit} and the theoretical one, which is evaluated 
%from Eqs.~(\ref{shiftA}) and (\ref{dreminscale}) 
by varying the scale parameter
%$\xi$ in the range $1/4\le\xi\le4$.
between $Q/2$ and $2Q$.

\begin{figure}
\centering
\includegraphics[width=0.85\textwidth]{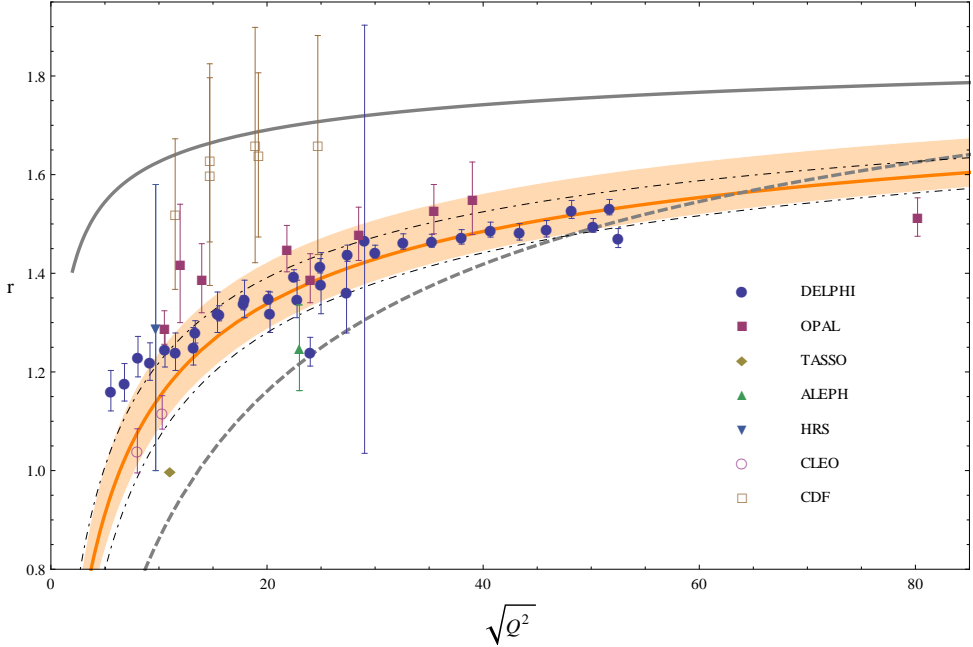}
\caption{\footnotesize{%
The average gluon-to-quark jet multiplicity ratio evaluated from Eq. (5.12)
%(\ref{ratiogen}) 
in the $\mathrm{LO}+\mathrm{NNLL}$ (dashed/gray lines) and 
$\mathrm{N}^3\mathrm{LO}_\mathrm{approx}+\mathrm{NLO}+\mathrm{NNLL}$ 
(solid/orange lines) approximations using the corresponding fit results for
$\langle n_h(Q_0^2)\rangle_g$ and $\langle n_h(Q_0^2)\rangle_q$ from Table 3
%\ref{tab:fit} 
are compared with experimental data.
The experimental and theoretical uncertainties in the
$\mathrm{N}^3\mathrm{LO}_\mathrm{approx}+\mathrm{NLO}+\mathrm{NNLL}$ result are
indicated by the shaded/orange bands and the bands enclosed between the
dot-dashed curves, respectively.
The prediction given by Eq. (5.8)
%(\ref{dreminscaleplus}) 
\cite{Capella:1999ms} 
is indicated by the continuous/gray line.}}
\label{Fig:ratio}
\end{figure}

While our fits rely on individual measurements of the 
%average 
gluon and quark AJMs,
%jet multiplicities, 
the experimental literature also reports determinations of
their ratio; see
Refs.~\cite{Abreu:1999rs,Abdallah:2005cy,Abbiendi:1999pi,Siebel:2003zz,Alam:1997ht},
%\cite{Alam:1997ht}-\cite{Abbiendi:2003gh},
%,Albrecht:1991vp,Alam:1992ir,Acosta:2004js,Derrick:1985du,%
%Braunschweig:1989um,Alexander:1991ce,Acton:1993jm,OPAL:1995ab,Biebel:1996mc,%
%Buskulic:1995sw,Abreu:1995hp,Alexander:1996qr,Ackerstaff:1997xg,%
%Abbiendi:2003gh}, 
which essentially cover all the available measurements.
In order to find out how well our fits describe the latter and thus to test
the global consistency of the individual measurements, we compare in
Fig.~\ref{Fig:ratio} the experimental data on the 
%average 
gluon-to-quark AJM
%jet multiplicity 
ratio with our evaluations of Eq.~(\ref{ratiogen}) in the
$\mathrm{LO}+\mathrm{NNLL}$ and 
$\mathrm{N}^3\mathrm{LO}_\mathrm{approx}+\mathrm{NLO}+\mathrm{NNLL}$
approximations using the corresponding fit results from Table~\ref{tab:fit}.
As in Fig.~\ref{Fig:plotmult}, we present in Fig.~\ref{Fig:ratio} also the
experimental and theoretical uncertainties in the
$\mathrm{N}^3\mathrm{LO}_\mathrm{approx}+\mathrm{NLO}+\mathrm{NNLL}$ result.
%As in Figs.~\ref{Fig:gluon_unc} and \ref{Fig:quark_unc}, they are
%represented relative to the default result, with $\xi=1$, in
%Fig.~\ref{Fig:ratio_unc}.
For comparison, we include in Fig.~\ref{Fig:ratio} also the prediction of
Ref.~\cite{Capella:1999ms} given by Eq.~(\ref{dreminscaleplus}).

Looking at Fig.~\ref{Fig:ratio}, we observe that the experimental data are
very well described by the 
$\mathrm{N}^3\mathrm{LO}_\mathrm{approx}+\mathrm{NLO}+\mathrm{NNLL}$ result for
$Q$ values above 10~GeV, while they somewhat overshoot it below.
This discrepancy is likely to be due to the fact that, following
Ref.~\cite{Abdallah:2005cy}, we excluded the older data from 
Ref.~\cite{Abreu:1999rs} from our fits because they are inconsistent with the
experimental data sample compiled in Ref.~\cite{Abdallah:2005cy}.

The Monte Carlo analysis of Ref.~\cite{Eden:1998ig} suggests that the average
gluon and quark jet multiplicities should coincide at about $Q=4$~GeV.
As is evident from Fig.~\ref{Fig:ratio}, this agrees with our
$\mathrm{N}^3\mathrm{LO}_\mathrm{approx}+\mathrm{NLO}+\mathrm{NNLL}$ result
reasonably well given the considerable uncertainties in the small-$Q^2$ range
discussed above.

\begin{figure}
\centering
\includegraphics[width=0.85\textwidth]{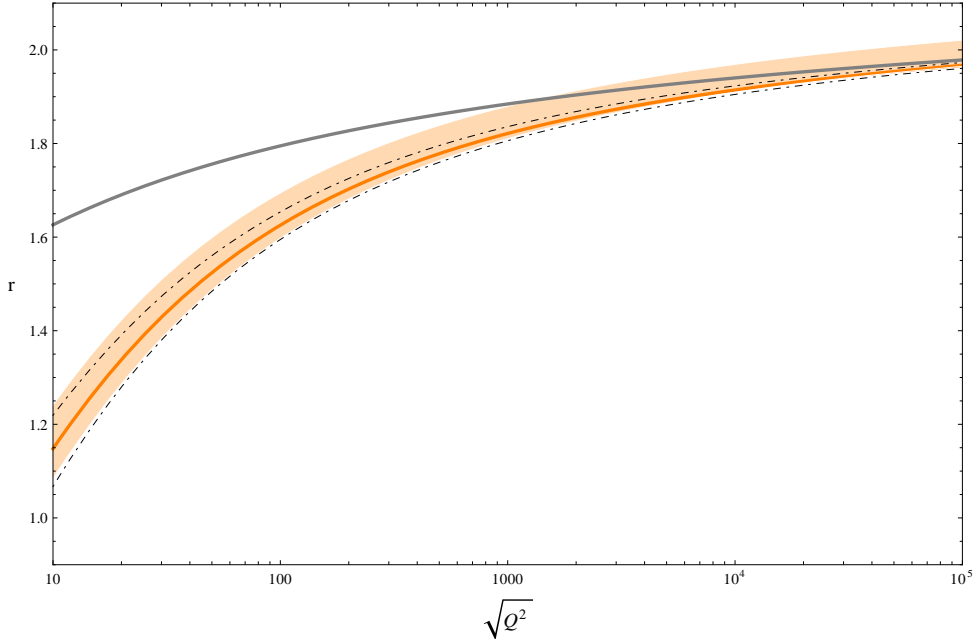}
\caption{\footnotesize{
High-$Q$ extension of Fig. 5.
%~\ref{Fig:ratio}.
}}
\label{Fig:high_en_ratio}
\end{figure}

As is obvious from Fig.~\ref{Fig:ratio}, the approximation of $r(Q^2)$ by
$r_+(Q^2)$ given in Eq.~(\ref{dreminscaleplus}) \cite{Capella:1999ms} leads to
a poor approximation of the experimental data, which reach up to $Q$ values of
about 50~GeV.
It is, therefore, interesting to study the high-$Q^2$ asymptotic behavior of
the average gluon-to-quark jet ratio.
This is done in Fig.~\ref{Fig:high_en_ratio}, where the
$\mathrm{N}^3\mathrm{LO}_\mathrm{approx}+\mathrm{NLO}+\mathrm{NNLL}$ result
including its experimental and theoretical uncertainties is compared with the
approximation by Eq.~(\ref{dreminscaleplus}) way up to $Q=100$~TeV.
We observe from Fig.~\ref{Fig:high_en_ratio} that the approximation
approaches the
$\mathrm{N}^3\mathrm{LO}_\mathrm{approx}+\mathrm{NLO}+\mathrm{NNLL}$ result
rather slowly.
Both predictions agree within theoretical errors at $Q=100$~TeV, which is one
order of magnitude beyond LHC energies, where they are still about 10\% below
the asymptotic value $C_A/C_F=2.25$.
%Figure~\ref{Fig:high_en_ratio} also nicely illustrates how, as a 
%consequence of
%the asymptotic freedom of QCD, the theoretical uncertainty decreases with
%increasing value of $Q^2$ and thus becomes considerably smaller than the
%experimental error.

\subsubsection{Determination of strong-coupling constant from average multiplicity} 
\label{coupling}

\begin{table}
\centering
\begin{tabular}{|c|c|c|}
\hline
 & $\mathrm{N}^3\mathrm{LO}_\mathrm{approx}+\mathrm{NNLL}$ &
$\mathrm{N}^3\mathrm{LO}_\mathrm{approx}+\mathrm{NLO}+\mathrm{NNLL}$ \\
\hline
$\langle n_h(Q_0^2)\rangle_g$ & $24.18\pm0.32$ & $24.22\pm 0.33$ \\
$\langle n_h(Q_0^2)\rangle_q$ & $15.86\pm0.37$ & $15.88\pm 0.35$ \\
$\alpha_s^{(5)}(m_Z^2)$ & $0.1242\pm0.0046$ & $0.1199\pm0.0044$ \\
$\chi_\mathrm{dof}^2$ & {\color{red} 2.84} & {\color{red} 2.85} \\
\hline
\end{tabular}
\caption{\footnotesize%
Fit results for $\langle n_h(Q_0^2)\rangle_g$ and $\langle n_h(Q_0^2)\rangle_q$
at $Q_0=50$~GeV and for $\alpha_s^{(5)}(m_Z^2)$ with 90\% CL errors and minimum
values of $\chi_\mathrm{dof}^2$ achieved in the
$\mathrm{N}^3\mathrm{LO}_\mathrm{approx}+\mathrm{NNLL}$ and 
$\mathrm{N}^3\mathrm{LO}_\mathrm{approx}+\mathrm{NLO}+\mathrm{NNLL}$
approximations.}
\label{tab:fit2}
\end{table}

In the previous Section,
%~\ref{analysis}, 
we took $\alpha_s^{(5)}(m_Z^2)$ to be a fixed input
parameter for our fits.
Motivated by the excellent goodness of our
$\mathrm{N}^3\mathrm{LO}_\mathrm{approx}+\mathrm{NNLL}$ and
$\mathrm{N}^3\mathrm{LO}_\mathrm{approx}+\mathrm{NLO}+\mathrm{NNLL}$ fits, we
now include it among the fit parameters, the more so as the fits should be
sufficiently sensitive to it in view of the wide $Q^2$ range populated by the
experimental data fitted to.
We fit to the same experimental data as before and again put $Q_0=50$~GeV. 
The fit results are summarized in Table~\ref{tab:fit2}.
We observe from Table~\ref{tab:fit2} that the results of the
$\mathrm{N}^3\mathrm{LO}_\mathrm{approx}+\mathrm{NNLL}$ \cite{Bolzoni:2012cv}
and $\mathrm{N}^3\mathrm{LO}_\mathrm{approx}+\mathrm{NLO}+\mathrm{NNLL}$ fits
for $\langle n_h(Q_0^2)\rangle_g$ and $\langle n_h(Q_0^2)\rangle_q$ are
mutually consistent.
They are also consistent with the respective fit results in
Table~\ref{tab:fit}.
As expected, the values of $\chi_\mathrm{dof}^2$ are reduced by relasing
$\alpha_s^{(5)}(m_Z^2)$ in the fits, from 3.71 to 2.84 in the
$\mathrm{N}^3\mathrm{LO}_\mathrm{approx}+\mathrm{NNLL}$ approximation and
from 2.95 to 2.85 in the
$\mathrm{N}^3\mathrm{LO}_\mathrm{approx}+\mathrm{NLO}+\mathrm{NNLL}$ one.
The three-parameter fits strongly confine $\alpha_s^{(5)}(m_Z^2)$, within an
error of 3.7\% at 90\% CL in both approximations.
The inclusion of the $r_-(Q^2)$ term has the beneficial effect of shifting
$\alpha_s^{(5)}(m_Z^2)$ closer to the world average, $0.1184\pm0.0007$
\cite{Beringer:1900zz}.
In fact, our
$\mathrm{N}^3\mathrm{LO}_\mathrm{approx}+\mathrm{NLO}+\mathrm{NNLL}$ value,
$0.1199\pm0.0044$ at 90\% CL, which corresponds to $0.1199\pm0.0026$ at 68\%
CL, is in excellent agreement with the former.
Note that similar $\alpha_s^{(5)}(m_Z^2)$ valu has been otained recently \cite{Perez-Ramos:2013eba}
in an extension of the MLLA approach.

\section{Conclusions} \indent

We have shown
%studied 
the $Q^2$-dependences of the SF
%structure function 
$F_2$ at small-$x$ values and of AJMs
%average jet multiplicities
%and $F_2^{cc}$ and of the slope 
%$\lambda^{\rm eff}_{F_2}=\partial \ln F_2/\partial \ln (1/x)$ 
 in the framework of perturbative QCD. 
We would like to stress that a good agreement wit the experimental data for the variables 
cannot be obtained without a proper consideration of the cotributions of both the ``$+$'' and ``$-$'' components.

The ``$+$'' components contain all large logarithms $\ln (1/x)$ as far as DIS SF 
$F_2$ and also for the average jet multiplicities. The large logarithms are resummed using famous BFKL approach
\cite{BFKL} in the PDF case and another famous MLLA approach \cite{Dokshitzer:1991wu} in the FF case.
\footnote{
Note, however, that in the case of DIS SF $F_2$ we use obly the first two orders of the perturbation theory
and our ``$+$'' component resum by DGLAP equation \cite{DGLAP}. The resummation leads to the Bessel-like form
of the ``$+$'' component. Including all orders of the perturbation theory should lead to a power-like form
as it was predicted in the framework of BFKL approach \cite{BFKL} (see discussion in \cite{Kotikov:2012ad}).}
Nevertheless,
the contributions of the ``$-$'' components are very important to have a good agreement with experimental 
data: they come with the additional free parameters. Moreover, the ``$-$'' components have other 
shapes to compare with the  ``$+$'' ones. For example, in the AJM case the ``$-$'' component is responsable for 
the difference in the $Q^2$-dependences of quark and gluon multiplicities. Indeed, the 
``$-$'' component gives essential contribution to the quark AJM
%average jet multiplicity 
but not to the gluon one.

In the case of DIS SF $F_2$,
our results 
%for the DIS SF $F_2$} 
%Our twist-two results 
are in very good agreement with 
%new 
precise HERA data at $Q^2 \geq 2\div 3$~GeV$^2$,
where perturbative theory can be applicable.
The application of the ``frozen'' and analytic coupling constants 
$\alpha_{\rm fr}(Q^2)$
and $\alpha_{\rm an}(Q^2)$ improves
%coupling constant $a_{fr}(Q^2)$
%(\ref{Intro:2}) 
%leads to good 
the agreement with the recent HERA data \cite{Aaron:2009aa}
%\cite{Surrow,H1slo,DIS02}
%for the slope $\lambda^{\rm eff}_{F_2}(x,Q^2)$ 
for small $Q^2$ values,
$Q^2 \geq 0.5$~GeV$^2$.

%%%%%%%%%%%%%%%%%%%%%%%%%%%%%%%%%%%%%%%%%%%%\\\\\\

Prior to our analysis in Ref.~\cite{Bolzoni:2012ii,Bolzoni:2013rsa}, experimental data on the
%average 
gluon and quark AJMs
%jet multiplicities 
could not be simultaneously
described in a satisfactory way mainly because the theoretical formalism failed
to account for the difference in hadronic contents between gluon and quark
jets, although the convergence of perturbation theory seemed to be well under
control \cite{Capella:1999ms}.
This problem was solved by including the ``$-$'' components governed by
$\hat{T}_-^\mathrm{res}(0,Q^2,Q_0^2)$ in Eqs.~(\ref{quarkgen}) and
(\ref{ratiogen}).
This was done for the first time in Ref.~\cite{Bolzoni:2012ii}.
%, albeit in connection with the LO result $r_-(Q^2)=0$.
The quark-singlet ``$-$'' component comes with an arbitrary normalization and has
a slow $Q^2$ dependence.
Consequently, its numerical contribution may be approximately mimicked by a
constant introduced to the average quark jet multiplicity as in
Ref.~\cite{Abreu:1999rs}.

Motivated by the goodness of our
%$\mathrm{N}^3\mathrm{LO}_\mathrm{approx}+\mathrm{NNLL}$ and
%$\mathrm{N}^3\mathrm{LO}_\mathrm{approx}+\mathrm{NLO}+\mathrm{NNLL}$ 
fits in \cite{Bolzoni:2012ii,Bolzoni:2013rsa}
with
fixed value of $\alpha_s^{(5)}(m_Z^2)$,
% in Ref.~\cite{Bolzoni:2012ii} and here,
we then included $\alpha_s^{(5)}(m_Z^2)$ among the fit parameters, which
yielded a further reduction of $\chi_\mathrm{dof}^2$.
The fit results are listed in Table~\ref{tab:fit2}.
%Also here, the inclusion of the NLO correction to $r_-(Q^2)$ is beneficial;
%it shifts $\alpha_s^{(5)}(m_Z^2)$ closer to the world average to become
%$0.1199\pm0.0026$.

\indent
%\vspace{1cm} \hspace{1cm} 
{\Large {} {\bf Acknowledgments} %\vspace{0.5cm}
}\\
\indent
This work was supported by 
RFBR grant 13-02-01005-a.
Author 
%A.V.K. 
thanks the Organizing Committee of 
XXII International Baldin Seminar on High Energy Physics Problems
for invitation.
% and Paolo Bolzoni for discussions.
%\section{...}

\section{Appendix A}
\label{App:A}
\def\theequation{A\arabic{equation}}
\setcounter{equation}{0}

For diagonalization of quark and gluon interaction it is neseccary
to introduce the corresponding matrix $U$, which diagonalize exactly the
LO AD
%has the following property
%We note that the Eqs.(\ref{3.31})-(\ref{3.4}) and (\ref{fu.3})-(\ref{fu.4})
%come from diagonalization of the anomalous dimesion matrix. So we can 
%rewrite Eqs. (\ref{fu.3})-(\ref{fu.4}) as 
\bea
\hat{U}^{-1} \left(\begin{array}{cc}  \gamma^{(0)}_{qq}(n)&
\gamma^{(0)}_{gq}(n) \\
\gamma^{(0)}_{qg}(n) & \gamma^{(0)}_{gg}(n) \end{array} \right)
\hat{U} ~=~ \left(\begin{array}{cc}  \gamma^{(0)}_{-}(n)&
0 \\
0 & \gamma^{(0)}_{+}(n) \end{array} \right)
~~~~~ \left(\hat{U}^{-1} \hat{U}=1 \right),
\label{A.1}
\eea
where
\bea
\hat{U} 
~=~ \left(\begin{array}{cc} \alpha_n &  \alpha_n    -1 \\ \beta_n &  \beta_n
 \end{array} \right),~~
\hat{U}^{-1} ~=~ \left(\begin{array}{cc}  
1 & \frac{1-\alpha_n}{\beta_n} \\ -1 & \frac{1\alpha_n}{\beta_n}  
\end{array} \right)
\label{A.2}
\eea
and $\alpha_n$ and $\beta_n$ have been defined in the main text, 
in Eq. (\ref{3S.4}).

At higher orders
the anomalous dimensions are transformed as follows
%in the following form
\bea
\hat{U}^{-1} \left(\begin{array}{cc}  \gamma^{(i)}_{qq}(n)&
\gamma^{(i)}_{gq}(n) \\
\gamma^{(i)}_{qg}(n) & \gamma^{(i)}_{gg}(n) \end{array} \right)
\hat{U} ~=~ \left(\begin{array}{cc}  \gamma^{(n)}_{--}(j)&
\gamma^{(i)}_{-+}(n) \\
\gamma^{(i)}_{+-}(n) & \gamma^{(i)}_{++}(n) \end{array} \right)
~,
%~~~~ \left(U^{-1} U=1 \right),
\label{A.3}
\eea
where exact representations for $\gamma^{(i)}_{\pm\pm}(n)$ and 
$\gamma^{(i)}_{\pm\mp}(n)$ were given in the main text, in Eq. (\ref{3S.7a}).

\subsection{Diagonalization of the renormalization group exponent}

Consider the renormalization group exponent (hereafter in the Appendix A
$\overline{a}_s = a_s(Q^2)$ and $a_s = a_s(Q^2_0)$)
\be
\hat{W}(a_s,\overline{a}_s) \equiv T_{a_s} \,
\exp \, \left[\int^{a_s}_{\overline{a}_s} \, \frac{da'}{a'} 
\frac{\hat{\gamma}(a')}{2\beta(a')} \right],
\label{A.4}
\ee
in the following form
\be
\hat{W}(a_s,\overline{a}_s) =  \hat{V}(a_s) \,
\exp \, \left[\frac{\hat{\gamma}^{(0)}(n)}{2\beta_0} 
\ln \frac{\overline{a}_s}{a_s} \right] \,
\hat{V}^{-1}(\overline{a}_s) \equiv  \hat{V}(a_s) \, 
\hat{W}^{(0)}(a_s,\overline{a}_s) \,
\hat{V}^{-1}(\overline{a}_s),
\label{A.5}
\ee
where the matrix $\hat{V}(a_s)$ contains high order coefficients.

To find the matrix $\hat{V}(a_s)$, it is better to find the derivation
%a differential
\be
\frac{d}{d a_s} \, \hat{W}(a_s,\overline{a}_s) \,.
\label{A.6}
\ee

The l.h.s. of (\ref{A.5}) leads to
\be
\frac{d}{d a_s} \, \hat{W}(a_s,\overline{a}_s) =
\frac{\hat{\gamma}(a_s)}{2\beta(a_s)} \,   \hat{V}(a_s) \, 
\hat{W}^{(0)}(a_s,\overline{a}_s) \,
\hat{V}^{-1}(\overline{a}_s).
\label{A.7}
\ee

For the r.h.s. of (\ref{A.5}), we have
\be
\frac{d}{d a_s} \, \hat{W}(a_s,\overline{a}_s) =
\left[\frac{d}{d a_s} \,  \hat{V}(a_s) -  \hat{V}(a_s) \, 
\frac{\hat{\gamma}^{(0)}(n)}{2\beta_0} \, \frac{1}{a_s} \right] \,
\hat{W}^{(0)}(a_s,\overline{a}_s) \,
\hat{V}^{-1}(\overline{a}_s).
\label{A.8}
\ee

Thus, the  matrix $\hat{V}(a_s)$ obeys the following equation
\be
\frac{d}{d a_s} \,  \hat{V}(a_s) + \frac{1}{a_s} \, 
\left[ \frac{\hat{\gamma}^{(0)}(n)}{2\beta_0}, \hat{V}(a_s) \right] =
\left(\frac{\hat{\gamma}(a_s)}{2\beta(a_s)} +  
\frac{\hat{\gamma}^{(0)}(n)}{2\beta_0} \, \frac{1}{a_s} \right) \,
\hat{V}(a_s) \, ,  
\label{A.9}
\ee
where the second term in the l.h.s. is the commutator of the matrices
$\hat{\gamma}^{(0)}(n)$ and $\hat{V}(a_s)$.

Now we consider LO, NLO and NNLO approximataions separately.

\subsubsection{LO}

At LO, the  matrix $\hat{V}(a_s)=I$ and the renormalization group exponent
have the form
\bea
\hat{W}(a_s,\overline{a}_s) = \hat{W}^{(0)}(a_s,\overline{a}_s)
= \left(\begin{array}{cc}  
{\left(\frac{\overline{a}_s}{a_s}\right)}^{d_-(n)} &
0 \\
0 & {\left(\frac{\overline{a}_s}{a_s}\right)}^{d_+(n)} \end{array} \right),
%~~~~~ \left(\hat{U}^{-1} \hat{U}=1 \right),
\label{A.10}
\eea
where
\be
d_{\pm}(n) = \frac{\gamma^{(0)}_{\pm}(n)}{2\beta_0}
\label{A.11}
\ee

\subsubsection{NLO}

At NLO, the  matrices $\hat{V}(a_s)$ and $\hat{V}^{-1}(a_s)$ has the form
\be
\hat{V}(a_s) = I + a_s \hat{V}^{(1)},~~ \hat{V}^{-1}(a_s) = I - 
a_s \hat{V}^{(1)},
\label{A.12}
\ee
and the Eq. (\ref{A.9}) can be replaced by one
\be
2\hat{V}^{(1)}(n) + \, 
\left[ \frac{\hat{\gamma}^{(0)}(n)}{\beta_0}, \hat{V}^{(1)}(n) \right] =
- \frac{\hat{\gamma}^{(1)}(n)}{\beta_0} +  
\frac{\hat{\gamma}^{(0)}(n)\beta_1}{\beta^2_0}
\label{A.13}
\ee

Applying the matrices $\hat{U}^{-1}$ and $\hat{U}$ to left and right sides
of above equation, respectively, and using Eq. (\ref{A.3}) for $i=1$ and
the representation 
\bea
\hat{U}^{-1} \hat{V}^{(i)}(n) \hat{U}  ~=~ 
\left(\begin{array}{cc}  V^{(i)}_{--}(n)& V^{(i)}_{-+}(n) \\
V^{(i)}_{+-}(n) & V^{(i)}_{++}(n) \end{array} \right)
~,
%~~~~ \left(U^{-1} U=1 \right),
\label{A.14}
\eea
for  $i=1$, we have the following matrix equation
\bea
%&&
 \left(\begin{array}{cc}  
2V^{(1)}_{--}(n) & V^{(1)}_{-+}(n)  \cdot
\frac{2\beta_0+\gamma^{(0)}_{-}(n)- \gamma^{(0)}_{+}(n)}{\beta_0} \\
V^{(1)}_{+-}(n) \cdot
\frac{2\beta_0+\gamma^{(0)}_{+}(n)- \gamma^{(0)}_{-}(n)}{\beta_0}  & 
2V^{(1)}_{++}(n) \end{array} \right) 
%\nonumber \\ && 
~=~ - \frac{1}{\beta_0} \,
 \left(\begin{array}{cc} \Gamma^{(1)}_{--}(n)
% - \frac{\gamma^{(1)}_{--}(j)}{\beta_0} +
% \frac{\gamma^{(0)}_{-}(j)\beta_1}{\beta^2_0}  
&   
\Gamma^{(1)}_{-+}(n) \\ \Gamma^{(1)}_{+-}(n)
%- \frac{\gamma^{(1)}_{-+}(j)}{\beta_0} \\  
%- \frac{\gamma^{(1)}_{+-}(j)}{\beta_0} 
& \Gamma^{(1)}_{++}(n)
%- \frac{\gamma^{(1)}_{++}(j)}{\beta_0} +
% \frac{\gamma^{(0)}_{+}(j)\beta_1}{\beta^2_0} 
 \end{array} \right) \, ,
%~~~~~ \left(\hat{U}^{-1} \hat{U}=1 \right) \, , 
\label{A.15}
\eea
where
\be
\Gamma^{(1)}_{\pm\pm}(n) = \gamma^{(1)}_{\pm\pm}(n) - 
\gamma^{(0)}_{\pm}(n) b_1,~~ \Gamma^{(1)}_{\pm\mp}(n) = 
\gamma^{(1)}_{\pm\mp}(n)
\label{A.16}
\ee

The Eq. (\ref{A.15})
%which 
leads to the results
\be
 V^{(1)}_{\pm\pm}(n) ~=~ - \frac{\Gamma^{(1)}_{\pm\pm}(n)}{2\beta_0},~~~
% + \frac{\gamma^{(0)}_{\pm}(j)\beta_1}{2\beta^2_0}, 
%\nonumber \\
 V^{(1)}_{\pm\mp}(n) ~=~ - 
\frac{\gamma^{(1)}_{\pm\mp}(n)}{2\beta_0+\gamma^{(0)}_{\pm}(n)-
\gamma^{(0)}_{\mp}(n)}
\label{A.17}
\ee

\subsubsection{NNLO}

At NNLO, the  matrices $\hat{V}(a_s)$ and $\hat{V}^{-1}(a_s)$ has the form
\be
\hat{V}(a_s) = I + a_s \hat{V}^{(1)} + a^2_s \hat{V}^{(2)},~~ 
\hat{V}^{-1}(a_s) = I - a_s \hat{V}^{(1)} -  a^2_s \hat{\tilde{V}}^{(2)},
~~  \hat{\tilde{V}}^{(2)} = \hat{V}^{(2)} -  \hat{V}^{(1)} \cdot  
\hat{V}^{(1)}
\label{A.18}
\ee
and the Eq. (\ref{A.9}) can be replaced by one
\be
%\bea&&
4\hat{V}^{(2)}(n) + \, 
\left[ \frac{\hat{\gamma}^{(0)}(n)}{\beta_0}, \hat{V}^{(2)}(n) \right] 
%\nonumber \\ &&
= - \frac{1}{\beta_0} \, \left[\hat{\gamma}^{(2)}(n)
+ \hat{\gamma}^{(1)}(n) \left(\hat{V}^{(1)}(n)-b_1\right)
- \hat{\gamma}^{(0)}(n) \left(b_1\hat{V}^{(1)}(n)+\tilde{b}_2\right) \right]
\label{A.19}
%\eea
\ee

Applying the matrices $\hat{U}^{-1}$ and $\hat{U}$ to left and right sides
of above equation, respectively, and using Eqs. (\ref{A.3}) and (\ref{A.14}) 
for $i=1$ and $i=2$, we have the following matrix equation
\bea
%&& 
\left(\begin{array}{cc}  
4V^{(2)}_{--}(n) & V^{(2)}_{-+}(n)  \cdot
\frac{4\beta_0+\gamma^{(0)}_{-}(n)- \gamma^{(0)}_{+}(n)}{\beta_0} \\
V^{(2)}_{+-}(n) \cdot
\frac{4\beta_0+\gamma^{(0)}_{+}(n)- \gamma^{(0)}_{-}(n)}{\beta_0}  & 
4V^{(2)}_{++}(n) \end{array} \right) 
%\nonumber \\ && 
~=~ - \frac{1}{\beta_0} \,
 \left(\begin{array}{cc} \Gamma^{(2)}_{--}(n) &   
\Gamma^{(2)}_{-+}(n) \\ \Gamma^{(1)}_{+-}(n)
& \Gamma^{(2)}_{++}(n)
 \end{array} \right) \, ,
\label{A.20}
\eea
where
\bea
\Gamma^{(2)}_{\pm\pm}(n) &=& \gamma^{(2)}_{\pm\pm}(n) 
+ \sum_{i=\pm} \gamma^{(1)}_{\pm i}(n) V^{(1)}_{i\pm}(n)
- b_1 \left( \gamma^{(1)}_{\pm\pm}(n) + \gamma^{(0)}_{\pm}(n)
V^{(1)}_{\pm\pm}(n) \right) 
%\nonumber \\ &&
- \left(b_2 - b_1^2 \right)
%\tilde{b}_2 
\gamma^{(0)}_{\pm}(n), \nonumber \\
\Gamma^{(2)}_{\pm\mp}(n) &=&  \gamma^{(2)}_{\pm\mp}(n) 
+ \sum_{i=\pm} \gamma^{(1)}_{\pm i}(n) V^{(1)}_{i\mp}(n)
- b_1 \left( \gamma^{(1)}_{\pm\mp}(n) + \gamma^{(0)}_{\pm}(n) V^{(1)}_{\pm\mp}(n) 
 \right)
\label{A.21}
\eea

The Eq. (\ref{A.20})
%which 
leads to the results
\be
 V^{(2)}_{\pm\pm}(n) ~=~ - \frac{\Gamma^{(2)}_{\pm\pm}(n)}{4\beta_0},~~~
 V^{(2)}_{\pm\mp}(n) ~=~ - 
\frac{\Gamma^{(2)}_{\pm\mp}(n)}{4\beta_0+\gamma^{(0)}_{\pm}(n)-
\gamma^{(0)}_{\mp}(n)}
\label{A.22}
\ee

Taking in brascets the relation between $\hat{\tilde{V}}^{(2)}$
and $\hat{V}^{(2)}$ given in the last relation of (\ref{A.18}), we have
\be
\tilde{V}^{(2)}_{\pm\pm}(n) ~=~ V^{(2)}_{\pm\pm}(n) - 
 \sum_{i=\pm} V^{(1)}_{\pm i}(n) V^{(1)}_{i\pm}(n),~~~
 \tilde{V}^{(2)}_{\pm\mp}(n) ~=~ V^{(2)}_{\pm\mp}(n) - 
 \sum_{i=\pm} V^{(1)}_{\pm i}(n) V^{(1)}_{i\mp}(n),
\label{A.23}
\ee

\subsection{$Q^2$ evolution of parton distributions}

In the matrix form, the $Q^2$ evolution of parton distributions 
\be
\left[{\bf f}_q(Q^2), {\bf f}_g(Q^2)\right] = \left[{\bf f}_q(Q^2_0), 
{\bf f}_g(Q^2_0)\right]
\cdot \hat W(a_s,\overline{a}_s)
\label{B.1}
\ee
can be represented in the form
\be
\left[{\bf f}_q(Q^2), {\bf f}_g(Q^2)\right] = \left[{\bf f}_q(Q^2_0), 
{\bf f}_g(Q^2_0)\right]
\hat{U} \cdot \left(\hat{U}^{-1} \hat W(a_s,\overline{a}_s)\hat{U} \right)
\hat{U}^{-1} \,.
\label{B.2}
\ee

The first part in the r.h.s. is
%\bea
\be
\left[{\bf f}_q(Q^2_0), {\bf f}_g(Q^2_0)\right] \hat{U} =
%&=&
\left[{\bf f}_q(Q^2_0)\alpha_n + {\bf f}_g(Q^2_0)\beta_n, 
{\bf f}_q(Q^2_0)\left(\alpha_n-1\right) + {\bf f}_g(Q^2_0)\beta_n \right] 
%\nonumber \\ &\equiv&
\equiv
\left[{\bf f}_S^-(Q^2_0), -{\bf f}_S^+(Q^2_0)\right]
\,,
\label{B.3}
\ee
where (see also (\ref{3S.3}) in the main text)
\be
{\bf f}_q^-(Q^2_0) ~=~ {\bf f}_q(Q^2_0)\alpha_n + {\bf f}_g(Q^2_0)\beta_n, ~~
{\bf f}_q^+(Q^2_0) ~=~ {\bf f}_q(Q^2_0)\left(1-\alpha_n\right) - 
{\bf f}_g(Q^2_0)\beta_n 
\,.
\label{B.4}
\ee

\subsubsection{LO}

At the LO, the renormalization group exponent 
$\hat{U}^{-1} \hat W(a_s,\overline{a}_s)\hat{U}$ has the diagonal form
(\ref{A.10}) and, thus, we have
\be
\left[{\bf f}_q(Q^2_0), {\bf f}_g(Q^2_0)\right] \hat{U} \cdot 
\left( \hat{U}^{-1} \hat W(a_s,\overline{a}_s)\hat{U} \right) ~=~
\left[{\bf f}_q^-(Q^2_0){\left(\frac{\overline{a}_s}{a_s}\right)}^{d_-(n)}, 
-{\bf f}_q^+(Q^2_0){\left(\frac{\overline{a}_s}{a_s}\right)}^{d_+(n)}\right]
\,,
\label{B.5}
\ee

Then, for the $Q^2$ evolution of parton distributions we have
\bea
\z
\left[{\bf f}_q(Q^2), {\bf f}_g(Q^2)\right] ~=~ 
\left[{\bf f}_q^-(Q^2_0){\left(\frac{\overline{a}_s}{a_s}\right)}^{d_-(n)}, 
-{\bf f}_q^+(Q^2_0)
{\left(\frac{\overline{a}_s}{a_s}\right)}^{d_+(n)}\right] \cdot
\left(\begin{array}{cc}  
1 & \frac{1-\alpha_n}{\beta_n} \\ -1 & \frac{\alpha_n}{\beta_n}  
\end{array} \right) \nonumber \\
\z
= \left[{\bf f}_q^-(Q^2_0){\left(\frac{\overline{a}_s}{a_s}\right)}^{d_-(n)}+ 
{\bf f}_q^+(Q^2_0){\left(\frac{\overline{a}_s}{a_s}\right)}^{d_+(n)},
{\bf f}_q^-(Q^2_0)\frac{1-\alpha_n}{\beta_n} 
{\left(\frac{\overline{a}_s}{a_s}\right)}^{d_-(n)}
{\bf f}_q^+(Q^2_0) \frac{\alpha_n}{\beta_n}
{\left(\frac{\overline{a}_s}{a_s}\right)}^{d_+(n)}\right] \nonumber \\
\z
 \equiv
\left[\sum_{i=\pm} {\bf f}_q^i(Q^2_0)
{\left(\frac{\overline{a}_s}{a_s}\right)}^{d_i(n)}, \sum_{i=\pm} 
{\bf f}_q^i(Q^2_0)
{\left(\frac{\overline{a}_s}{a_s}\right)}^{d_i(n)}\right]
\,,
\label{B.6}
\eea
where (see also (\ref{3S.3}) in the main text)
\bea
{\bf f}_g^-(Q^2_0) &=& {\bf f}_q^-(Q^2_0) \, \frac{1-\alpha_n}{\beta_n} ~=~ 
{\bf f}_q(Q^2_0)\epsilon_n + {\bf f}_g(Q^2_0)
\left(1-\alpha_n\right), \nonumber \\
{\bf f}_g^+(Q^2_0) &=& {\bf f}_q^+(Q^2_0) \, \frac{\alpha_n}{\beta_n} ~=~
 -{\bf f}_q(Q^2_0)\epsilon_n + {\bf f}_g(Q^2_0)\alpha_n 
\,,
\label{B.7}
\eea
because
\be
\epsilon_n ~=~ \frac{\alpha_n(1-\alpha_n)}{\beta_n} \, .
\label{B.7a}
\ee

\subsubsection{NLO}

At the NLO, the renormgroup exponent 
$\hat{U}^{-1} \hat W(a_s,\overline{a}_s)\hat{U}$ has the form
\be
\hat{U}^{-1} \hat W(a_s,\overline{a}_s)\hat{U} =
\left(I+a_s\hat{V}^{(1)}\right) \cdot 
\hat{U}^{-1} \hat{W}^{(0)}(a_s,\overline{a}_s)\hat{U} \cdot 
\left(I-\overline{a}_s\hat{V}^{(1)}\right)
\label{B.8}
\ee

%(\ref{A.10}) and, 
Thus, it is convenient to consider firstly the part
\be
\left[{\bf f}_q(Q^2_0), {\bf f}_g(Q^2_0)\right] \hat{U} \cdot 
\left(I+a_s\hat{V}^{(1)}\right)
\label{B.9}
\ee

Following to (\ref{B.3}) we can rewrite it as
%\bea
\be
\left[{\bf f}_q(Q^2_0), {\bf f}_g(Q^2_0)\right] \hat{U} \cdot 
\left(I+a_s\hat{V}^{(1)}\right) =
%&=&
\left[{\bf f}_q^-(Q^2_0), -{\bf f}_q^+(Q^2_0)\right] \cdot 
\left(I+a_s\hat{V}^{(1)}\right)
%\nonumber \\ &=& 
= \left[{\bf \tilde{f}}_S^-(Q^2_0), -{\bf \tilde{f}}_S^+(Q^2_0)\right] 
\,,
\label{B.10}
\ee
where (see also (\ref{3S.5}) in the main text)
\be
{\bf \tilde{f}}_q^{\pm}(n,Q^2_0) ~=~ {\bf f}_q^{\pm}(n,Q^2_0) 
\left(1+a_sV^{(1)}_{\pm\pm}(n)\right) -  {\bf f}_q^{\mp}(n,Q^2_0) 
a_s V^{(1)}_{\mp\pm,S}(n) 
\label{B.11}
\ee
and
\be
V^{(i)}_{\mp\pm,S} ~=~ V^{(i)}_{\mp\pm},~~
\tilde{V}^{(2)}_{\mp\pm,S} ~=~ \tilde{V}^{(2)}_{\mp\pm}
\label{B.11a}
\ee

We introduce notations $V^{(i)}_{\mp\pm,S}$ and $\tilde{V}^{(2)}_{\mp\pm,S}$
in (\ref{B.11a}),
because the corresponding ones in the gluon case are different 
(see Eq.(\ref{B.19a}) below).

The $\overline{a}_s$-part of (\ref{B.8}) has the form
\bea
%\z
\hat{U}^{-1} \hat{W}^{(0)}(a_s,\overline{a}_s)\hat{U} 
%\cdot 
\left(I-\overline{a}_s\hat{V}^{(1)}\right)
= \left(\begin{array}{cc}  
{\left(\frac{\overline{a}_s}{a_s}\right)}^{d_-(n)}\left(1-
\overline{a}_sV^{(1)}_{--}\right) & - \overline{a}_s
{\left(\frac{\overline{a}_s}{a_s}\right)}^{d_-(n)} V^{(1)}_{-+}
 \\  - \overline{a}_s
{\left(\frac{\overline{a}_s}{a_s}\right)}^{d_+(n)} V^{(1)}_{+-}
 & {\left(\frac{\overline{a}_s}{a_s}\right)}^{d_+(n)}\left(1-
\overline{a}_sV^{(1)}_{++}\right) \end{array} \right)
\label{B.12}
\eea

%which has the forllowing
Thus, we have
\bea
\z \left[{\bf f}_q(Q^2_0), {\bf f}_g(Q^2_0)\right] \hat{U} \cdot 
\left( \hat{U}^{-1} \hat W(a_s,\overline{a}_s)\hat{U} \right) 
\nonumber \\ \z~=~
\Biggl[{\bf \tilde{f}}_q^-(Q^2_0)
{\left(\frac{\overline{a}_s}{a_s}\right)}^{d_-(n)}
\left(1-\overline{a}_sV^{(1)}_{--}\right) 
%\nonumber \\ \z
+ \overline{a}_s 
{\bf \tilde{f}}_q^+(Q^2_0){\left(\frac{\overline{a}_s}{a_s}\right)}^{d_+(n)}
V^{(1)}_{+-}, 
%~~
\nonumber \\\z ~~~
-\left\{ {\bf \tilde{f}}_q^+(Q^2_0)
{\left(\frac{\overline{a}_s}{a_s}\right)}^{d_+(n)}
\left(1-\overline{a}_sV^{(1)}_{++}\right) + \overline{a}_s 
{\bf \tilde{f}}_q^-(Q^2_0){\left(\frac{\overline{a}_s}{a_s}\right)}^{d_-(n)}
V^{(1)}_{-+}\right\}\Biggr]
\,.
\label{B.13}
\eea

Then, to obtain the $Q^2$ evolution of parton distributions 
${\bf f}_S(Q^2)$ and ${\bf f}_G(Q^2)$ we should product the r.h.s. of 
(\ref{B.13})
on the matrix $\hat{U}^{-1}$. By analogy with the calculations at LO, we 
have for quark density
\bea
%\z 
{\bf f}_q(Q^2) &=& {\bf \tilde{f}}_q^-(Q^2_0)
{\left(\frac{\overline{a}_s}{a_s}\right)}^{d_-(n)}
\left(1-\overline{a}_sV^{(1)}_{--}\right) + \overline{a}_s 
{\bf \tilde{f}}_q^+(Q^2_0){\left(\frac{\overline{a}_s}{a_s}\right)}^{d_+(n)}
V^{(1)}_{+-} \nonumber \\
&& + {\bf \tilde{f}}_q^+(Q^2_0) 
{\left(\frac{\overline{a}_s}{a_s}\right)}^{d_+(n)}
\left(1-\overline{a}_sV^{(1)}_{++}\right) + \overline{a}_s 
{\bf \tilde{f}}_q^-(Q^2_0){\left(\frac{\overline{a}_s}{a_s}\right)}^{d_-(n)}
V^{(1)}_{-+}
\,.
\label{B.13a}
\eea

Taking together terms in the front of $(\overline{a}_s/a_s)^{d_{\pm}}$,
we have
%\bea
\be
{\bf  f}_q(Q^2) =
%&=& 
{\bf \tilde{f}}_q^-(Q^2_0)
{\left(\frac{\overline{a}_s}{a_s}\right)}^{d_-(n)}
\left(1-\overline{a}_s\left[V^{(1)}_{--}-V^{(1)}_{-+}\right] \right) 
%\nonumber \\ && 
+ {\bf \tilde{f}}_q^+(Q^2_0) 
{\left(\frac{\overline{a}_s}{a_s}\right)}^{d_+(n)}
\left(1-\overline{a}_s\left[V^{(1)}_{++}-V^{(1)}_{+-}\right]\right) 
\,,
\label{B.14}
\ee
or in more compact form
\bea
{\bf f}_q(n,Q^2) &=& \sum_{i=\pm} {\bf f}_q^i(n,Q^2), ~~ 
%\nonumber \\
 {\bf f}_q^i(n,Q^2) ~=~ {\bf \tilde{f}}_q^i(n,Q^2_0)
{\left(\frac{\overline{a}_s}{a_s}\right)}^{d_i(n)} H_q^i(n,Q^2),  \nonumber \\
H_q^{\pm}(n,Q^2) &=& 1-\overline{a}_s\left[V^{(1)}_{\pm\pm}(n)-
V^{(1)}_{\pm\mp,q}(n)\right],
%~~~  V^{(1)}_{\pm\mp,S} = V^{(1)}_{\pm\mp}
\,.
\label{B.15}
\eea
%where
%\be
%V^{(1)}_{\pm\mp,S} = V^{(1)}_{\pm\mp}
%\label{B.16}
%\ee

For gluon  density we have
\bea
{\bf f}_g(Q^2) &=& {\bf \tilde{f}}_q^-(Q^2_0) \, \frac{1-\alpha_n}{\beta_n} \,
{\left(\frac{\overline{a}_s}{a_s}\right)}^{d_-(n)}
\left(1-\overline{a}_sV^{(1)}_{--}\right) + \overline{a}_s 
{\bf \tilde{f}}_q^+(Q^2_0) \, \frac{1-\alpha_n}{\beta_n} \,
{\left(\frac{\overline{a}_s}{a_s}\right)}^{d_+(n)}
V^{(1)}_{+-} \nonumber \\
&& - {\bf \tilde{f}}_q^+(Q^2_0) \, \frac{\alpha_n}{\beta_n} \,
{\left(\frac{\overline{a}_s}{a_s}\right)}^{d_+(n)}
\left(1-\overline{a}_sV^{(1)}_{++}\right) - \overline{a}_s 
{\bf \tilde{f}}_q^-(Q^2_0) \, \frac{\alpha_n}{\beta_n} \,
{\left(\frac{\overline{a}_s}{a_s}\right)}^{d_-(n)}
V^{(1)}_{-+}
\,.
\label{B.17}
\eea

Taking together terms in the front of $(\overline{a}_s/a_s)^{d_{\pm}}$,
we have
\bea
{\bf  f}_g(Q^2) &=& {\bf \tilde{f}}_q^-(Q^2_0) \, \frac{1-\alpha_n}{\beta_n} \,
{\left(\frac{\overline{a}_s}{a_s}\right)}^{d_-(n)}
\left(1-\overline{a}_s\left[V^{(1)}_{--}- \frac{\alpha_n}{\alpha_n-1}
V^{(1)}_{-+}\right] \right) 
\nonumber \\
&& - {\bf \tilde{f}}_q^+(Q^2_0) \, \frac{\alpha_n}{\beta_n} \,
{\left(\frac{\overline{a}_s}{a_s}\right)}^{d_+(n)}
\left(1-\overline{a}_s\left[V^{(1)}_{++}- \frac{\alpha_n-1}{\alpha_n}
V^{(1)}_{+-}\right]\right) 
\,,
\label{B.18}
\eea
or in more compact form
\bea
\z  {\bf f}_g(n,Q^2) ~=~ \sum_{i=\pm} {\bf f}_g^i(n,Q^2), ~~ 
%\nonumber \\
 {\bf f}_g^i(n,Q^2) ~=~ {\bf \tilde{f}}_g^i(n,Q^2_0)
{\left(\frac{\overline{a}_s}{a_s}\right)}^{d_i(n)} H_g^i(n,Q^2),  \nonumber \\
\z H_g^{\pm}(n,Q^2) = 1-\overline{a}_s\left[V^{(1)}_{\pm\pm}(n)-
V^{(1)}_{\pm\mp,g}\right](n),
\label{B.19}
\eea
where
\be
V^{(1)}_{-+,g}(n) = V^{(1)}_{-+}(n) 
\frac{\alpha_n}{\alpha_n-1},~ V^{(1)}_{+-,g}(n) = V^{(1)}_{+-}(n) 
\frac{\alpha_n-1}{\alpha_n}
%\,,
\label{B.19a}
\ee
%where 
and (see also (\ref{3S.5}) in the main text)
\bea
{\bf \tilde{f}}_g^-(n,Q^2_0) &=& {\bf \tilde{f}}_q^-(n,Q^2_0)\, 
\frac{1-\alpha_n}{\beta_n} \,
= {\bf f}_q^{-}(n,Q^2_0) \, \frac{1-\alpha_n}{\beta_n} \,
\left(1+a_sV^{(1)}_{--}(n)\right) 
%\nonumber \\ && 
-  {\bf f}_q^{+}(n,Q^2_0) a_s V^{(1)}_{+-}(n)
\, \frac{1-\alpha_n}{\beta_n} \nonumber \\ 
&=&
{\bf f}_g^{-}(n,Q^2_0) \left(1+a_sV^{(1)}_{--}(n)\right) - 
{\bf f}_g^{+}(n,Q^2_0) a_s 
V^{(1)}_{+-}(n) \, \frac{\alpha_n-1}{\alpha_n} \nonumber \\
&=&   {\bf f}_g^{-}(n,Q^2_0) \left(1+a_sV^{(1)}_{--}(n)\right) - 
{\bf f}_g^{+}(Q^2_0) a_s 
V^{(1)}_{+-,g}(n) \, , \nonumber \\
{\bf \tilde{f}}_g^+(n,Q^2_0) &=& -{\bf \tilde{f}}_q^+(n,Q^2_0)\, 
\frac{\alpha_n}{\beta_n} \,
= -{\bf f}_q^{+}(n,Q^2_0) \, \frac{\alpha_n}{\beta_n} \,
\left(1+a_sV^{(1)}_{++}9n0\right) 
%\nonumber \\ && 
+  {\bf f}_q^{-}(n,Q^2_0) a_s V^{(1)}_{-+}(n)
\, \frac{\alpha_n}{\beta_n} \nonumber \\ 
&=&
 {\bf f}_g^{+}(n,Q^2_0) \left(1+a_sV^{(1)}_{++}(n)\right) - 
{\bf f}_g^{-}(n,Q^2_0) a_s 
V^{(1)}_{-+}(n) \, \frac{\alpha_n}{\alpha_n-1} \nonumber \\
&=&   {\bf f}_g^{+}(n,Q^2_0) \left(1+a_sV^{(1)}_{++}(n)\right) 
- {\bf f}_g^{-}(n,Q^2_0) a_s 
V^{(1)}_{-+,g}(n) \, ,
%\,,
\label{B.20}
\eea
or by analogy with (\ref{B.11}), in the general form,
\be
{\bf \tilde{f}}_g^{\pm}(n,Q^2_0) = {\bf f}_g^{\pm}(n,Q^2_0) 
\left(1+a_sV^{(1)}_{\pm\pm}(n)\right) 
- {\bf f}_g^{\mp}(n,Q^2_0) a_s V^{(1)}_{\mp\pm,G}(n) \, ,
%\,,
\label{B.21}
\ee

\subsubsection{NNLO}

At the NNLO, we can perform an analysis, which is very similar to the one 
in the previous subsection for the NLO approfimation. The one difference is 
the terms $\sim a^2_s$ and $\sim \overline{a}^2_s$ for the matrices
$\hat{V}$ and $\hat{V}^{-1}$ (see Eq. (\ref{A.18})).

So, the final resuls have the form $(a=q,g)$
\bea
{\bf f}_a(n,Q^2) &=& \sum_{i=\pm} {\bf f}_a^i(n,Q^2), ~~ 
%\nonumber \\
 {\bf f}_a^i(n,Q^2) ~=~ {\bf \tilde{f}}_a^i(n,Q^2_0)
{\left(\frac{\overline{a}_s}{a_s}\right)}^{d_i(n)} H_a^i(n,Q^2),  \label{B.22} \\
%\nonumber \\
H_a^{\pm}(n,Q^2) &=& 1-\overline{a}_s\left[V^{(1)}_{\pm\pm}(n)-
V^{(1)}_{\pm\mp,a}(n)\right] -\overline{a}^2_s\left[
\tilde{V}^{(2)}_{\pm\pm}(n)-
\tilde{V}^{(2)}_{\pm\mp,a}(n)\right], \nonumber \\
{\bf \tilde{f}}_a^{\pm}(n,Q^2_0) &=& {\bf f}_a^{\pm}(n,Q^2_0) 
\left(1+a_sV^{(1)}_{\pm\pm}(n)+a^2_sV^{(2)}_{\pm\pm}(n)\right) 
%\nonumber \\ && 
- {\bf f}_a^{\mp}(n,Q^2_0) \left(a_s V^{(1)}_{\mp\pm,a}(n) 
+a^2_sV^{(2)}_{\mp\pm,a}(n)\right)
\,. ~~
%\label{B.22}
\nonumber
\eea

\subsection{$Q^2$-dependence of Mellin moments}

The $Q^2$-dependence of the singlet part $M_n^{S}(Q^2)$
of the Mellin moments can be obtained using the PDF $Q^2$-dependence
 (see the previous subsection of the Appendix)
and the relation between the
parton densities and the (singlet part of) the Mellin moments given by Eqs. (\ref{3.a}) and
(\ref{3.ab}).

Sometimes, it is convenient to obtain directly the $Q^2$-dependence of the
Mellin moments $M_n^{S}(Q^2)$. In the matrix form, it has the form (\ref{B.2})
\be
%\bea
M_n^{S}(Q^2) =
%&=&
\left[{\bf f}_q(Q^2), {\bf f}_g(Q^2)\right]  \left(\begin{array}{c}
C_q(n,\overline{a}_s) \\
C_g(n,\overline{a}_s) \end{array} \right) 
%\nonumber \\ &=& 
= \left[{\bf f}_q(Q^2_0),
{\bf f}_g(Q^2_0)\right]
\hat{U} \cdot \left(\hat{U}^{-1} \hat W(a_s,\overline{a}_s)\hat{U} \right)
\hat{U}^{-1}
 \left(\begin{array}{c}
C_q(n,\overline{a}_s) \\
C_g(n,\overline{a}_s) \end{array} \right) \, ,
\label{C.1}
%\eea
\ee
where
\bea
\hat{U}^{-1}
 \left(\begin{array}{c}
C_q(n,\overline{a}_s) \\
C_g(n,\overline{a}_s) \end{array} \right) ~=~  \left(\begin{array}{c}
C_-(n,\overline{a}_s) \\
-C_+(n,\overline{a}_s) \end{array} \right)
\label{C.2}
\eea
and
\be
C_{\pm}(n,\overline{a}_s)~=~ 1 + \overline{a}_s B^{(1)}_{\pm}(n) +
\overline{a}^2_s B^{(2)}_{\pm}(n)
\label{C.3}
\ee
with
\bea
B^{(i)}_{+}(n) &=& B_{q}^{(i)}(n)-\frac{\alpha_n}{\beta_n} \,
B_{g}^{(i)}(n)\, , ~~~(i=1,2)
\nonumber \\
B^{(i)}_{-}(n) &=& B_{q}^{(i)}(n)+\frac{1-\alpha_n}{\beta_n} \, B_{g}^{(i)}(n)
~=~ B_{+}^{(i)}(n) + \frac{1}{\beta_n} \, B_{g}^{(i)}(n) \,  ; \label{C.4a} 
%B_{+}^{(2)}(n) &=& B_{q}^{(2)}(n)-\frac{\alpha_n}{\beta_n} \,
%B_{g}^{(2)}(n)\, ,
%\nonumber \\
%B_{-}^{(2)}(n) &=& B_{q}^{(2)}}(n)+\frac{1-\alpha_n}{\beta_n} \,
%B_{g}_{NNLO}(n) ~=~ B^{+}_{NNLO}(n) + \frac{1}{\beta_n} \,
%B^{g}_{NNLO}(n) \, . ~~~~ \label{C.4b}
\eea

The basic idea is to split the $Q^2_0$-dependence to the initial conditions
${\bf f}_q^{\pm}(Q^2_0)$ (\ref{B.4}) and above LO to
${\bf \tilde{f}}_q^{\pm}(n,Q^2_0)$ (\ref{B.11}) and (\ref{B.22}). The
$Q^2$-dependence combines the PDF one from the previous subsection and
the one in (\ref{C.3}) and (\ref{C.4a}). As it was above, we will consider LO, NLO
and NNLO cases separately.

Mote here that  the
Mellin moments $M_n^{S}(Q^2_0)$ can be easy extracted from above equation
 (\ref{C.1})
\bea
M_n^{S}(Q^2_0) &=&
\left[{\bf f}_q(Q^2_0), {\bf f}_g(Q^2_0)\right]  \left(\begin{array}{c}
C_q(n,a_s) \\
C_g(n,a_s) \end{array} \right)
%\nonumber \\
~=~ \left[{\bf f}_q(Q^2_0),
{\bf f}_g(Q^2_0)\right]
\hat{U} \cdot \hat{U}^{-1}
 \left(\begin{array}{c}
C_q(n,a_s) \\
C_g(n,a_s) \end{array} \right) \nonumber \\
 &=&
\left[{\bf f}_q^-(Q^2_0), -{\bf f}_q^+(Q^2_0)\right]  \left(\begin{array}{c}
C_-(n,a_s) \\
-C_+(n,a_s) \end{array} \right)
~=~ \sum_{i=\pm} \, {\bf f}_q^i(Q^2_0)C_i(n,a_s)\, ,
\label{C.5}
\eea
where $C_i(n,a_s)$ are given by (\ref{C.3}).

\subsubsection{LO}

Here
\be
M_n^{S}(Q^2) ~=~
\left[{\bf f}_q^-(Q^2), {\bf f}_q^+(Q^2)\right]
\, \left(\begin{array}{cc}
{\left(\frac{\overline{a}_s}{a_s}\right)}^{d_-(n)} & 0
 \\ 0
 & {\left(\frac{\overline{a}_s}{a_s}\right)}^{d_+(n)} \end{array} \right)
\, \left(\begin{array}{c}
1 \\
-1  \end{array} \right) \nonumber \\
~=~ \sum_{i=\pm} \, M_n^{(S,i)}(Q^2) \, ,
\label{C.6}
\ee
where
\be
M_n^{(S,\pm)}(Q^2) ~=~ {\bf f}_q^i(Q^2_0)
{\left(\frac{\overline{a}_s}{a_s}\right)}^{d_{\pm}(n)}
\label{C.7}
\ee

\subsubsection{NLO}

Here we have Eq. (\ref{B.8})
\be
\hat{U}^{-1} \hat W(a_s,\overline{a}_s)\hat{U} =
\left(I+a_s\hat{V}^{(1)}\right) \cdot
\hat{U}^{-1} \hat{W}^{(0)}(a_s,\overline{a}_s)\hat{U} \cdot
\left(I-\overline{a}_s\hat{V}^{(1)}\right) \, ,
%\label{B.8}
\nonumber
\ee
which leads to $Q_0^2$-part  (\ref{B.10}))
\be
\left[{\bf f}_q(Q^2_0), {\bf f}_g(Q^2_0)\right] \hat{U} \cdot
\left(I+a_s\hat{V}^{(1)}\right)
~=~ \left[{\bf \tilde{f}}_q^-(Q^2_0), -{\bf \tilde{f}}_q^+(Q^2_0)\right]
\,,
%\label{B.10}
\nonumber
\ee
%where (see also (\ref{3S.5}) in the main text)
with ${\bf \tilde{f}}_q^{\pm}(n,Q^2_0)$ given by (\ref{B.11}).

The $Q^2$-dependent part consists from
\be
\hat{U}^{-1} \hat{W}^{(0)}(a_s,\overline{a}_s)\hat{U}
%\cdot
\left(I-\overline{a}_s\hat{V}^{(1)}\right)
\nonumber
\ee
given by the r.h.s. of Eq. (\ref{B.12}) and
$\hat{U}^{-1}\hat{C}(n,\overline{a}_s)$ give by the Eqs. (\ref{C.2}) with the
NLO coefficints (\ref{C.3}).

So, we have
\bea
%\z
&&\hat{U}^{-1} \hat{W}^{(0)}(a_s,\overline{a}_s)\hat{U}
%\cdot
\left(I-\overline{a}_s\hat{V}^{(1)}\right) \hat{U}^{-1}\hat{C}(n,\overline{a}_s)
\nonumber \\
&&= \left(\begin{array}{cc}
{\left(\frac{\overline{a}_s}{a_s}\right)}^{d_-(n)}\left(1-
\overline{a}_sV^{(1)}_{--}\right) & - \overline{a}_s
{\left(\frac{\overline{a}_s}{a_s}\right)}^{d_-(n)} V^{(1)}_{-+}
 \\  - \overline{a}_s
{\left(\frac{\overline{a}_s}{a_s}\right)}^{d_+(n)} V^{(1)}_{+-}
 & {\left(\frac{\overline{a}_s}{a_s}\right)}^{d_+(n)}\left(1-
\overline{a}_sV^{(1)}_{++}\right) \end{array} \right)
\, \left(\begin{array}{c}
1+ \overline{a}_s B_{-}^{(1)}\\
-\Bigl(1+ \overline{a}_s B_{+}^{(1)}\Bigr)  \end{array} \right) \nonumber \\
&&=\left(\begin{array}{c}
{\left(\frac{\overline{a}_s}{a_s}\right)}^{d_-(n)} \left(
1+ \overline{a}_s \left[B_{-}^{(1)}-V^{(1)}_{--}+V^{(1)}_{-+}\right]\right)\\
-{\left(\frac{\overline{a}_s}{a_s}\right)}^{d_+(n)}
\left(1+ \overline{a}_s \left[B_{+}^{(1)}-V^{(1)}_{++}+V^{(1)}_{+-}\right]\right)
 \end{array} \right)
\label{C.8}
\eea

For the Mellin moments $M_n^S(Q^2)$ we have product $Q^2$ and $Q_0^2$ parts:
\be
M_n^{S}(Q^2) ~=~ \sum_{i=\pm} \, M_n^{(S,i)}(Q^2) \, ,
\label{C.9}
\ee
where
\be
M_n^{(S,\pm)}(Q^2) ~=~ {\bf  \tilde{f}}_q^i(Q^2_0)
{\left(\frac{\overline{a}_s}{a_s}\right)}^{d_{\pm}(n)}
\left(1+ \overline{a}_s R_{\pm}^{(1)}
%\left[B^{+}_{NLO}-V^{(1)}_{++}+V^{(1)}_{+-}\right]
\right)
\label{C.10}
\ee
and
\be
R_{\pm}^{(1)} ~=~ B_{\pm}^{(1)} -V^{(1)}_{\pm \pm}+V^{(1)}_{\pm \mp}
\label{C.11}
\ee

Note that the eqs (\ref{C.10})--(\ref{C.11}) can be rewriten as
\be
M_n^{(S,\pm)}(Q^2) ~=~ M_n^{(S,\pm)}(Q^2_0)
{\left(\frac{\overline{a}_s}{a_s}\right)}^{d_{\pm}(n)}
\frac{\left(1+ \overline{a}_s R_{\pm}^{(1)}\right)}{\left(1+
a_s R_{\pm}^{(1)}\right)} \, ,
\label{C.12}
\ee
where
\bea
M_n^{(S,\pm)}(Q^2_0) &=& {\bf  \tilde{f}}_q^{\pm}(Q^2_0)
\left(1+ a_s
\left[B_{\pm}^{(1)}-V^{(1)}_{\pm \pm}+V^{(1)}_{\pm \mp}\right]\right)
\nonumber \\
&=&  {\bf f}_q^{\pm}(Q^2_0)
\left(1+ a_s
\left[B_{\pm}^{(1)}+V^{(1)}_{\pm \mp}\right]\right)
+ {\bf f}_S^{\mp}(Q^2_0) a_s V^{(1)}_{\mp \pm}
\label{C.13}
\eea

\subsubsection{NNLO}

Repeating the calculations in the previous case, we have
\be
M_n^{S}(Q^2) ~=~ \sum_{i=\pm} \, M_n^{(S,i)}(Q^2) \, ,
\label{C.14}
\ee
where
\be
M_n^{(S,\pm)}(Q^2) ~=~ {\bf  \tilde{f}}_q^i(Q^2_0)
{\left(\frac{\overline{a}_s}{a_s}\right)}^{d_{\pm}(n)}
\left(1+ \overline{a}_s R_{\pm}^{(1)}+ \overline{a}^2_s R_{\pm}^{(2)}
%\left[B^{+}_{NLO}-V^{(1)}_{++}+V^{(1)}_{+-}\right]
\right) \, ,
\label{C.15}
\ee
where ${\bf  \tilde{f}}_q^i(Q^2_0)$ and $R_{\pm}^{(1)}$ are given by (\ref{B.22})
and (\ref{C.11}), respectively, and
\be
R_{\pm}^{(2)} ~=~ B_{\pm}^{(2)} - B_{\pm}^{(1)}\Bigl(V^{(1)}_{\pm \pm}-
V^{(1)}_{\pm \mp}\Bigr)
-\tilde{V}^{(2)}_{\pm \pm}+\tilde{V}^{(2)}_{\pm \mp}
\label{C.16}
\ee
with $\tilde{V}^{(2)}_{\pm \pm}$ and $\tilde{V}^{(2)}_{\pm \mp}$ given by Eqs.
(\ref{A.23}).

The eqs (\ref{C.15})--(\ref{C.16}) can be rewriten as
\be
M_n^{(S,\pm)}(Q^2) ~=~ M_n^{(S,\pm)}(Q^2_0)
{\left(\frac{\overline{a}_s}{a_s}\right)}^{d_{\pm}(n)}
\frac{\left(1+ \overline{a}_s R_{\pm}^{(1)}+ \overline{a}^2_s R_{\pm}^{(2)}
\right)}{\left(1+
a_s R_{\pm}^{(1)}+ a^2_s R_{\pm}^{(2)}\right)} \, ,
\label{C.17}
\ee
where
\bea
&& M_n^{(S,\pm)}(Q^2_0) = {\bf  \tilde{f}}_q^{\pm}(Q^2_0)
\left(1+ a_s R_{\pm}^{(1)}+ a^2_s R_{\pm}^{(2)}\right)
%\nonumber \\
\label{C.18} \\
&&=  {\bf f}_q^{\pm}(Q^2_0)
\left(1+ a_s \left[B_{\pm}^{(1)}+V^{(1)}_{\pm \mp}\right]
+ a_s^2\left[B_{\pm}^{(2)}+B_{\pm}^{(1)} V^{(1)}_{\pm \mp}+
\tilde{V}^{(2)}_{\pm \mp}\right]\right) 
%\nonumber \\ && 
+ {\bf f}_S^{\mp}(Q^2_0) \left[a_s V^{(1)}_{\mp \pm}+ a_s^2
\tilde{V}^{(2)}_{\pm \mp}\right]
%\label{C.18}
\nonumber 
\eea

\end{document}